\documentclass[ejs,preprint]{imsart}

\usepackage{amsmath}
\RequirePackage[numbers]{natbib}
\RequirePackage[colorlinks,citecolor=blue,urlcolor=blue]{hyperref}

\doi{10.1214/154957804100000000}
\pubyear{0000}
\volume{0}
\firstpage{1}
\lastpage{1}

\startlocaldefs
\usepackage{amssymb}
\usepackage{amsfonts}
\usepackage[hyperref,thmmarks]{ntheorem} 

\usepackage{graphicx}
\usepackage[labelsep=newline,font=small,labelfont=bf,singlelinecheck=false]{caption}

\usepackage{xspace}

\usepackage{dsfont}
\usepackage{subcaption}
\usepackage{enumitem}

\newcommand{\parsign}[1]{#1^{\text{\tiny P}}} 

\newcommand{\mydim}{k} 
\newcommand{\Mydim}{j-1} 
\newcommand{\myidx}{k} 
\newcommand{\Myidx}{n} 
\newcommand{\oneToMyIdx}{1:\Myidx}

\newcommand{\vineDim}{d}
\newcommand{\idxIterator}{e}
\newcommand{\idxset}{{\cal E}_{1}^d}
\newcommand{\idxsett}{{\cal E}_{2}^d}

\newcommand{\iGroup}{l}
\newcommand{\nGroups}{L}
\newcommand{\partElement}{\Lambda}

\newcommand{\iPartition}{m}
\newcommand{\nPartitions}{M}

\newcommand{\pvc}[1]{#1^{\text{\tiny PVC}}} 
\newcommand{\pvcI}[2]{#1^{\text{\tiny PVC},#2}}
\newcommand{\pvcIdx}[1]{\pvcI{#1}{\myidx}}

\newcommand{\svc}[1]{#1^{\text{\tiny SVC}}}
\newcommand{\svcI}[2]{#1^{\text{\tiny SVC},#2}}
\newcommand{\svcIdx}[1]{\svcI{#1}{\myidx}}

\newcommand{\indu}{\ensuremath{h_{e}}}
\newcommand{\indv}{\ensuremath{i_{e}}}
\newcommand{\indk}{\ensuremath{l_{e}}}
\newcommand{\Pitu}{U_{\indu|\condset}}
\newcommand{\Pitv}{U_{\indv|\condset}}
\newcommand{\pitu}{u_{\indu|\condset}}
\newcommand{\pitv}{u_{\indv|\condset}}
\newcommand{\Ppitu}{\pvc{U}_{\indu|\condset}}
\newcommand{\Ppitv}{\pvc{U}_{\indv|\condset}}
\newcommand{\PpituIdx}{\pvcIdx{U}_{\indu|\condset}}
\newcommand{\PpitvIdx}{\pvcIdx{U}_{\indv|\condset}}
\newcommand{\ppitu}{\pvc{u}_{\indu|\condset}}
\newcommand{\ppitv}{\pvc{u}_{\indv|\condset}}
\newcommand{\CondVar}{U_{\condset}}
\newcommand{\condVar}{u_{\condset}}

\newcommand{\R}{\mathbb{R}}

\newcommand{\cs}{;\kern 0.08em}
\newcommand{\ps}{;\kern 0.08em}
\newcommand{\condset}{D_{e}}
\newcommand{\condsetDvine}{i+1:i+j-1}

\newcommand{\ivb}[1]{\mathds{1}_{\{#1\}}} 

\newcommand{\Cov}{\mathrm{Cov}}
\newcommand{\Corr}{\mathrm{Corr}}
\newcommand{\Var}{\mathrm{Var}}

\renewcommand{\P}{\mathbb{P}}
\newcommand{\expec}{\mathbb{E}}
\newcommand{\PVCA}{partial vine copula approximation}
\renewcommand{\PVCA}{PVC}

\newcommand{\SA}{simplifying assumption}

\newcommand{\indep}{\perp}

\newcommand{\plim}[1]{\mathrm{plim}_{n\to\infty}} 

\newcommand{\domainOfPi}{(0,1)^{L}}

\newcommand{\anuncop}{an unconditional copula}
\newcommand{\uncops}{unconditional copulas}

\newcommand{\condcops}{bivariate conditional distribution functions with uniform margins}

\renewcommand{\uncops}{bivariate distribution functions with uniform margins}
\renewcommand{\anuncop}{a bivariate distribution function with uniform margins}
\newcommand{\uncopstwo}{bivariate (unconditional) distribution functions with uniform margins}

\renewcommand{\condcops}{bivariate conditional copulas}

\renewcommand{\uncops}{pair-copulas}
\renewcommand{\anuncop}{a pair-copula}
\renewcommand{\uncopstwo}{pair-copulas}

\DeclareMathOperator*{\supp}{supp}

\let\orgautoref\autoref

         \providecommand{\appref}
{\def\subsectionautorefname{Appendix}%
         \orgautoref}
\renewcommand{\autoref}
        {\def\mydefautorefname{Definition}%
         \def\figureautorefname{Figure}%
         \def\subfigureautorefname{Figure}%
         \def\sectionautorefname{Section}%
         \def\subsectionautorefname{Section}%
         \def\mypropautorefname{Proposition}%
         \def\mycorautorefname{Corollary}%
         \def\subtableautorefname{Table}%
         \def\tableautorefname{Table}%
         \def\myexautorefname{Example}%
         \def\myalgoautorefname{Algorithm}%
         \def\mylemautorefname{Lemma}%
         \def\myremautorefname{Remark}%
         \def\mytheautorefname{Theorem}%
         \def\myassautorefname{Assumption}%
         \def\itemautorefname{}%
         \orgautoref}
\makeatletter
\let\@autoref=\autoref
\renewcommand*{\autoref}[2][]{\ifthenelse{\equal{#1}{}}{\@autoref{#2}}{\hyperref[#1]{\begin{NoHyper}\@autoref{#2}~\subref{#1}\end{NoHyper}}}\xspace}
\makeatother

\theoremstyle{break} 
\theorembodyfont{\itshape} 
\theoremseparator{} 
\newtheorem{myprop}{Proposition}
\newtheorem{mylem}{Lemma}
\newtheorem{mythe}{Theorem}

\theoremstyle{plain}
\theorembodyfont{\itshape}
\theoremseparator{}
\newtheorem{myproplongtitle}[myprop]{Proposition} 
\newtheorem{mythelongtitle}[mythe]{Theorem} 

\theoremstyle{break}
\theorembodyfont{\normalfont}
\theoremseparator{}
\newtheorem{myex}{Example}
\newtheorem{mydef}{Definition}
\newtheorem{myass}{Assumption}

\theoremstyle{plain}
\theorembodyfont{\normalfont}
\theoremseparator{}
\newtheorem{mydeflongtitle}[mydef]{Definition} 
\newtheorem{myasslongtitle}[myass]{Assumption} 

\theorembodyfont{\normalfont}
\theoremstyle{nonumberplain}
\theoremseparator{.}
\theoremsymbol{\ensuremath{\blacksquare}}
\newtheorem{myproof}{Proof}

\numberwithin{equation}{section}

\usepackage{array}
\usepackage{ragged2e}

\endlocaldefs

\begin{document}

\begin{frontmatter}

\title{Testing the simplifying assumption in high-dimensional vine copulas}
\runtitle{Testing the simplifying assumption}

\author{\fnms{Malte S.} \snm{Kurz}\corref{}\ead[label=e1]{malte.kurz@tum.de}}
\address{TUM School of Management, Technical University of Munich,\\ Arcisstr.\ 21, 80333 Munich, Germany.\\ \printead{e1}}

\author{\fnms{Fabian} \snm{Spanhel}\ead[label=e2]{spanhel@stat.uni-muenchen.de}}
\address{Department of Statistics, Ludwig-Maximilians-Universit{\"a}t M{\"u}nchen,\\ Akademiestr. 1, 80799 Munich, Germany.\\ \printead{e2}}

\runauthor{M.\ S.\ Kurz and F.\ Spanhel}

\begin{abstract}
Testing the simplifying assumption in high-dimensional vine copulas is a difficult task. 
Tests must be based on estimated observations and check constraints on high-dimensional distributions.
So far, corresponding tests have been limited to single conditional copulas with a low-dimensional set of conditioning variables.
We propose a novel testing procedure that is computationally feasible for high-dimensional data sets and that exhibits a power that decreases only slightly with the dimension.
By discretizing the support of the conditioning variables and incorporating a penalty in the test statistic, we mitigate the curse of dimensionality by looking for the possibly strongest deviation from the simplifying assumption. The use
of a decision tree renders the test computationally feasible for large dimensions.
We derive the asymptotic distribution of the test and analyze its finite sample performance in an extensive simulation study.
An application of the test to four real data sets is provided.

\end{abstract}


\received{\smonth{3} \syear{2021}}

\end{frontmatter}

\tableofcontents{}

\section{Introduction}
Vine copulas \citep{Joe1997, Bedford2002, Aas2009} are a popular tool to model multivariate dependence.
An extensive literature concerning vine copulas has been based on the simplifying assumption
\citep{Dissmann2013, Grothe2013, Joe2010, Kurowicka2011, Nagler2016}.
This assumption states that every conditional copula in the vine copula does not vary in its conditioning arguments \citep{HobakHaff2010}.
Multivariate distributions which can be represented as simplified vine copulas have been identified in early papers \citep{HobakHaff2010, Stoeber2013}.
More recently, the simplifying assumption has again attracted a lot of attention
\citep{Derumigny2016, Gijbels2016, Gijbels2017, Kraus2017, Mroz2020, Nagler2016, Spanhel2016}.
In the context of (bivariate) conditional copulas non- and semiparametric tests for the simplifying assumption have been developed \citep{Acar2013, Gijbels2016, Gijbels2017}.
See \citet{Derumigny2016} for a survey.
In these studies the simplifying assumption is tested for one single conditional copula with a low-dimensional conditioning vector.
The assumption is not tested for a vine copula where several conditional copulas with a possibly high-dimensional conditioning vector need to be checked.

We propose a framework for testing the simplifying assumption in high-dimensional vine copulas.
By means of the partial vine copula, we introduce a novel stochastic interpretation of the simplifying assumption which is useful for testing it in high dimensions.
We test the null hypothesis that the conditional correlation of the partial probability integral transforms associated with an edge of a vine is constant w.r.t. the conditioning variables.
A rejection of this hypothesis implies that the \SA{} can be rejected as well. 
To obtain a test which is still powerful in high dimensions, we discretize the support of the conditioning variables into a finite number of subsets and incorporate a penalty in the test statistic.
To render the test computationally feasible in high dimensions, we apply a decision tree to find the possibly largest difference in the set of conditional correlations.
The test can be applied in high dimensions which is demonstrated in real data applications and in simulation studies with up to 12-dimensional data sets.
An accompanying \texttt{R}-package \texttt{pacotest} \citep{Kurz2017} is publicly available and has  been applied to even higher-dimensional data sets \citep{Kraus2017}.
The test can be used to detect building blocks of a vine copula where the simplifying assumption does not seem to be adequate
and the estimation of a conditional copula that is varying in its conditioning variables can improve the modeling \citep{Schellhase2017}.
It can also be applied to construct new methods for the structure selection of vine copulas \citep{Kraus2017}.

The organization of the paper is as follows.
The partial vine copula and stochastic interpretations of the simplifying assumption are discussed in \autoref{secPvcStochRep}.
In \autoref{secTestPartialCopula} we present the test for constant conditional correlations
and derive its asymptotic distribution.
A decision tree algorithm for searching for the largest deviation from the \SA{} 
is proposed in \autoref{secAlgDecTree}.
An extensive analysis of the finite sample performance of the test is provided in \autoref{secSimStudy}.
In \autoref{secSeqProcedure} a hierarchical procedure to test the simplifying assumption in vine copulas is presented and applied to simulated and real data.
Concluding remarks are given in \autoref{secConclusion}.

Throughout the paper we use the following notation and assumptions.
The cdf of a $d$-dimensional random vector $X_{1:\vineDim}$ is denoted by $F_{X_{1:\vineDim}} := \mathbb{P}(X_1 \leq x_1, \ldots, X_\vineDim \leq x_\vineDim)$.
The distribution function or copula of a random vector $U_{1:\vineDim}$ with uniform margins is denoted by $F_{1:d} = C_{1:d}$.
For simplicity, we assume that all random variables are real-valued and absolutely continuous.
If $X$ and $Y$ are stochastically independent we write $X \perp Y$.
For the indicator function we use $\ivb{ A} = 1$ if $A$ is true, and $\ivb{ A} = 0$ otherwise.
$\partial_{\theta}g(\theta)$ denotes the gradient w.r.t. $\theta$ and if $h(\gamma)$ is a $d$-dimensional function then $\partial_ih(\gamma)$ is the partial derivative w.r.t. the $i$-th element.
All proofs are deferred to the Appendix.

\section{The partial vine copula and stochastic interpretations of the simplifying assumption}\label{secPvcStochRep}
In this section, we first introduce concepts that are required for the remainder of this paper. 
Thereafter, we establish a new stochastic interpretation of the \SA{} to check its validity.

\newcommand{\A}{A}
\newcommand{\B}{B}
\newcommand{\C}{\boldsymbol{Z}}
\newcommand{\ccc}{\boldsymbol{z}}
\newcommand{\fgm}[1]{#1^{\text{\tiny FGM}}}

Let $(\A, \B, \C) \sim F_{\A, \B, \C}$ be a random vector where $\A$, $\B$ are random variables and
$\C$ is a vector of dimension at least 1.
For  $\ccc\in \text{supp}(\C)$ let $F_{\A|\C}(\cdot|\ccc)$, $F_{\B|\C}(\cdot|\ccc)$, and $F_{\A, \B|\C}(\cdot, \cdot|\ccc)$ be the corresponding conditional distributions of $\A$, $\B$, and $(\A, \B)$ given $\C=\ccc$, respectively.

\pagebreak[4]
\begin{mydef}
[CPIT, bivariate conditional and partial copula] 
\label{testSA_def_conditional_copula}
\begin{enumerate}[label=(\roman*)]
\item 
$U_{\A|\C} := F_{\A|\C}(\A|\C)$ is the conditional probability integral transform (CPIT) of $\A$ w.r.t. $\C$.
\item For  $\ccc \in \supp(\C)$, the bivariate conditional copula
$C_{\A,\B\cs \C}(\cdot, \cdot \cs \ccc)$ of the bivariate conditional distribution $F_{\A,\B|\C}(\cdot, \cdot | \ccc) $ (\citet{Patton2006}) is defined as
\begin{align*}
C_{\A,\B\cs \C}(a,b \cs \ccc)  &:=
\P(U_{\A|\C} \leq a, U_{\B|\C} \leq b|\C = \ccc).
\end{align*}
\item
The bivariate partial copula  $\parsign{C}_{\A,\B\ps \C}$ of $F_{\A,\B|\C}$  \citep{Bergsma2004, Gijbels2015b, Spanhel2016b} is 
defined as 
\begin{align*}
\parsign{C}_{\A,\B\ps \C}(a,b) & := \mathbb{P}(U_{\A|\C} \leq a, U_{\B|\C} \leq b).
\end{align*}
\end{enumerate}
\end{mydef}

While the bivariate conditional copula is the conditional distribution of a pair of uniform CPITs, the partial copula is the bivariate unconditional distribution of a pair of uniform CPITs.
Sometimes, the function $C_{\A,\B\cs \C}(\cdot, \cdot \cs \ccc)$ is constant for all $\ccc \in \text{supp}(\C)$, e.g., if the distribution of $(\A, \B, \C)$ is multivariate normal. 
In this case, all conditional copulas $C_{\A,\B\cs \C}(\cdot, \cdot \cs \ccc), \ccc \in \text{supp}(\C)$, are equal to the partial copula $\parsign{C}_{\A,\B\ps \C}$. 
The bivariate conditional copula $C_{\A,\B\cs \C}(\cdot, \cdot \cs \ccc), \ccc \in \text{supp}(\C)$, 
of the bivariate conditional distribution $F_{\A,\B|\C}(\cdot, \cdot | \ccc)$ arises if one expresses the cdf $F_{\A, \B, \C}$ as follows
\begin{align}
    F_{\A, \B, \C}(a, b, \ccc) & = \int_{\boldsymbol{t}\leq \ccc} F_{\A, \B|\C}(a, b|\boldsymbol{t})dF_{\C}(\boldsymbol{t}) \notag \\
    &= \int_{\boldsymbol{t}\leq \ccc}C_{\A, \B\cs \C}\big(F_{A|\C}(a|\boldsymbol{t}), F_{B|\C}(b|\boldsymbol{t})\cs \boldsymbol{t}\big)dF_{\C}(\boldsymbol{t}). \label{decomp_cond}
\end{align}

A complete decomposition of the corresponding multivariate copula $C_{\A, \B, \C}$ into bivariate conditional copulas of bivariate conditional distributions is given by an R-vine copula. 
The underlying graphical structure of an R-vine copula is an R-vine. 

\begin{mydef}
[R-vine   --  \citet{Bedford2001}]
\label{testSA_rvinedef}
The sequence of trees
$\mathcal{V} := (T_{1}, \ldots, T_{\vineDim-1})
:= \left( (N_{1}, E_{1}), \ldots, (N_{\vineDim-1}, E_{\vineDim-1}) \right)$ is an R-vine on $\vineDim$ elements if
\begin{enumerate}[label=(\roman*)]
\item $T_{1}$ is a tree with
nodes $N_{1} = \lbrace 1, \ldots, \vineDim \rbrace$
and set of edges $E_{1}$.
\item For $j = 2, \ldots, \vineDim - 1$: $T_j$ is a tree with
nodes $N_{j} = E_{j-1}$
and set of edges $E_{j}$.
\item Proximity condition for $j = 2, \ldots, \vineDim - 1$:
If two nodes in $T_{j}$ are joined by an edge, the nodes (being edges in $T_{j-1}$) must share a common node in $T_{j-1}$.
\end{enumerate}
The complete union associated with the edge $e = \lbrace a, b \rbrace \in E_{j}$ in tree $T_j$ is defined as
\[
\mathcal{U}_{e} := \left\lbrace n \in N_1: \exists e_1 \in E_1, \ldots, \exists e_{j-1} \in E_{j-1},\text{ with } n \in e_1 \in \ldots \in e_{j-1} \in e \right\rbrace.
\]
The conditioning set $\condset$ of the edge $e = \lbrace a, b \rbrace \in E_{j}$ in tree $T_j$ is given by $\condset := \mathcal{U}_{a} \cap \mathcal{U}_{b}$ and always consists of $(j-1)$ elements.
The conditioned sets associated with the edge $e = \lbrace a, b \rbrace \in E_{j}$ in tree $T_j$ are defined as $\indu := \mathcal{U}_{a} \setminus \condset$ and $\indv := \mathcal{U}_{b} \setminus \condset$ and are by construction singleton indices.
We denote the set of all edges of the R-vine
$\mathcal{V}$
from tree $T_l$ on as
$\mathcal{E}_l^d := \bigcup_{j=l}^{d-1} E_{j}$
so that the constraint set of the R-vine $\mathcal{V}$ is given as
$\mathcal{CV} := \lbrace (\indu,\indv\cs\condset): \idxIterator \in \idxset \rbrace$.
\end{mydef}

The representation of a multivariate copula $C_{1:d}$ in terms of an R-vine copula arises if one assigns 
the corresponding bivariate copulas $C_{\indu, \indv}, e \in E_1$, of the bivariate distributions $F_{\indu, \indv}$ to the edges of the first tree and the 
corresponding bivariate conditional copulas $C_{\indu,\indv\cs\condset}, e\in E_j, j=2,\ldots, d-1$, of the bivariate conditional distributions 
$F_{\indu, \indv|\condset}$
to the edges of higher trees \citep{Bedford2001}.
Using \eqref{decomp_cond} for uniform margins and multiple times for different dimensions of $\C$ and taking all partial derivatives to get the density, the following \autoref{testSA_dvinedef} can be shown.

\begin{myproplongtitle}
\label{testSA_dvinedef}
{\bf\rmfamily
(R-vine copula representation   --  \citet{Bedford2001})
\\}
Let $\vineDim\geq 3$ and $U_{1:\vineDim}$ be a uniform random vector with cdf \mbox{$F_{1:\vineDim} = C_{1:\vineDim}$}.
Consider an R-vine structure $\mathcal{V}$ (\autoref{testSA_rvinedef}).
Define \mbox{$u_{\indk|\condset}:= F_{\indk|\condset}(u_{\indk}|u_{\condset})$} for \mbox{$\idxIterator \in \idxsett$}, $\indk=\indu,\indv$ and denote the conditional copula of $F_{\indu,\indv|\condset}(\cdot, \cdot| u_{\condset})$ by $C_{\indu,\indv\cs \condset}(\cdot, \cdot \cs u_{\condset})$ (\autoref{testSA_def_conditional_copula}).
For $e \in E_1$ we set $u_{\indk|\condset} = u_{\indk}$ with $\indk=\indu,\indv$ and \mbox{$C_{\indu,\indv\cs \condset} = C_{\indu, \indv}$}.
The density of $C_{1:d}$ can be expressed as
\begin{align*}
c_{1:\vineDim}(u_{1:\vineDim}) & = 
\prod_{\idxIterator \in \idxset} c_{\indu,\indv\cs \condset}
\big( \pitu, \pitv \cs u_{\condset}\big).
\end{align*}
\end{myproplongtitle} 

In general, the estimation of an R-vine copula density is a difficult task if the dimension is not low.
In order to simplify the modeling process and to overcome the curse of dimensionality, it is commonly assumed that the \SA{} holds for the data generating vine copula.

\begin{myasslongtitle}
\label{testSA_simplidef}
{\bf\rmfamily
(The simplifying assumption -- \citet{HobakHaff2010})
\\}
{The R-vine copula in \autoref{testSA_dvinedef} satisfies the \SA{} if
\linebreak$c_{\indu,\indv\cs \condset}
( \cdot,\cdot \cs u_{\condset})$ { does not depend on} $u_{\condset}$
for all $\idxIterator \in \idxsett$.}
\end{myasslongtitle}

\autoref{testSA_simplidef} characterizes the \SA{} in terms of restrictions that are placed on the functional form of bivariate conditional copulas of 
bivariate conditional distributions. 
That is, the \SA{} holds if each $(j+1)$-dimensional%
\footnote{For $e \in E_j$, $c_{h_e, i_e; D_e}$ is a $(j + 1)$-dimensional function because it maps $(\pitu, \pitv, u_{D_e})$ to a value of the density of the conditional distribution of $(\Pitu, \Pitv)$ given $U_{D_e} = u_{d_e}$.}
function $c_{\indu,\indv\cs \condset}(\cdot,\cdot \cs u_{\condset})$ only depends on its first two arguments, but the other $(j-1)$ arguments $u_{\condset}$ have no effect for all edges $\idxIterator \in E_{j}$, $j = 2, \ldots, \vineDim - 1$.

Note that the building blocks of the R-vine copula in \autoref{testSA_dvinedef} are determined by $C_{1:d}$. 
The other way round, we can assign arbitrary \condcops\ to the edges of an R-vine to construct a multivariate copula.
It is also possible to assign bivariate copulas to the edges of an R-vine to construct a multivariate copula. 
In this case, we call the resulting construction a simplified vine copula which is given as follows.

\begin{mydeflongtitle}
\label{testSA_svc}
{\bf\rmfamily
(Simplified vine copula (SVC) or pair-copula construction -- \citet{Joe1996}, \citet{Aas2009})
\\}
{
Let $\vineDim\geq 3$ and consider an R-vine structure $\mathcal{V}$.
For $e \in {\cal E}_{1}^d$, let $\svc{C}_{\indu, \indv\cs D_e}$ be a bivariate copula which we call \anuncop.%
\footnote{We denote the building blocks
$\svc{C}_{\indu, \indv\cs D_e}, e\in E_j, j=1,\ldots,d-1$, of a simplified vine copula as \uncops\
to avoid any confusion with the bivariate copulas $C_{\indu, \indv}, e \in E_1$, in the first tree of the vine, or the 
bivariate conditional copulas $C_{\indu, \indv\cs D_e}, e\in E_j, j=2,\ldots, d-1,$ of bivariate conditional distributions $F_{\indu, \indv|D_e}$ in higher trees,
where $F_{\indu, \indv|D_e}$ is a conditional distribution of $U_{1:d}\sim \svc{C}_{1:d}$.
} 
Let $\svc{u}_{\indk|\condset} := u_{\indk}$ for $e \in E_1, \indk=\indu,\indv$.
For $\idxIterator \in E_{j}$, $j = 2, \ldots, \vineDim - 1$, define
$\svc{u}_{\indk|\condset}$
recursively as
\begin{align}
\svc{u}_{\indk|\condset}
&:= \partial_2\svc{C}_{l_{\bar{e}}, m_{\bar{e}} \ps D_{\bar{e}}} (\svc{u}_{l_{\bar{e}}|D_{\bar{e}}},
\svc{u}_{m_{\bar{e}}|D_{\bar{e}}}), \label{eqPpit1WithSmallU}
\end{align}
with $\bar{e} \in E_{j-1}$ and
where $m_{\bar{e}} \in \condset$ is selected such that
$(l_{\bar{e}}, m_{\bar{e}} \ps D_{\bar{e}}) = 
(l_{e}, m_{\bar{e}} \ps \condset \setminus m_{\bar{e}}) \in \mathcal{CV}$.\footnote{
By the proximity condition of the R-vine $\mathcal{V}$ (\autoref{testSA_rvinedef}) there exists a $m_{\bar{e}} \in \condset$ such that either
$(l_{e}, m_{\bar{e}} \ps \condset \setminus m_{\bar{e}}) \in \mathcal{CV}$
or $(m_{\bar{e}}, l_{e} \ps \condset \setminus m_{\bar{e}}) \in \mathcal{CV}$
with $\bar{e} \in E_{j-1} = N_{j}$
and w.l.o.g. the definition in equation~\eqref{eqPpit1WithSmallU} is restricted to the first case
--
otherwise one can use the identity
$\svc{C}_{\indv,\indu\ps\condset}(a,b) = \svc{C}_{\indu,\indv\ps\condset}(b,a)$.
}
The density of the resulting simplified vine copula is given by
\begin{align*}
\svc{c}_{1:\vineDim}(u_{1:\vineDim}) &:= \prod_{\idxIterator\in\idxset}
\svc{c}_{\indu,\indv\ps \condset}(\svc{u}_{h_e|D_e}, \svc{u}_{i_e|D_e}).
\end{align*}
}
\end{mydeflongtitle}

Note that $\svc{u}_{\indk|\condset}$ for $\idxIterator \in E_{j}$, $j = 2, \ldots, \vineDim - 1$, in \autoref{testSA_svc} is a function of $(u_{l_e}, u_{D_e})$%
\footnote{It would be more explicit to write $\svc{u}_{\indk|\condset}(u_{l_e}, u_{D_e})$ instead of $\svc{u}_{\indk|\condset}$. However, the shortened notation is commonly used in the vine copula literature and simplifies the notation.
} and that the density is specified by $j(j-1)/2$ \uncopstwo\ $\svc{C}_{h_e, i_e\cs D_e}, e\in E_j, j=1, \ldots, d-1$.
From the second tree on, each of these \uncops, i.e., $\svc{C}_{h_e, i_e\cs D_e}$ for $e\in E_j, j=2, \ldots, d-1$, together with 
$(\svc{u}_{h_e|\condset}, \svc{u}_{i_e|\condset})$, specifies the conditional distribution $F_{h_e, i_e|D_e}$ of $U_{1:\vineDim} \sim \svc{C}_{1:\vineDim}$.
Moreover, $\svc{C}_{h_e, i_e\cs D_e}$ determines the bivariate conditional copula $C_{h_e, i_e\cs D_e}$ of the bivariate conditional distribution
$F_{h_e, i_e|D_e}$ which is given by
\begin{align*}
\mathbb{P}(U_{h_e|{D_e}}\leq a, U_{i_e|{D_e}} \leq b| U_{D_e} = u_{D_e}) =: C_{h_e, i_e\cs D_e}(a, b \cs u_{D_e}) &= \svc{C}_{h_e, i_e\cs D_e}(a, b).
\end{align*}

If the data generating R-vine copula does not satisfy the simplifying assumption, a simplified vine copula can be used as an approximation. 
The partial vine copula is a special simplified vine copula which minimizes the Kullback-Leibler divergence from the data generating copula in a tree-wise fashion (\citet{Spanhel2016}).
Moreover, it gives rise to a stochastic interpretation of the simplifying assumption which is useful for testing the assumption in high dimensions.
In order to define the partial vine copula (PVC) we have to construct partial probability integral transforms and higher-order partial copulas.

\pagebreak[4]
\begin{mydeflongtitle}
\label{testSA_def_hopartial}
{\bf\rmfamily
(Partial probability integral transforms and higher-order partial copulas -- \citet{Spanhel2016})
\\}
Let $\vineDim\geq 3$ and $U_{1:\vineDim}$ be a uniform random vector with cdf \mbox{$F_{1:\vineDim} = C_{1:\vineDim}$}.
Consider an R-vine structure $\mathcal{V}$ (\autoref{testSA_rvinedef}).
For $e \in E_{2}$ we define the \textbf{first-order partial copula} as
\begin{align}
\pvc{C}_{\indu,\indv; \condset}&:= \parsign{C}_{\indu,\indv; \condset}. \label{first-order}
\end{align}
Define the \textbf{partial probability integral transforms (PPITs)} as
$\pvc{U}_{\indk|\condset} := U_{\indk|\condset}$ for $e \in E_{2}$ with $\indk=\indu,\indv$
and for $\idxIterator \in E_{j}$, $j = 3, \ldots, \vineDim - 1$, as
\begin{align}
\pvc{U}_{\indk|\condset}
&:= \partial_2\pvc{C}_{l_{\bar{e}}, m_{\bar{e}} \ps D_{\bar{e}}} (\pvc{U}_{l_{\bar{e}}|D_{\bar{e}}},
\pvc{U}_{m_{\bar{e}}|D_{\bar{e}}}), \label{eqPpit1}
\end{align}
with $\bar{e} \in E_{j-1}$ and
where $m_{\bar{e}} \in \condset$ is selected such that
$(l_{\bar{e}}, m_{\bar{e}} \ps D_{\bar{e}}) = 
(l_{e}, m_{\bar{e}} \ps \condset \setminus m_{\bar{e}}) \in \mathcal{CV}$.

\noindent For $\idxIterator \in E_{j}$, $j = 3, \ldots, \vineDim - 1$, the \textbf{$\boldsymbol{(j-1)}$-th order partial copula} is defined as
\begin{align}
 \pvc{C}_{\indu,\indv\ps\condset}(a,b) &:= 
\mathbb{P}(\pvc{U}_{\indu|\condset}\leq a,\pvc{U}_{\indv|\condset}\leq b). \label{eqPpit2}
\end{align}
\end{mydeflongtitle}

Note that a first-order partial copula $\pvc{C}_{\indu,\indv\ps\condset}$ in \eqref{first-order} equals the corresponding partial copula $\parsign{C}_{\indu,\indv; \condset}$ for $e \in E_{2}$. 
However, a higher-order partial copula in \eqref{eqPpit2} is, in general, not equal to the corresponding partial copula for $\idxIterator \in E_{j}$, $j = 3, \ldots, \vineDim - 1$.
The partial vine copula is now constructed by assigning the specific first-order and higher-order partial copulas to the edges of an R-vine.

\begin{mydef}
[Partial vine copula (\PVCA) -- \citet{Spanhel2016}]

Let $\vineDim\geq 3$ and $U_{1:\vineDim}$ be a uniform random vector with cdf \mbox{$F_{1:\vineDim} = C_{1:\vineDim}$}.
Consider an R-vine structure $\mathcal{V}$ (\autoref{testSA_rvinedef}).
$\pvc{C}_{1:\vineDim}$ denotes the \textbf{partial vine copula (PVC)} of $(C_{1:\vineDim}, \mathcal{V})$ and its density is given by 
\begin{align*}
\pvc{c}_{1:\vineDim}(u_{1:\vineDim}) &:= \prod_{\idxIterator\in\idxset}
\pvc{c}_{\indu,\indv\ps \condset}(\ppitu,
\ppitv),
\end{align*}
where $\pvc{c}_{\indu,\indv\ps \condset}$ is the density of $\pvc{C}_{\indu,\indv\ps \condset}$ defined in \eqref{first-order} for $e \in E_{2}$ and \eqref{eqPpit2} for $\idxIterator \in E_{j}$, $j = 3, \ldots, \vineDim - 1$.
For $e \in E_{1}$ we set $\pvc{C}_{\indu,\indv; \condset} = C_{\indu,\indv}$.
\end{mydef}

\begin{figure}[ht]
\centering
\begin{subfigure}[b]{0.5\textwidth}
\centering
  \includegraphics[width=0.95\textwidth]{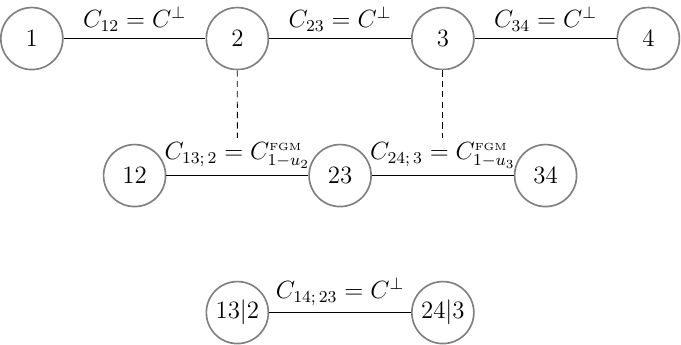}
  \label{svc}
  \caption{Vine copula \autoref{exFgm}.}
  \end{subfigure}%
\begin{subfigure}[b]{0.5\textwidth}
  \centering
  \includegraphics[width=0.95\textwidth]{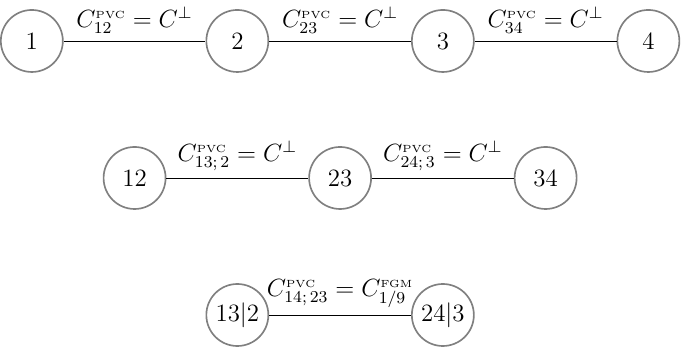}
  \label{vc}
  \caption{PVC of \autoref{exFgm}.}
\end{subfigure}
\caption{The non-simplified vine copula in \autoref{exFgm} and its PVC.
The influence of conditioning variables on the conditional copulas is indicated by dashed lines.}
\label{fig_vine}
\end{figure}

To illustrate non-simplified vine copulas, simplified vine copulas and the partial vine copula, we consider a special case of Example 4.2 in \citet{Spanhel2016}.
For this purpose, let us introduce the following notation.
If $D_\theta$ denotes a parametric copula family with dependence parameter $\theta \in \Theta$ and cdf $D_\theta(a, b)$ we write $C_{\A,\B\cs \C} = D_{g(\ccc)}$, where $g(\ccc) \in \Theta$, if $C_{\A,\B\cs \C}(a, b \cs \ccc) = D_{g(\ccc)}(a, b)$ for all $(a, b, \ccc)$.
Moreover, $C_{\A,\B\cs \C} = D$, where $D$ is a bivariate copula with cdf $D(a,b)$, if $C_{\A,\B\cs \C}(a, b \cs \ccc) = D(a, b)$ for all $(a, b, \ccc)$.
\begin{myex}\label{exFgm}
Let $C^{\perp}$ be the product copula with cdf $C^{\perp}(a, b) = ab$
and $\fgm{C}_{\theta}$ be the FGM copula with parameter $\theta \in [-1, 1]$
and cdf $\fgm{C}_{\theta}(a, b) := ab[1 + \theta(1-a)(1-b)]$.
The building-blocks of the four-dimensional vine copula are chosen to be
\begin{align*}
C_{12} = C_{23} = C_{34} = C^{\perp}, \quad
C_{13\cs 2} = \fgm{C}_{1-u_2}, \quad C_{24\cs 3} = \fgm{C}_{1-u_3}, \quad
C_{14\cs 23} = C^{\perp}.
\end{align*}
\end{myex}
The density of the data generating R-vine copula defined in \autoref{exFgm} is given by
\begin{align*}
c_{1:4}(u_{1:4}) 
& = 
\underbrace{c_{12}(u_1, u_2)\ c_{23}(u_2, u_3)\ c_{34}(u_3, u_4)}_{\text{first tree }T_1} \notag
\\
& \quad \times
\underbrace{c_{13\cs 2}(u_{1|2}, u_{3|2} \cs u_2)\ c_{24\cs 3}(u_{2|3}, u_{4|3} \cs u_3)}_{\text{second tree }T_2}
\times \underbrace{c_{14\cs 23}(u_{1|23}, u_{4|23} \cs u_{2:3})}_{\text{third tree }T_3} \notag
\\ &
= \underbrace{c_{13 \cs 2}(u_1, u_3 \cs u_2) \ c_{24 \cs 3}(u_2, u_4 \cs u_3)}_{\text{second tree }T_2} \notag
\\ & = \fgm{c}_{1-u_2}(u_1, u_3) \ \fgm{c}_{1-u_3}(u_2, u_4),
\end{align*}
where the second equality holds because the copulas in the first and third tree are product copulas.
For this data generating process, all building blocks of the vine copula are product copulas
except for the second tree $T_2$. 
In tree $T_2$, both $c_{13\cs 2}(\cdot, \cdot \cs u_2)$ and $c_{24\cs 3}(\cdot, \cdot \cs u_3)$ depend on the conditioning variables $u_2$ and $u_3$, respectively. 
Thus, the simplifying assumption does not hold.
The building blocks of the non-simplified vine copula defined in \autoref{exFgm} are illustrated in the left panel of \autoref{fig_vine}.

In order to approximate the non-simplified copula by a simplified vine copula one could assign arbitrary \uncops\ to the edges of the vine. 
For example the corresponding partial copulas.
One can easily check that the partial copulas in the second tree are given by product copulas. 
Moreover, the partial copulas in the other trees are also given by product copulas.
Thus, the resulting simplified vine copula model is the four-dimensional product copula with density given by
\begin{align*}
\svc{c}_{1:4}(u_{1:4}) 
& = \underbrace{\parsign{c}_{13 \cs 2}(u_{1|2}, u_{3|2}) \ \parsign{c}_{24 \cs 3}(u_{2|3}, u_{4|3})}_{\text{second tree }T_2}
 = 1.
\end{align*}

A better approximation, in terms of Kullback-Leibler divergence minimization, is given by the PVC (see Example 4.2 in \citet{Spanhel2016}) with density
\begin{align*}
\pvc{c}_{1:4}(u_{1:4}) &  
= \underbrace{\pvc{c}_{14 \cs 23}(\pvc{u}_{1|23}, \pvc{u}_{4|23})}_{\text{third tree }T_3}
= \pvc{c}_{14 \cs 23}(u_1, u_4)
= \fgm{c}_{1/9}(u_1, u_4),
\end{align*}
where the second equality holds because the copulas in the first and second tree of the PVC are product copulas. 
Note that the PVC approximates $C_{13\cs 2}$ and $C_{24\cs 3}$ in the second tree by the first-order partial copula which, by definition, always equals the partial copula. 
Thus, just like the previous approximation, the second tree is modeled by product copulas. 
However, the second-order partial copula $\pvc{C}_{14 \cs 23}$ in the third tree $T_3$ is a FGM copula with positive dependence and not equal to the corresponding partial copula $\parsign{C}_{14 \ps 23}$, which is the product copula.
Consequently, the four-dimensional approximation given by the PVC is not the four-dimensional product copula. 
The building blocks of the PVC corresponding to \autoref{exFgm} are illustrated in the right panel of \autoref{fig_vine}.

The partial copula as well as the PVC give rise to the following stochastic interpretations of the simplifying assumption.

\begin{mythelongtitle}
\label{ProbStochInterpretation}
{\bf\rmfamily
(Stochastic interpretations of the simplifying assumption)
\\}
Let $\vineDim\geq 3$ and $U_{1:\vineDim}$ be a uniform random vector.
Consider a fixed R-vine structure $\mathcal{V}$ and the corresponding R-vine copula decomposition stated in \autoref{testSA_dvinedef}.
The following statements are equivalent:
\begin{enumerate}[label=(\roman*)]
\item The \mbox{R-vine} copula 
satisfies the simplifying assumption (\autoref{testSA_simplidef}). \label{SI1}
\item $\forall \idxIterator \in \idxsett:
\qquad
(\Pitu, \Pitv)\perp U_{\condset}$ \label{SI4}
\item $\forall \idxIterator \in \idxsett:
\qquad
(\Ppitu, \Ppitv) \perp U_{\condset}$ \label{SI3}
\end{enumerate}
\end{mythelongtitle}

\autoref{ProbStochInterpretation} highlights that the simplifying assumption is equivalent to $(\vineDim-1)(\vineDim-2)/2$ vectorial independence assumptions. 
Note that $(\Pitu, \Pitv)\perp U_{\condset}$ in \ref{SI4}  can be replaced by $C_{\indu,\indv\cs \condset} = \parsign{C}_{\indu,\indv\ps \condset}$ and that
$(\Ppitu, \Ppitv) \perp U_{\condset}$ in \ref{SI3} can be replaced by
$C_{\indu,\indv\cs \condset} = \pvc{C}_{\indu,\indv\ps \condset}$. 
While the different stochastic interpretations \ref{SI4} and \ref{SI3} are equivalent in theory, \ref{SI3} is much more useful for testing the \SA{}. 
In practice, observations from the pair of CPITs $(\Pitu,\Pitv)$ or the pair of PPITs
 $(\Ppitu,\Ppitv)$ are not observable and have to be estimated from data. 
Observations from the CPIT $U_{\indk|\condset}$ can be obtained by estimating the $j$-dimensional conditional distribution function $F_{\indk|\condset}$ of $U_{\indk}$ given $U_{\condset}$, which is not an easy task.
For $\idxIterator \in E_{j}$, $j = 2, \ldots, \vineDim - 1$, the PPIT $U_{\indk|\condset}^{\text{\tiny PVC}}$ is a composition of $j(j-1)/2$ \uncops\ which belong to the building blocks of the corresponding PVC given in equation~\eqref{eqPpit1} of \autoref{testSA_def_hopartial}.
Thus, for observations from a PPIT one can sequentially estimate a sequence of \uncops\ which is much simpler. 
Therefore, we use \ref{SI3} to construct a test which is based on pseudo-observations from the PPITs.

To illustrate the practical advantage of testing the simplifying assumption with PPITs (\ref{SI3} in \autoref{ProbStochInterpretation}) instead of CPITs (\ref{SI4} in \autoref{ProbStochInterpretation}) let us reconsider \autoref{exFgm}.
In this case, the simplifying assumption can be formulated either as
\begin{align*}
\ref{SI4}\!: &\quad 
	\begin{aligned}
\text{Tree }T_2: & \quad (U_{1|2}, U_{3|2}) \perp U_2 \quad \text{and} \quad (U_{2|3}, U_{4|3}) \perp U_3
\\
\text{Tree }T_3: & \quad (U_{1|23}, U_{4|23}) \perp U_{2:3},
	\end{aligned}
\\
&\quad\text{or}
\\
\ref{SI3}\!: &\quad 
	\begin{aligned}
\text{Tree }T_2: & \quad (U_{1|2}, U_{3|2}) \perp U_2 \quad \text{and} \quad (U_{2|3}, U_{4|3}) \perp U_3
\\
\text{Tree }T_3: & \quad (\pvc{U}_{1|23}, \pvc{U}_{4|23}) \perp U_{2:3}.
	\end{aligned}
\end{align*}
For the four-dimensional \autoref{exFgm}, both formulations \ref{SI4} and \ref{SI3} consist of three vectorial independencies, all of which must be true to be equivalent to the simplifying assumption, see \autoref{ProbStochInterpretation}.
In tree $T_2$, the required conditions are identical for both \ref{SI4} and \ref{SI3}. Since the copulas in tree $T_2$ are given by bivariate FGM copulas with a varying parameter, these vectorial independencies do not hold. Thus, the simplifying assumption does not hold according to both \ref{SI4} and \ref{SI3}.
Note that the difference between \ref{SI4} and \ref{SI3} becomes only apparent after the second tree.
In the third tree, the vectorial independencies stated in \ref{SI4} and \ref{SI3} are quite different. 
For the data generating vine copula (\autoref{exFgm}) the statement in tree $T_3$ is obviously true for \ref{SI4}.
However, it is false for \ref{SI3} because  $\pvc{U}_{1|23} = U_1$ and, as can be readily verified, $U_1 \perp U_{2:3}$ is false. 

If the data generating vine copula is unknown, it is more practical to base a test on \ref{SI3} than on \ref{SI4}.
In order to check whether the pair of CPITs $(U_{1|23}, U_{4|23})$ are jointly independent from $U_{2:3}$ in \ref{SI4}, one has to estimate the unknown conditional distribution functions $F_{1|23}$ and $F_{4|23}$.
The specification of a flexible parametric model for such conditional distributions is difficult.
Non-parametric estimation might be sensible if the conditioning vector is very low-dimensional but suffers from the curse of dimensionality if the conditioning vector is high-dimensional. 
On the other side, \ref{SI3} just requires the estimation of \uncops, irrespective of the dimension of the conditioning vector. 
That is because both PPITs can be written as a composition of \uncops. 
For instance, 
\begin{align*}
\pvc{U}_{1|23} = \partial_2 \pvc{C}_{13\cs 2}\big(\partial_2 C_{12}(U_1, U_2), \partial_1 C_{23}(U_2, U_3)\big),
\end{align*}
so that only the estimation of $C_{12}, C_{23}$ and $\pvc{C}_{13\ps 2}$ are required to estimate $\pvc{U}_{1|23}$. 
In many applications, it might be feasible to find good parametric models for these pair-copulas in order to obtain an appropriate estimate of $\pvc{U}_{1|23}$.

\section{Constant conditional correlation (CCC) test for $H_0 \colon (\Ppitu, \Ppitv) \perp U_{\condset}$}\label{secTestPartialCopula}
Before we consider testing the simplifying assumption, we first develop in this and the next section tests for the null hypothesis
$
H_0 \colon (\Ppitu, \Ppitv) \perp U_{\condset}
$
for some $e\in E_j, j=2,\ldots,d-1$.
The main challenge is that the dimension $(j-1)$ of $U_{\condset}$ can be rather large so that the power of consistent tests is not satisfying in higher trees.
For instance, a consistent test could be obtained using a \mbox{Cram\'{e}r-von} Mises test for vectorial independence  (\citet{Kojadinovic2009} and \citet{Quessy2010}). However, as it is pointed out by \citet{Gijbels2016} and shown in our simulation, the power of such a consistent test rapidly approaches the significance level if the dimension increases.
Our focus is on a test that exhibits a power that is high for alternatives that one encounters in practical applications and that is quite robust to the dimension of the data set. 
For this reason, we consider the null hypothesis that the conditional correlation of the PPITs $(\Ppitu, \Ppitv)$ associated with one edge of a vine is constant w.r.t. the conditioning variables $ U_{\condset}$. 
A rejection of this hypothesis implies that the simplifying assumption can be rejected as well.
To obtain a test whose power does not collapse substantially with the dimension of the conditioning vector, we now discretize the support of the conditioning vector into a finite number of subsets.

\subsection{CCC test with known observations}\label{subsecCCC}
For the ease of exposition, we introduce the main idea of the test without a reference to vine copulas.
Let $\A$, $\B$ be uniform random variables and $\C$ be a random vector with uniform margins.
{%
	\renewcommand{\Ppitu}{\A}%
	\renewcommand{\Ppitv}{\B}%
	\renewcommand{\CondVar}{\C}%
	We want to check the null hypothesis $H_0 \colon (\Ppitu, \Ppitv) \perp \CondVar$.
	Let $\partElement_0 := \text{supp}(\CondVar)$,  $\partElement_1,\partElement_2 \subset 				\partElement_0$ with $\partElement_1 \cap \partElement_2 = \emptyset$, and $\mathbb{P}(\CondVar \in \partElement_1 		\cup \partElement_2) = 1$ with $\mathbb{P}(\CondVar \in \partElement_1), \mathbb{P}(\CondVar \in \partElement_2) > 		0$.
	We call $\Gamma:= \lbrace \partElement_1, \partElement_2\rbrace$ a partition of the support $\partElement_0$ of $			\CondVar$ into two disjoint subsets.%
	\footnote{The idea of discretization has some similarity to the boxes approach of \citet{Derumigny2016} but differs 				substantially. We only discretize the support of the conditioning vector and the rejection of our null hypothesis is still a 			rejection of the \SA{} which is not always true for the approach of \citet{Derumigny2016}. Moreover, we present a data-%
	driven approach to select the partition so that the idea of discretization can also be applied in high-dimensional settings 			without the need to impose strong  a priori assumptions on the form of the partition.
	} 
	We are interested in the correlation between $\Ppitu$ and $\Ppitv$ in the two subgroups determined by $\Gamma$, i.e.,
	\begin{align*}
	r_{\iGroup}
	&:= \Corr( \Ppitu, \Ppitv | \CondVar \in \partElement_\iGroup)  
	= \frac{\Cov( \Ppitu, \Ppitv | \CondVar \in \partElement_\iGroup)}
	 {\sqrt{ \Var( \Ppitu | \CondVar \in \partElement_\iGroup) \Var( \Ppitv | \CondVar \in \partElement_\iGroup) }},
	\end{align*}
	for $\iGroup=1,2$. 
	Under $H_0: (\Ppitu,\Ppitv) \perp \CondVar$, it follows that
	\[
	\Corr( \Ppitu, \Ppitv) = r_1
	= r_2,
	\]
	i.e., the conditional correlations are constant w.r.t. the conditioning event.
	As estimate for the correlation in the $\iGroup$-th group we use the sample version of the conditional correlation which we denote by $\hat{r}_{\iGroup}$.
	
	A statistic for testing the equality of the correlations in the two samples is given by
	\[
	T_{\Myidx}^\star(\Gamma) = \Myidx \frac{(\hat{r}_{1} - \hat{r}_{2})^2}
	{\hat{\sigma}^2(\hat{r}_{1})
	+ \hat{\sigma}^2(\hat{r}_{2})},
	\]
	where $\hat{\sigma}^2(\hat{r}_{\iGroup}), \iGroup = 1,2,$ is a consistent estimator (see \appref{proof_proposition1}) for the asymptotic variance of $\sqrt{\Myidx}(\hat{r}_{\iGroup} - r_{\iGroup})$.
	By construction of the estimators
	, the asymptotic covariance between $\hat{r}_{1}$ and $\hat{r}_{2}$ is zero.
	Thus, under regularity conditions and $H_0$ it can be readily verified that $T_{\Myidx}^\star(\Gamma) \stackrel{d}{\rightarrow} \chi^2(1)$.
	
	In a more general setting, one can also use a partition of the support $\partElement_0$ into $\nGroups \in \mathbb{N}$ pairwise disjoint subsets  $\Gamma := \lbrace \partElement_1, \ldots, \partElement_\nGroups\rbrace$ and test whether
	\begin{align*}
	H_{0}: r_{1} = \ldots = r_{\nGroups}
	\quad\text{ vs. } \quad H_{A}: \exists \iGroup_1,\iGroup_2 \in \lbrace 1,\ldots,\nGroups \rbrace, \iGroup_1\neq \iGroup_2: r_{\iGroup_1} \neq r_{\iGroup_2}.
	\end{align*}
	For this purpose, denote the vector of sample correlations in the groups by $\hat{R}_{\Gamma}^\star = (\hat{r}_{1}, \ldots, \hat{r}_{\nGroups})^{T}$.
	Further, define the diagonal $\nGroups \times \nGroups$ matrix $\hat{\Sigma}_{R_{\Gamma}}^\star$, with diagonal elements $\hat{\Sigma}_{R_{\Gamma},\iGroup,\iGroup}^\star = \hat{\sigma}^2(\hat{r}_{\iGroup})$
	and a $(\nGroups-1) \times \nGroups$ first-order difference matrix $A$
	so that $(A \hat{R}_{\Gamma}^\star)^{T} A \hat{R}_{\Gamma}^\star
	= \sum_{l=1}^{L-1} (\hat{r}_{l} - \hat{r}_{l+1})^2$.
	A statistic to test the equality of correlations in $\nGroups$ groups is then defined by the quadratic form\footnote{
	The statistic $T_{\Myidx}^\star(\Gamma)$ also follows from 
	$(B \hat{R}_{\Gamma}^\star)^{T} B \hat{R}_{\Gamma}^\star
	= \sum_{l=1}^{L-1} \hat{\pi}_l (\hat{r}_{l} - \bar{r})^2$,
	where $\bar{r}$ is the average correlation,
	$\hat{\pi}_l$ is defined in \appref{proof_proposition1} as the fraction of data corresponding to the subset $\partElement_\iGroup$
	and with an appropriately defined matrix $B$.
	}
	\begin{align*}
	T_{\Myidx}^\star(\Gamma) = \Myidx (A \hat{R}_{\Gamma}^\star)^{T} (A\hat{\Sigma}_{R_{\Gamma}}^\star A^{T})^{-1} A \hat{R}_{\Gamma}^\star,
	\end{align*}
	with asymptotic distribution
	$T_{\Myidx}^\star(\Gamma) \stackrel{d}{\rightarrow} \chi^2(\nGroups-1)$
	under \mbox{$H_0: (\Ppitu,\Ppitv) \perp \CondVar$}.
}

\subsection{CCC test with estimated pseudo-observations from the PPITs}
In order to test the simplifying assumption, we have to test null hypotheses of the form $H_0 \colon (\Ppitu, \Ppitv) \perp U_{\condset}$.
In practice, if $X_{1:\vineDim} \sim F_{X_{1:\vineDim}}$ is the data generating process, we cannot directly observe a sample of $(\Ppitu, \Ppitv, \CondVar)$ but have to estimate such observations on the basis of a sample of 
$(X_{h_e}, X_{i_e}, X_{D_e})$. To obtain these pseudo-observations, we use a semi-parametric approach. 
First, we use normalized ranks to obtain pseudo-observations from $U_i = F_{X_i}(X_i)$ for $i=1,\ldots,\vineDim$.
We then consider a fixed R-vine structure $\mathcal{V}$ and assume that there is a parametric simplified vine copula model $\{\svc{C}_{1:\vineDim;\,\theta_{1:d-1}}\colon\theta_{1:d-1}\in \Upsilon\}$ for the PVC $\pvc{C}_{1:d}$ of $(C_{1:d}, \mathcal{V})$ with
$\theta_{1:d-1;\,0}\in \Upsilon$ so that 
$\svc{C}_{1:d;\, \theta_{1:d-1;\,0}} = \pvc{C}_{1:d}$.
Finally, we use the common stepwise ML estimator \citep{HobakHaff2013} to construct pseudo-observations from $(\Ppitu, \Ppitv, \CondVar)$.
The asymptotic distribution of the resulting test statistic with these estimated pseudo-observations is stated in \autoref{Proposition2}. 
\begin{myprop}
\label{Proposition2}
Let 
$(X_{1:d}^k)_{k=1,\ldots,n}$ be $n$ independent copies of $X_{1:d}$ and $C_{1:d}$ be the copula of $X_{1:d}$. 
Consider a fixed R-vine structure $\mathcal{V}$ (\autoref{testSA_rvinedef}) and assume there is a parametric simplified vine copula model $\{\svc{C}_{1:\vineDim;\,\theta_{1:d-1}}\colon\theta_{1:d-1}\in \Upsilon\}$ for the PVC $\pvc{C}_{1:d}$ of $(C_{1:\vineDim}, \mathcal{V})$ with
$\theta_{1:d-1;\,0}\in \Upsilon$ so that 
$\svc{C}_{1:d;\, \theta_{1:d-1;\,0}} = \pvc{C}_{1:d}$.
Let $\idxIterator \in \idxsett$ be fixed.
Assume that the regularity conditions stated in Theorem 1 in \citet{HobakHaff2013} hold\,\footnote{
Note that the results in \citet{HobakHaff2013} are written up for D-vine copulas but can be generalized to R-vine copulas \citep{HobakHaff2013}.
The entries of the matrices needed to compute the standard errors of the sequentially estimated parameters of R-vine copulas can for example be found in \citet{Stoeber2013c}.}%
and that the partition  $\Gamma := \lbrace \partElement_1, \ldots, \partElement_\nGroups\rbrace$, where $\partElement_1,\ldots,\partElement_\nGroups \subset \partElement_0  = \supp(\CondVar)$, satisfies
\begin{enumerate}[label=(\roman*)]
\item $\partElement_{\iGroup_1} \cap \partElement_{\iGroup_2} = \emptyset$, for $\iGroup_1 \neq \iGroup_2$ with $1 \leq \iGroup_1,\iGroup_2 \leq \nGroups$,
\item $\mathbb{P}(\CondVar \in \bigcup_{\iGroup = 1}^{\nGroups} \partElement_\iGroup) = 1$ with
$\mathbb{P}(\CondVar \in \partElement_\iGroup) > 0$ for all $1 \leq \iGroup \leq \nGroups$.
\end{enumerate}
Let {$\hat{R}_{\Gamma}$} be the vector of sample correlations that are computed using the pseudo-observations from the PPITs
and $\hat{\Sigma}_{R_{\Gamma}} := 
\hat{\Sigma}_{R_{\Gamma}}^\star + \hat{\Sigma}_{\text{\tiny PVC}}+
\hat{\Sigma}_r,
$ 
where $\hat{\Sigma}_{R_{\Gamma}}$ is defined in equation~\eqref{eq_sigma_hat} in \appref{proof_proposition1}.
Construct the test statistic
\begin{align}
T_{\Myidx}(\Gamma) = \Myidx (A \hat{R}_{\Gamma})^{T} (A\hat{\Sigma}_{R_{\Gamma}} A^{T})^{-1} A \hat{R}_{\Gamma}.
\label{eq_def_test_stat}
\end{align}
Under \mbox{$H_0: (\Ppitu,\Ppitv) \perp \CondVar$} it holds that
\[
T_{\Myidx}(\Gamma) \stackrel{d}{\rightarrow} \chi^2(\nGroups-1).
\]
\end{myprop}

The matrices $\Sigma_{\text{\tiny PVC}} = \plim{\Myidx} \hat{\Sigma}_{\text{\tiny PVC}}$ and 
$\Sigma_r = \plim{\Myidx} \hat{\Sigma}_{r}$ quantify the change in the asymptotic covariance due to the estimation of pseudo-observations from the PPITs.
If the marginal distributions are known and we don't have to estimate ranks to obtain pseudo-observations from $\pvc{C}_{1:d}$ it follows that $\Sigma_r = 0$. 
If the PVC $\pvc{C}_{1:d}$ of $(C_{1:d}, \mathcal{V})$ is known it follows that $\Sigma_{\text{\tiny PVC}}=0$.
Note that, in general,
the off-diagonal elements of $\Sigma_{R_{\Gamma}} = \plim{\Myidx} \hat{\Sigma}_{R_{\Gamma}}$, i.e., the asymptotic covariances between estimated correlations in different groups, are no longer zero if observations from the PPITs are estimated.

\subsection{CCC test with a combination of partitions}
The selected partition $\Gamma$ has an influence on how well a varying conditional correlation is detected by the test. 
To illustrate this and to motivate a test based on the combination of several partitions, we use the following \autoref{exSimStudy}.
\begin{myex}\label{exSimStudy}
Let $C^{\text{\tiny Cl}}_\theta$ and $C^{\text{\tiny Fr}}_\theta$ be the Clayton and Frank copula with parameter $\theta$, respectively.
The building-blocks of the four-dimensional D-vine copula are chosen to be
\begin{align*}
C_{12} &= C_{23} = C_{34} = C^{\text{\tiny Cl}}_{\theta_1}, \\
C_{13\cs 2} &= \pvc{C}_{13\ps 2} = C_{24\cs 3} = \pvc{C}_{24\ps 3} = C^{\text{\tiny Cl}}_{\theta_2}, \\
C_{14\cs 23} &= C^{\text{\tiny Fr}}_{\alpha(u_{2:3}; \lambda)}, \\
\alpha(u_{2:3}; \lambda) &= 1 + 2.5 \lambda (1-1.5(u_2+u_3))^2,
\end{align*}
with $\theta_1 := \frac{2\tau}{1-\tau}$ and $\theta_2 := \frac{\theta_1}{1 + \theta_1}$, where $\tau$ is the value of Kendall's $\tau$ of the bivariate margins in the first tree.\footnote{
The parameter function is a generalization of the function $\theta(X) = 1 + 2.5(3-X)^2$ used by \citet{Acar2013} for a Frank copula with a one-dimensional conditioning set, where the conditioning variable $X$ is assumed to be uniformly distributed on the interval $[2,5]$.
The parameters of the bivariate Clayton copulas in the first and second tree are specified in a way that \autoref{exSimStudy} can be considered as a four-dimensional Clayton copula where the copula in the last tree is replaced by a Frank copula with varying parameter.}
\end{myex}

\begin{figure}
\begin{center}
\includegraphics[width=\textwidth]{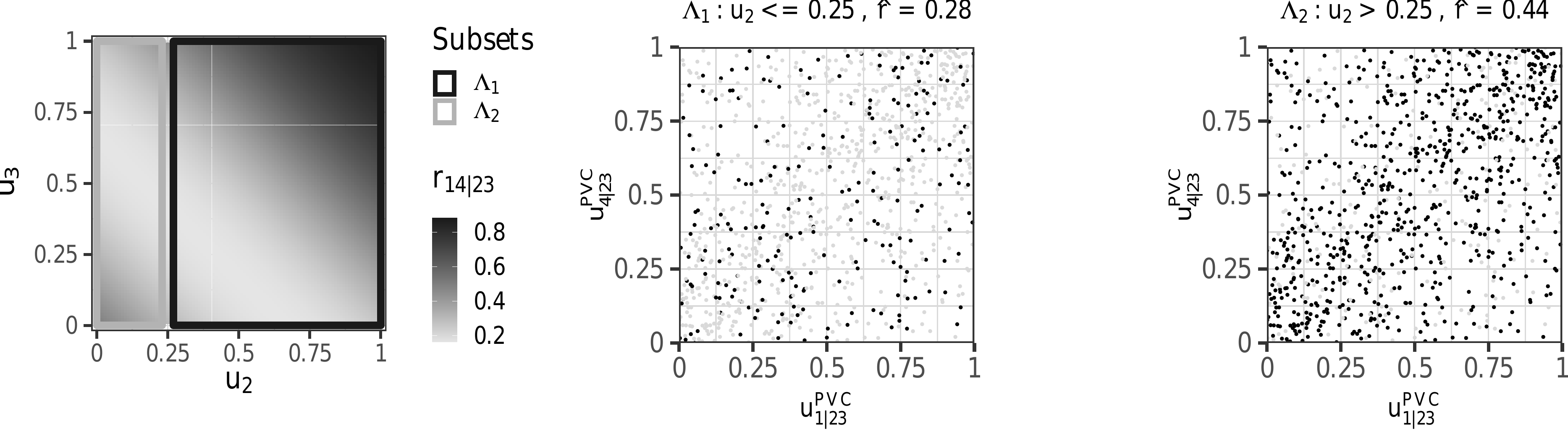}
\includegraphics[width=\textwidth]{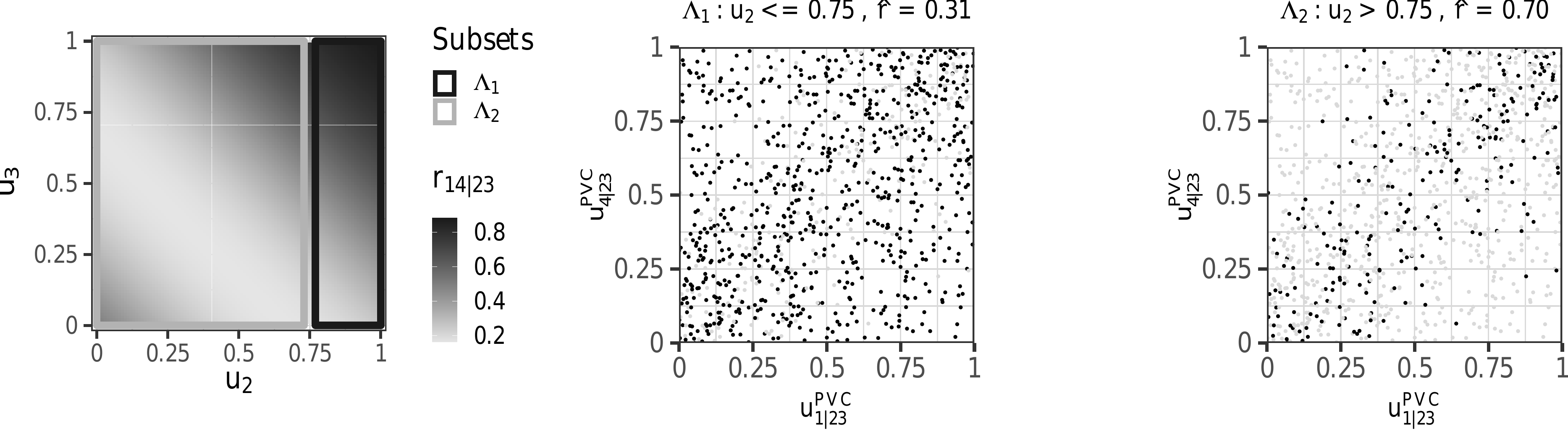}
\end{center}
\caption{The plots show $1000$ random draws from \autoref{exSimStudy}.
On the left hand side the shaded background visualizes the conditional correlation $r_{14|23}$ of $C_{14\cs 23}$ as a function of $U_2$ and $U_3$. 
In the middle and on the right hand side, realizations from the PPITs $(\pvc{U}_{1|23}, \pvc{U}_{4|23})$
grouped according to $\Gamma_1$ in the upper row and $\Gamma_2$ in the lower row are shown.
The black points are observations being assigned to the corresponding subset and the light-gray points are observations which have been assigned to the other subset.
}\label{figDiffParts}
\end{figure}

For the illustration we set in \autoref{exSimStudy}  $\tau = 0.4$ and $\lambda = 1$ and simulate a sample of size $\Myidx = 1000$.
For instance, if we choose
$\partElement_1 = {\lbrace (u_2, u_3) \in [0,1]^2: u_3 \leq u_2 \rbrace}$
and $\partElement_2 = \partElement_0 \setminus \partElement_1$,
then $r_1 = r_2$ and the power of the test is asymptotically equal to the level of the test. 
Instead, we could use partitions such as
$\Gamma_1 :=
\lbrace \partElement_1 = [0, 0.25] \times [0,1],\, 
\partElement_2 = (0.25, 1] \times [0,1] \rbrace$
or $\Gamma_2 :=
\lbrace \partElement_1 = [0, 0.75] \times [0,1],\,
\partElement_2 = (0.75, 1] \times [0,1] \rbrace$.
In \autoref{figDiffParts}, we illustrate the resulting tests
$T_\Myidx(\Gamma_1)$
and 
$T_\Myidx(\Gamma_2)$.
The upper row corresponds to the first partition $\Gamma_1$ where the difference of the correlations in the two groups is $\hat{r}_2 - \hat{r}_1 = 0.16$, yielding a test statistic value of $T_\Myidx(\Gamma_1) = 5.41$.
In contrast, if we consider the second partition $\Gamma_2$ shown in the lower row of \autoref{figDiffParts}, we get $\hat{r}_2 - \hat{r}_1 = 0.39$ and $T_\Myidx(\Gamma_2) = 65.72$.

In order to increase the probability that the test detects a varying conditional correlation, it seems naturally to consider several partitions $\Gamma_0, \ldots, \Gamma_\nPartitions$, $\nPartitions \geq 1$, where each partition $\Gamma_\iPartition$ is a collection of $\nGroups_\iPartition \geq 2$ subsets of the support $\partElement_0$.
A test statistic using a combination of partitions is given by
\begin{align}
\Theta_{\Myidx} = \max \left\lbrace T_{\Myidx}(\Gamma_0) + \Myidx \lambda_{\Myidx}, T_{\Myidx}(\Gamma_1), \ldots, T_{\Myidx}(\Gamma_\nPartitions) \right\rbrace - \Myidx \lambda_{\Myidx}, \label{eqStatTheta}
\end{align}
where $\lambda_{\Myidx}$ is a penalty function and $\Gamma_0$ is the base partition whose corresponding test statistic $T_{\Myidx}(\Gamma_0)$ is the only one that is not penalized.
The construction of such a test has some similarity to the approach of \citet{Lavergne2008}.
The idea is that by choosing an appropriate penalty function, under $H_0$, the asymptotic distribution of $\Theta_\Myidx$ and $T_n(\Gamma_0)$ should be equivalent.
Moreover, if $H_0$ is not true, then $\Theta_\Myidx$ should have more power than $T_n(\Gamma_0)$ because, by construction,   $\Theta_{\Myidx} \geq T_n(\Gamma_0)$. 
Precise conditions are given in the following proposition. 

\begin{myprop}
\label{Proposition3}
Assume that the conditions stated in \autoref{Proposition2} hold and that the partitions $\Gamma_0, \ldots, \Gamma_\nPartitions$ fulfill the conditions stated for $\Gamma$ in \autoref{Proposition2}.
Additionally, let $\lambda_{\Myidx}: \mathbb{N} \rightarrow \mathbb{R}^{+}$ be a penalty function such that
\begin{enumerate}[label=(\roman*)]
\item $\Myidx\lambda_{\Myidx} \rightarrow \infty$ for $\Myidx \rightarrow \infty$,
\item $\lambda_{\Myidx}\to 0$ for $\Myidx \rightarrow \infty$.
\end{enumerate}
{ Set
$\Theta_{\Myidx} = \max \left\lbrace T_{\Myidx}(\Gamma_0) + \Myidx\lambda_{\Myidx}, T_{\Myidx}(\Gamma_1), \ldots, T_{\Myidx}(\Gamma_\nPartitions) \right\rbrace - \Myidx\lambda_{\Myidx}$, 
where $T_n(\Gamma_m),0\leq m \leq M,$ is defined like $T_n(\Gamma)$ in equation \eqref{eq_def_test_stat} in \autoref{Proposition2} with $\Gamma$ replaced by $\Gamma_m$. 
}
Under $H_0: (\Ppitu,\Ppitv) \perp \CondVar$ it holds that
\[
\Theta_{\Myidx} \stackrel{d}{\rightarrow} \chi^2(\nGroups_0-1),
\]
where $\nGroups_0$ is the number of subsets of $\Gamma_0$.
If there is a partition $\Gamma_{m^{\star}},m^{\star}\in\{0,\ldots,M\}$, such that 
$\plim{\Myidx} \frac{1}{\Myidx} T_\Myidx(\Gamma_{m^{\star}}) =: c>0$
it follows that 
\[
\Theta_{\Myidx} \stackrel{p}{\rightarrow} \infty.
\]
\end{myprop}

Thus, the critical value of $\Theta_\Myidx$ under $H_0$ asymptotically (as $n \rightarrow \infty$, for fixed $M$) only depends on the number of elements $L_0$ in $\Gamma_0$ but not on $\Gamma_1,\ldots,\Gamma_\nPartitions$.
In finite samples, however, the distribution of $\Theta_\Myidx$ under $H_0$ also depends on $\Gamma_1,\ldots,\Gamma_\nPartitions$ and the usage of asymptotic critical values derived from the limiting distribution $\chi^2(\nGroups_0-1)$ may result in an empirical size that is not close to the theoretical level.\footnote{
In extensive simulation studies in \autoref{secSimStudy} and \autoref{secSimStudyMultCop} it is demonstrated that the empirical size is often comparable to the theoretical level and the choice of the penalty function is analyzed in a simulation study in \appref{appPenalty}}
Moreover, if there is a partition $\Gamma_{m^{\star}}$ such that the corresponding correlations are not identical, i.e., $c>0$, the power of the test approaches 1 if the sample size goes to infinity.
Possible choices of the partitions $\Gamma_0, \ldots, \Gamma_\nPartitions$  and their selection will be discussed in the next section.

\section{A data-driven algorithm for testing with a high-dimensional conditioning vector}\label{secAlgDecTree}
Both constant conditional correlation (CCC) tests, $T_{\Myidx}(\Gamma)$ and $\Theta_{\Myidx}$, are based on partitions of $\partElement_0 := \supp(\CondVar)$.
We will first illustrate why an a priori determination of such partitions is problematic and then show how such partitions can be selected in a data-driven fashion.

\subsection{A priori defined partitions}\label{secNaiveGroup}
\newcommand{\qmed}[1]{q_{0.5}^{(#1)}}
As a naive approach one could use the sample median $\qmed{i}$ of each conditioning variable $u_i$ to split the observations
into groups and then consider the partition that results from the combinations of these groups.
For instance, for $C_{14;23}$ in the third tree, we obtain the following four subsets forming a partition
\begin{align*}
    \partElement_{(1,1)} &= \{(u_2, u_3)\in \partElement_0\colon u_2 \leq \qmed{2}, u_3 \leq \qmed{3}\}, \\
    \partElement_{(1,2)} &= \{(u_2, u_3)\in \partElement_0\colon u_2 > \qmed{2}, u_3 \leq \qmed{3}\}, \\ 
    \partElement_{(2,1)} &= \{(u_2, u_3)\in \partElement_0\colon u_2 \leq \qmed{2}, u_3 > \qmed{3}\}, \\
    \partElement_{(2,2)} &= \{(u_2, u_3)\in \partElement_0\colon u_2 > \qmed{2}, u_3 > \qmed{3}\}.
\end{align*}
Such an approach is however only feasible for a low-dimensional conditioning vector, since the number of subsets increases exponentially with the dimension of the conditioning vector. 
Moreover, the number of observations that are contained in a single subset might get too small.
Alternatively, one could use the mean $\bar{u}_{D_e}  = \frac{1}{\Mydim} \sum_{\mydim=1}^{\Mydim} u_{D_e^k}$ of the conditioning variables $u_{D_e}$ for $\idxIterator \in E_{j}$, $j = 2, \ldots, \vineDim - 1$,
and consider the sample median $q_{0.5}^{\bar{u}_{D_e}}$ of $\bar{u}_{D_e}$ as split point to obtain the following partition $\Gamma_{\text{med}} := \lbrace \partElement_1, \partElement_2 \rbrace$ of the support $\partElement_{0} = \text{supp}(\CondVar)$:
\begin{align}
\partElement_1 &:= \left\lbrace \condVar \in \partElement_0 : \bar{u}_{D_e} \leq q_{0.5}^{\bar{u}_{D_e}} \right\rbrace
\text{ and } \partElement_2 := \partElement_0 \setminus \partElement_1. \label{gamma_med}
\end{align}

However, in practice, such an a priori definition of the partition is arbitrary and might not detect a conditional variation.
Therefore, we introduce now a decision tree algorithm which selects the partitions in a more data-driven way and is computationally feasible in high dimensions.

\subsection{Data-driven choice of the partitions by means of a decision tree algorithm}\label{secDecTreeGroup}
\begin{figure}[t!]
\centering
\includegraphics[width=0.85\textwidth]{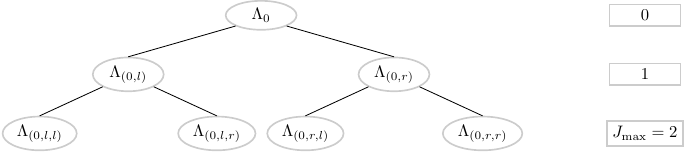}
\caption{Partitioning of the support $\partElement_{0} = \text{supp}(\CondVar)$ of the random vector $\CondVar$ into disjoint subsets $\lbrace \partElement_{(0,l,l)}, \partElement_{(0,l,r)}, \partElement_{(0,r,l)}, \partElement_{(0,r,r)} \rbrace$ 
using a decision tree algorithm with depth $J_{\text{max}} = 2$.}\label{FigDecTree}
\end{figure}

The test statistic $\Theta_{\Myidx}$ can be rewritten in the following way
\begin{align*}
\Theta_{\Myidx} &= 
\text{max} \left\lbrace T_{\Myidx}(\Gamma_0) + \Myidx\lambda_{\Myidx}, T_{\Myidx}(\Gamma_{\max}) \right\rbrace - \Myidx\lambda_{\Myidx},
\end{align*}
with $\Gamma_{\max} := \text{argmax}_{\Gamma_m \in \lbrace \Gamma_1, \ldots, \Gamma_M\rbrace} T_\Myidx(\Gamma_m)$.
The set $\Gamma_{\max}$ denotes the partition, excluding the base partition $\Gamma_0$, for which a possible violation of the $H_0$ is most pronounced.
As base partition we set $\Gamma_0 = \Gamma_{\text{med}}$ (see equation~\eqref{gamma_med}).
To find $\Gamma_{\max}$ in a data-driven and computationally efficient way we use decision trees with depth $J_{\max} = 2$ like the one shown in \autoref{FigDecTree}.
The decision tree recursively uses binary splits to partition the support $\partElement_0 = \text{supp}(U_{\condset})$ into disjoint subsets to obtain $\Gamma_{\max} : = \lbrace \partElement_{(0,l,l)}, \partElement_{(0,l,r)}, \partElement_{(0,r,l)}, \partElement_{(0,r,r)}\}$. 
The possible split points in the first level
are given by the empirical quartiles of each conditioning variable and by the empirical quartiles of the mean of the conditioning vector.
In the second level, the split points for each leaf are chosen similarly but conditional on the chosen subsets of the first level.
Among these possible splits the split is chosen that maximizes the statistic of the CCC test.
For algorithmic details of the decision tree with a general depth of $J_{\text{max}}$ we refer to \appref{appDecisionTreeAlgorithm}.

\begin{figure}[t!]
\centering
\includegraphics[width=\textwidth]{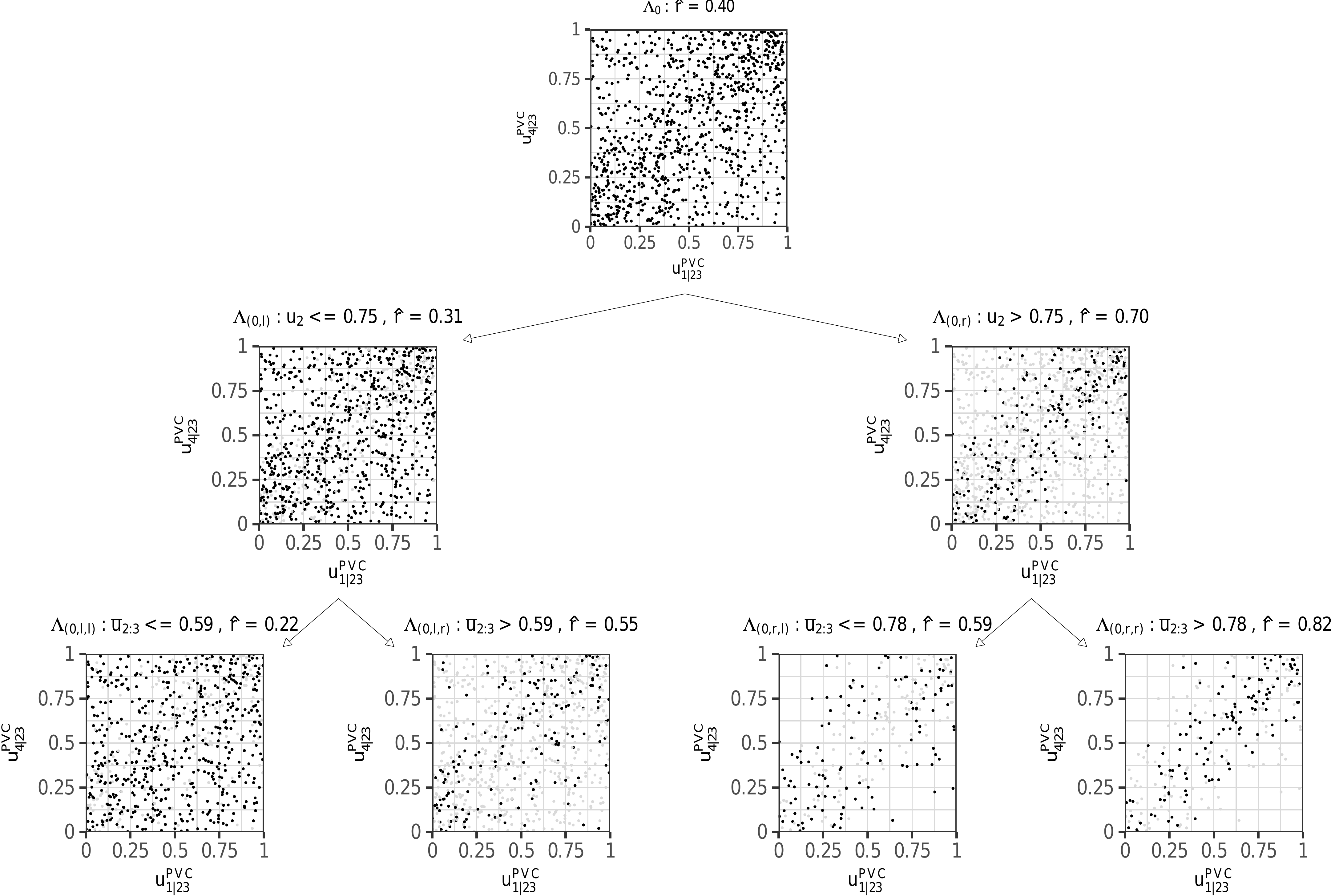}
\caption{
The plots show the same $1000$ observations from \autoref{exSimStudy} as \autoref{figDiffParts}.
Shown is the decision tree-based grouping of the PPITs $(\pvc{U}_{1|23}, \pvc{U}_{4|23})$ into four different groups according to the partition $\Gamma_{\text{max}} := \lbrace \partElement_{(0,l,l)}, \partElement_{(0,l,r)}, \partElement_{(0,r,l)}, \partElement_{(0,r,r)} \rbrace$.
The black points show observations being assigned to the corresponding subset of the support $\partElement_0$ and the light-gray points correspond to the observations which have been assigned to the other subset due to the binary split.
The mean of the conditioning variables $u_2$ and $u_3$ is denoted by $\bar{u}_{2:3}$.
}\label{figGrouped}
\end{figure}

We revisit \autoref{exSimStudy} to illustrate how the decision tree adapts to the variation in the conditional correlation.
For the same simulated sample that is used for \autoref{figDiffParts}, the decision tree is applied to test $H_{0}: (\pvc{U}_{1|23}, \pvc{U}_{4|23}) \perp U_{2:3}$.
The partitioning of $\partElement_0$ into $\Gamma_{\text{max}}$ is visualized in \autoref{figGrouped} which shows the grouping of the observations from the PPITs $(\pvc{U}_{1|23}, \pvc{U}_{4|23})$  according to $\Gamma_{\text{max}}$.
In each scatter plot the black observations have been assigned to this leaf while the observations in light gray have been assigned to the other leaf.
Furthermore, the estimated correlations in each group, which are used for the CCC test, are shown. 
We see that the decision tree chooses a partition with estimated correlations that are quite different and a maximal difference of
$\widehat{\Corr}(\pvc{U}_{1|23},\pvc{U}_{4|23}|U_{2:3}\in \Lambda_{(0,l,l)})-
\widehat{\Corr}(\pvc{U}_{1|23},\pvc{U}_{4|23}|U_{2:3}\in \Lambda_{(0,r,r)}) = 0.60.$

\begin{table}[ht]
{\scriptsize
\caption{
The table refers to the same $1000$ observations from \autoref{exSimStudy} as \autoref{figDiffParts}.
It lists the subsets formed with a decision tree to find $\Gamma_{\text{max}} := \lbrace \partElement_{(0,l,l)}, \partElement_{(0,l,r)}, \partElement_{(0,r,l)}, \partElement_{(0,r,r)} \rbrace$.
The first column contains the decision tree level $J$, the second column the subsets and the third column the estimated correlations for the subsets.
}\label{tableGammaMaxPartition}
\begin{center}
\begin{tabular}{lcc}
$\boldsymbol{J}$ &
\textbf{Subset} $\boldsymbol{\Lambda_\gamma}$ &
$\boldsymbol{\widehat{\Corr}(\pvc{U}_{1|23},\pvc{U}_{4|23}|U_{2:3}\in \Lambda_\gamma)}$
\\[2pt] \hline \\[-5pt]
0 & $\Lambda_0 = [0, 1]^2$ & $0.40$ \\[2pt] \hline \\[-5pt]
1 & $\Lambda_{(0,l)} = [0, 0.75] \times [0,1]$ & $0.31$ \\[2pt]
1 & $\Lambda_{(0,r)} = [0.75, 1] \times [0,1]$ & $0.70$ \\[2pt] \hline \\[-5pt]
2 & $\Lambda_{(0,l,l)} = \Lambda_{(0,l)} \cap \{(u_2, u_3): \frac{1}{2}(u_2 + u_3) \leq 0.59 \}$ & $0.22$ \\[2pt]
2 & $\Lambda_{(0,l,r)} = \Lambda_{(0,l)} \cap \{(u_2, u_3): \frac{1}{2}(u_2 + u_3) > 0.59 \}$ & $0.55$ \\[2pt]
2 & $\Lambda_{(0,r,l)} = \Lambda_{(0,r)} \cap \{(u_2, u_3): \frac{1}{2}(u_2 + u_3) \leq 0.78 \}$ & $0.59$ \\[2pt]
2 & $\Lambda_{(0,r,r)} = \Lambda_{(0,r)} \cap \{(u_2, u_3): \frac{1}{2}(u_2 + u_3) > 0.78 \}$ & $0.82$
\end{tabular}
\end{center}
}
\end{table}

\begin{figure}[ht]
\centering
\includegraphics[width=\textwidth]{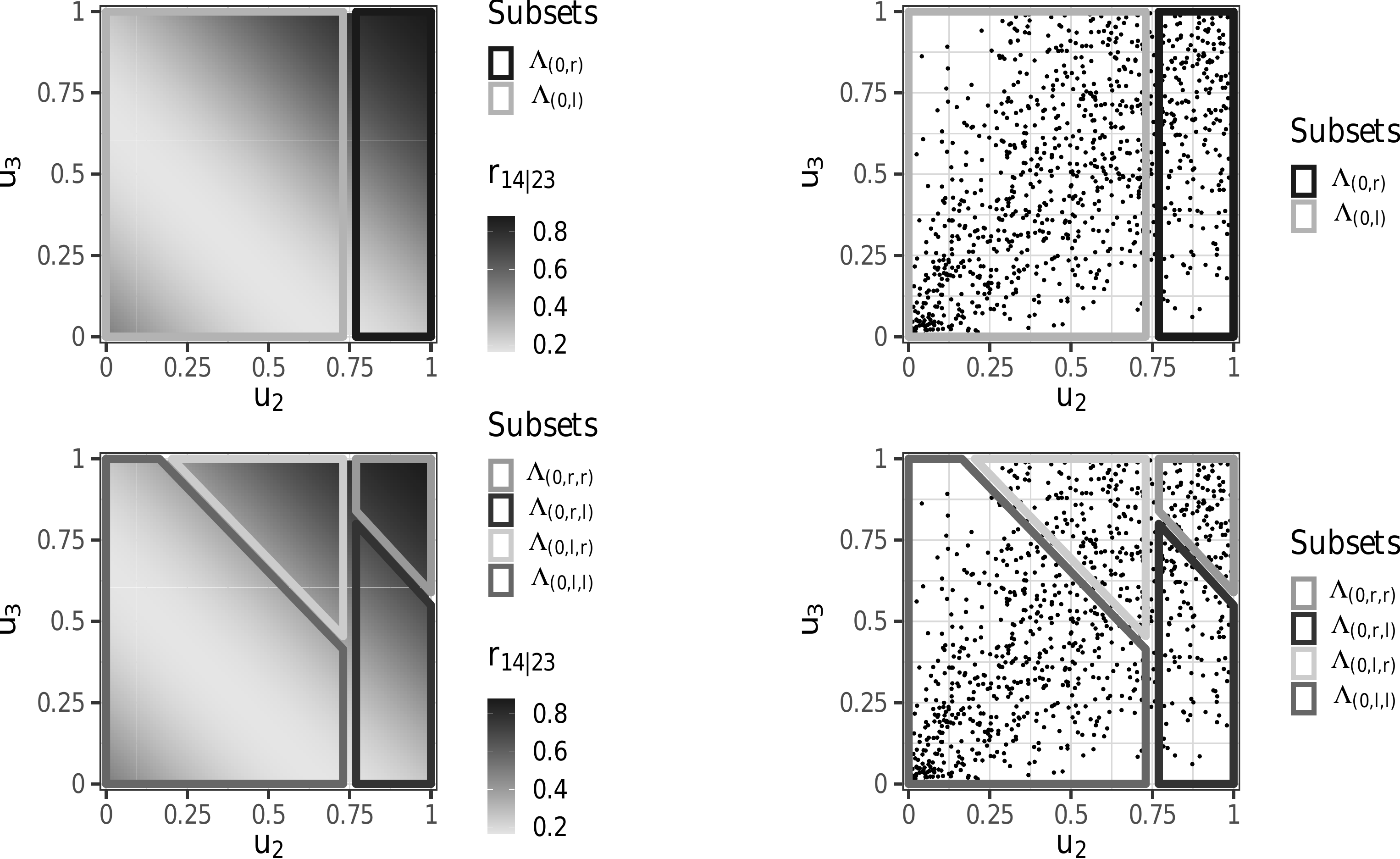}
\caption{
The plots show the same $1000$ observations from \autoref{exSimStudy} as \autoref{figDiffParts}.
On the left hand side the shaded background shows the conditional correlation $r_{14|23}$ of $C_{14\cs 23}$ as a function of $U_2$ and $U_3$. 
On the right hand side the realizations of the conditioning variables $U_2$ and $U_3$ are shown.
The upper plots correspond to the first binary split of the decision tree and the lower plots to the second and third binary split.
The colored frames show the different subsets $\lbrace \partElement_{(0,l)}, \partElement_{(0,r)}, \partElement_{(0,l,l)}, \partElement_{(0,l,r)}, \partElement_{(0,r,l)}, \partElement_{(0,r,r)} \rbrace$ of the support $\partElement_0 := [0,1]^2$.
}\label{figPartitioning}
\end{figure}

The subsets that are constructed by the decision tree
to obtain the selected $\Gamma_{\text{max}}$ partition are stated in \autoref{tableGammaMaxPartition} together with the corresponding estimated correlations.
The decision tree 
generates partitions which are no longer simple one-directional splits as in \autoref{figDiffParts}.
Instead we now get more complex polygons as subsets which are visualized as colored frames in \autoref{figPartitioning}.
On the left hand side of \autoref{figPartitioning}, the shaded area shows the variation in the correlation of $C_{14\cs 23}$ as a function of $U_2$ and $U_3$.
Areas with darker gray correspond to higher values of the conditional correlation $r_{14|23}$.
The right hand side of \autoref{figPartitioning} illustrates how the dependence within the conditioning set, determined by the Clayton copula $C_{23} = C^{\text{\tiny Cl}}_{\theta_1}$, influences the variation of $C_{14\cs 23}$.
Shown are the realized values of $(U_2,U_3)$ and their grouping into the subsets.
The two plots on top correspond to the first binary split, which is done according to the 75\% quartile of $U_2$.
The two plots at the bottom show the splits in the second level of the decision tree, which is done according to quartiles of the mean of the conditioning variables $U_2$ and $U_3$.
The adaption of the decision tree based partition to the variation in the conditional correlation becomes clear when looking at the shaded background which shows the conditional correlation $r_{14|23}$ of $C_{14\cs 23}$ as a function of $U_2$ and $U_3$.
We see how $\Lambda_0$ is partitioned into areas with relatively low ($\Lambda_{(0,l,l)}$), medium ($\Lambda_{(0,l,r)}$ and $\Lambda_{(0,r,l)}$) and high correlation ($\Lambda_{(0,r,r)}$).

In all simulations in \autoref{secSimStudy} and \autoref{secSimStudyMultCop}, and the real data applications in \autoref{secRealData}, we use decision trees with depth $J_{\max} = 2$ and  $\lambda_{\Myidx} = \frac{1}{\sqrt{\Myidx}}$ for the penalty.
The choice of the penalty function is analyzed and explained in \appref{appPenalty}.
We further set $\Gamma_0 = \Gamma_{\text{med}}$ (see equation~\eqref{gamma_med}) as base partition,
because we assume to have no a priori information about the relative importance of each conditioning variable
and because the median of the mean of the\pagebreak[4] conditioning vector as split point guarantees well-balanced sample sizes in the groups.%
\footnote{
If the application at hand suggests a priori that $H_0$ is likely to be false and a particular partition $\Gamma_{\star}$ would result in a more pronounced difference 
between the resulting conditional correlations, i.e., $T_{\Myidx}(\Gamma_{\star}) > T_{\Myidx}(\Gamma_{\text{med}})$, 
then setting $\Gamma_{0} = \Gamma_{\star}$ instead of $\Gamma_0 = \Gamma_{\text{med}}$ would not deteriorate but possibly 
improve the power of the test provided $H_0$ is actually false.
However, as it is shown in \autoref{secPowerGain} the choice of $\Gamma_0$ is not crucial if the alternative partitions are selected in a data-driven way using the decision tree.}

\section{CCC test: Simulation study}\label{secSimStudy}
In the following, the finite-sample performance of the CCC test $\Theta_\Myidx$ is analyzed and compared to the performance of the vectorial independence (VI) test of \citet{Kojadinovic2009}.
The simulation study is build around \autoref{exSimStudy} using a ceteris paribus setup in order to analyze the different key drivers of the empirical power.
We will analyze the power of the CCC and the VI test w.r.t. different variations of the conditional copula in \autoref{secFunForm},
illustrate the power gain of the CCC test due to the decision tree algorithm in \autoref{secPowerGain},
and investigate the performance of both tests w.r.t. the dimensionality of the testing problem in \autoref{secPowerDim}. 
The effect of misspecified copula families in lower trees is discussed in \autoref{secMisspec}.

All results for the CCC test are computed with estimated pseudo-observations from the PPITs.
Since the asymptotic distribution of the VI test with estimated pseudo-observations is unknown, we use the true observations from the PPITs for the VI test to compute p-values on the basis of 1000 bootstrap samples \citep{Quessy2010}.
Therefore, in simulations with possible misspecifications we do not show results for the VI test.

\subsection{Power study: The functional form of the conditional copula}\label{secFunForm}
\begin{figure}[htb]
\begin{center}
\includegraphics[width=\textwidth]{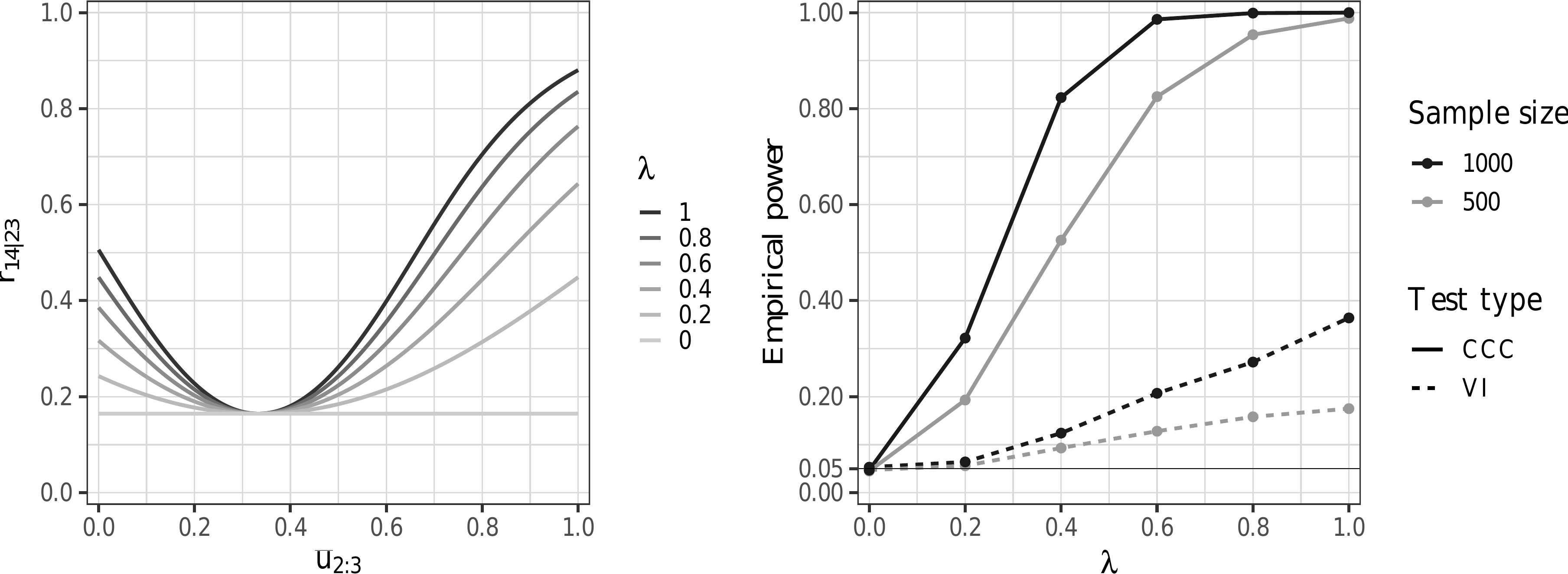}
\end{center}
\caption{
On the left hand side the conditional correlation $r_{14|23}$ of the Frank copula $C_{14\cs 23}$ as a function of the mean $\bar{u}_{2:3}$ of the conditioning variables $u_2$ and $u_3$ is shown for different values of the parameter $\lambda$.
The parameter function is stated in \eqref{eqParFunctional} and the data generating process is defined in \autoref{exSimStudy}.
The plot on the right hand side shows the empirical power for different values of $\lambda$ and the theoretical 5\% level of the tests.
Empirical power values are based on $1000$ samples.
The solid lines correspond to the CCC test $\Theta_\Myidx$ and the dashed lines to the VI test.
Different sample sizes are shown color coded. 
}\label{figPowerLambda}
\end{figure}

The conditional copula $C_{14\cs 23} = C^{\text{\tiny Fr}}_{\alpha(u_{2:3}; \lambda)}$ in \autoref{exSimStudy} varies in $u_{2:3}$.
To alter the variation, we choose values between zero and one for the parameter $\lambda$ in the function
\begin{align}
\alpha(u_{2:3}; \lambda) = 1 + 2.5 \lambda (1-1.5(u_2+u_3))^2. \label{eqParFunctional}
\end{align}
For $\lambda = 0$, $C_{14\cs 23}$ does not vary in $u_{2:3}$.
Recall that the simplifying assumption is satisfied in the second tree in \autoref{exSimStudy}, so that $H_{0}: (\pvc{U}_{1|23}, \pvc{U}_{4|23}) \perp U_{2:3}$ is true if $\lambda = 0$.
For $\lambda > 0$, $C_{14\cs 23}$ varies in $u_{2:3}$ and $H_{0}: (\pvc{U}_{1|23}, \pvc{U}_{4|23}) \perp U_{2:3}$ is false.
For $\lambda = 1$ the variation is most pronounced.
In \autoref{figPowerLambda}, the variation of the conditional correlation $r_{14|23}$ of $C_{14\cs 23}$ as a function of the mean $\bar{u}_{2:3} = \frac{1}{2} (u_2 + u_3)$ is shown on the left hand side.\footnote{
Note that we have already seen $r_{14|23}$ for $\lambda = 1$ as a function of $u_2$ and $u_3$ as shaded background in the plots on the left hand side of \autoref{figDiffParts} and \autoref{figPartitioning}.}

For the sample sizes $\Myidx=500, 1000,$ and $\lambda = 0, 0.2, 0.4, 0.6, 0.8, 1,$ we apply the CCC test $\Theta_\Myidx$ and the VI test \citep{Kojadinovic2009} for the hypothesis $H_{0}: (\pvc{U}_{1|23}, \pvc{U}_{4|23}) \perp U_{2:3}$.
On the right hand side of \autoref{figPowerLambda}, empirical power values are plotted for different values of the parameter $\lambda$.
The numbers are based on $1000$ samples for each combination of $\lambda$ and $\Myidx$.
The level of the tests is chosen to be 5\%.
For both tests and sample sizes the empirical size (i.e. the case $\lambda = 0$) is close to the theoretical level of the test.
The empirical power of both tests is clearly increasing for all values of $\lambda$ if one doubles the sample size from $500$ to $1000$ observations.
Furthermore, both tests are more powerful the more the variation in $C_{14\cs 23}$ is pronounced, i.e., the larger the parameter $\lambda$ is.
In terms of empirical power, the CCC test outperforms the VI test in all settings with
a relative improvement that often exceeds 300\%.
Both tests are implemented in an accompanying \texttt{R}-package \texttt{pacotest} \citep{Kurz2017} and are computationally feasible.
For $\lambda = 1$ and a sample size of $n=1000$, the computational time for the CCC test is $2.25$ seconds and for the VI test $0.88$ seconds.\footnote{
The reported computational times are the median from $20$ repetitions on a single core of an AMD Ryzen 7 PRO 4750U processor.
Note that a major part of the computational time for the CCC test is required for the computation of the asymptotic covariances that account for estimated pseudo-observations from the PPITs.
In a fair comparison of computational times, the CCC test can, like the VI test, also be computed with known observations from the PPITs.
The computational time of the CCC test then drops to $0.01$ seconds and is much shorter than the $0.88$ seconds of the VI test.}

\subsection{Power study: Improved power due to the decision tree algorithm}\label{secPowerGain}
We now compare the CCC test based on the decision tree approach $\Theta_\Myidx$ with the CCC test $T_n(\Gamma_0)$ which only considers the base partition $\Gamma_0$.
By construction, $\Theta_\Myidx \geq T_\Myidx(\Gamma_0)$ always holds, meaning that if we reject on the basis of $T_n(\Gamma_0)$, we also reject on the basis of $\Theta_n$.
As a consequence, the empirical power of $\Theta_\Myidx$ is never smaller than the empirical power of $T_\Myidx(\Gamma_0)$.
The improvement in power due to the use of $\Theta_\Myidx$ instead of $T_\Myidx(\Gamma_0)$ depends on the data generating process and will be investigated in the following.

\begin{figure}
\centering
\includegraphics[width=\textwidth]{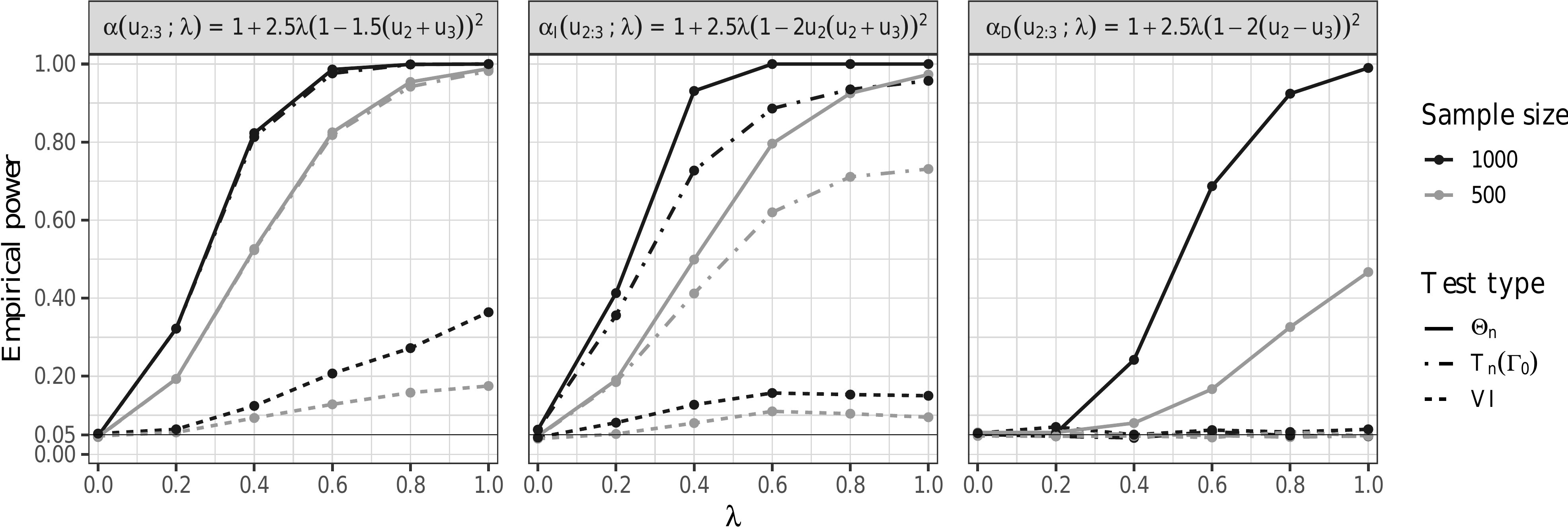}
\caption{
Empirical power of the CCC tests $\Theta_\Myidx$ and $T_\Myidx(\Gamma_0)$ and the VI test for the hypothesis \mbox{$H_{0}: (\pvc{U}_{1|23}, \pvc{U}_{4|23}) \perp U_{2:3}$} in \autoref{exSimStudy}.
The three panels correspond to the parameter functions $\alpha(\cdot)$, $\alpha_{I}(\cdot)$ and $\alpha_{D}(\cdot)$, respectively.
Empirical power values are based on $1000$ samples and plotted against the parameter $\lambda$.
The solid lines correspond to the CCC test $\Theta_\Myidx$, the dashdotted lines to the CCC test $T_\Myidx(\Gamma_0)$ and the dashed lines to the VI test.
Different sample sizes are shown color coded.
}\label{figCompToSumMedian}
\end{figure}

For the base partition we choose $\Gamma_0 = \Gamma_{\text{med}}$ as in \autoref{secAlgDecTree}.
As data generating processes we consider \autoref{exSimStudy} and the resulting vine copulas that arise if the 
parameter $\alpha(u_{2:3}; \lambda)$ of $C^{\text{\tiny Fr}}_{\alpha(u_{2:3}; \lambda)}$ in the edge of tree $T_3$ of \autoref{exSimStudy} is given by%
\begin{align}
\alpha_{I}(u_{2:3}; \lambda) &= 1 + 2.5 \lambda (1-2 u_2 (u_2+u_3))^2, \label{eqParFunctionalAlternative1}
\intertext{or}
\alpha_{D}(u_{2:3}; \lambda) &= 1 + 2.5 \lambda (1-2(u_2-u_3))^2. \label{eqParFunctionalAlternative2}
\end{align}
\autoref{figCompToSumMedian} shows the empirical power of the CCC tests $\Theta_\Myidx$ and $T_\Myidx(\Gamma_0)$ for the hypothesis $H_{0}: (\pvc{U}_{1|23}, \pvc{U}_{4|23}) \perp U_{2:3}$.
For the case of \autoref{exSimStudy} (left panel in \autoref{figCompToSumMedian}), the test with the fixed partition $\Gamma_0 = \Gamma_{\text{med}}$ delivers a test $T_\Myidx(\Gamma_0)$ which performs almost as good as $\Theta_n$. 
That is because the parameter $\alpha(u_{2:3}; \lambda)$ of $C^{\text{\tiny Fr}}_{\alpha(u_{2:3}; \lambda)}$ in \autoref{exSimStudy} can be written as a function of the mean of the conditioning variables $\bar{u}_{2:3} = \frac{1}{2} (u_2 + u_3)$.
Furthermore, the conditioning variables are positively associated due to the Clayton copula with $\tau_{23} = 0.4$.
As a result, the decision tree rarely finds a partition $\Gamma_{\max}$ which in terms of the test statistic $T_n(\Gamma_{\max})$ improves over the base partition $\Gamma_0 = \Gamma_{\text{med}}$ by an amount larger than the penalty $\lambda_n$.
Therefore, the test statistic $\Theta_n$ often coincides with the test statistic $T_n(\Gamma_0)$ of the base partition.

For the other two cases, the partition $\Gamma_{\text{med}}$ is not a good choice and the decision tree algorithm finds substantially better partitions in a data-driven way.
The varying parameter $\alpha_{I}(u_{2:3}; \lambda)$ of $C^{\text{\tiny Fr}}_{\alpha_{I}(u_{2:3}; \lambda)}$ (middle panel in \autoref{figCompToSumMedian}) introduces an interaction effect between the two conditioning variables.
Although the test with the fixed $\Gamma_{\text{med}}$ partition can detect some variation in $C^{\text{\tiny Fr}}_{\alpha_{I}(u_{2:3}; \lambda)}$, the decision tree finds better partitions which can increase the empirical power by more than 20 percentage points.
The gain of power is even more pronounced if 
the parameter $\alpha_{D}(u_{2:3}; \lambda)$ of $C^{\text{\tiny Fr}}_{\alpha_{D}(u_{2:3}; \lambda)}$ (right panel in \autoref{figCompToSumMedian})
 is a function of the difference of the conditioning variables. 
Even if $\lambda = 1$ and $C^{\text{\tiny Fr}}_{\alpha_{D}(u_{2:3}; \lambda)}$ is strongly varying in $u_{2:3}$, the test with the fixed partition $\Gamma_{\text{med}}$, that is based on the mean of the conditioning variables, cannot recognize the variation.
As a result, the empirical power is identical to the level of the test. 
On the contrary, the data-driven selection of the partition results in a substantial power increase.
For $\lambda = 1$ and $n = 1000$, the power increases from 5\% to 99\%. 

In summary, the choice of the base partition $\Gamma_0$ determines a lower bound for the empirical power of the test $\Theta_\Myidx$ and the decision tree can increase its power. 
The magnitude of the improvement depends on the data generating process and ranges from negligible, e.g., $\alpha(\cdot)$, to huge, e.g., $\alpha_{D}(\cdot)$. 
For all data generating processes, the power of the data-driven test $\Theta_\Myidx$ is much better than the power of the VI test. 
The difference is most pronounced for $\alpha_{D}(\cdot)$ where the empirical power of the VI test is always approximately 5\% while the empirical power of the CCC test $\Theta_\Myidx$ can be 99\%.
In \appref{appTypicalParts} we investigate what kind of partitions are typically selected by the decision trees and how they adapt to the
different variations in the conditional correlation determined by $\alpha(\cdot), \alpha_I(\cdot),$ and $\alpha_D(\cdot)$.

\subsection{Power study: The dimension of the conditioning set}\label{secPowerDim}
\begin{figure}[ht]
\begin{center}
\includegraphics[width=\textwidth]{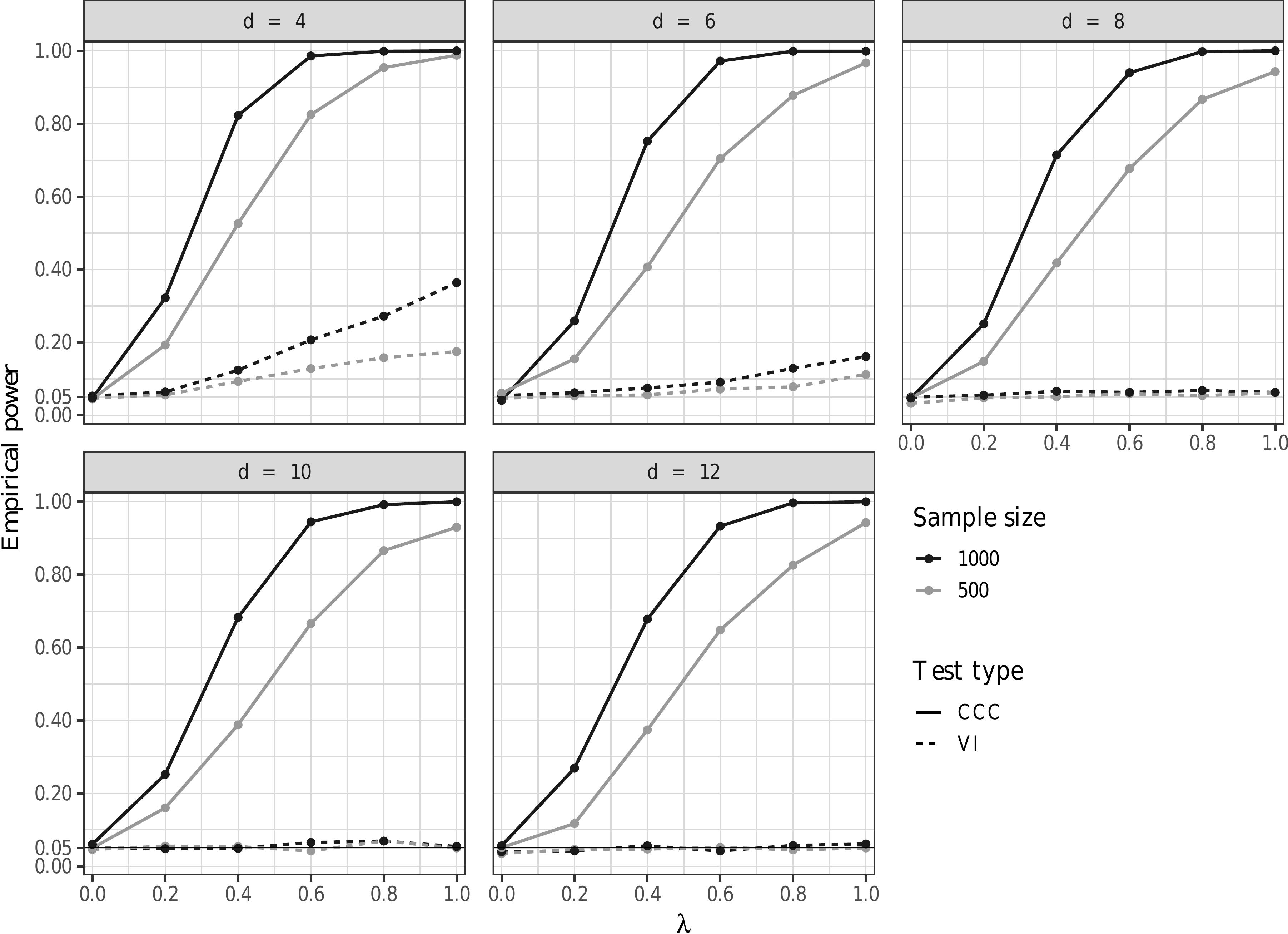}
\end{center}
\caption{
Empirical power of the CCC and VI test for the hypothesis $H_{0}: (\pvc{U}_{1|2:d-1}, \pvc{U}_{d|2:d-1}) \perp U_{2:d-1}$ in \autoref{exSimStudy2}.
Empirical power values are based on $1000$ samples and plotted against the parameter $\lambda$.
For $\lambda = 0$ the plots show the empirical size of the tests.
Each plot corresponds to a specific dimension $d$ of the D-vine copula (\autoref{exSimStudy2}).
The solid lines correspond to the CCC test $\Theta_\Myidx$ and the dashed lines to the VI test.
Different sample sizes are shown color coded.
}\label{figPowerDimension}
\end{figure}

For high-dimensional vine copulas, the dimension of the conditioning set of a conditional copula increases rapidly.
Therefore, it is substantial that a test still has power if the dimension of the conditioning set is not small. 
To investigate the performance of the CCC test w.r.t. the dimension of the conditioning set, we consider a up to twelve-dimensional Clayton copula where the Clayton copula in the edge of the last tree is replaced by a Frank copula with varying parameter.
\begin{myex}\label{exSimStudy2}
For $d \geq 4$, the building-blocks of the $d$-dimensional D-vine copula are chosen to be
\begin{align*}
C_{i,i+j\cs  \condsetDvine} &= C^{\text{\tiny Cl}}_{\theta_j}, \qquad 1\leq j \leq d-2, 1\leq i \leq d-j, \\
C_{1,d\cs 2:d-1} &= C^{\text{\tiny Fr}}_{\alpha(u_{2:3}; \lambda)}, \\
\alpha(u_{2:3}; \lambda) &= 1 + 2.5 \lambda (1-1.5(u_2+u_3))^2, \\
\theta_j &= \frac{\theta_1}{1 + (j-1) \theta_1}, \qquad 2\leq j \leq d-2, \\
\theta_1 &= \frac{2 \tau}{1-\tau}.
\end{align*}
\end{myex}
For $d=4$ \autoref{exSimStudy2} coincides with \autoref{exSimStudy} and as before we set $\tau=0.4$ and consider different values for $\lambda$.
For all dimensions, the functional form of the parameter only depends on the conditioning variables $U_2$ and $U_3$.
Therefore, the variation of $C_{1,d\cs 2:d-1} = C^{\text{\tiny Fr}}_{\alpha(u_{2:3}; \lambda)}$ in $u_{2:d-1}$
is always the same but the dimension of the testing problem increases with $d$.
Grouped by the dimension $d$, the empirical power and size of the VI and the CCC test $\Theta_\Myidx$ 
for the hypothesis $H_{0}: (\pvc{U}_{1|2:d-1}, \pvc{U}_{d|2:d-1}) \perp U_{2:d-1}$
are shown in \autoref{figPowerDimension}.

While the empirical power of the VI test decreases a lot if the dimension of the conditioning set is increased, the empirical power of the CCC test decreases only slightly for higher values of $d$.
In particular, for the setup $\lambda=1$ and $\Myidx = 1000$, the power of the VI test drops from 36\% to 5\% if the dimension is increased from $d=4$ to $d=12$. On the contrary, the power of the CCC test for this setup is always 100\%. 
Moreover, even when the power of the CCC test is not 100\% for $d=4$, the decrease in its power is still marginal. 
For instance, for $\lambda = 0.6$ and $\Myidx = 500$, the power of the CCC test only decreases from 83\% to 67\% while the power of the VI test quickly drops to approximately 5\% if the dimension is increased from $d=4$ to $d=12$.
Thus, the introduction of a penalty in the CCC test statistic $\Theta_\Myidx$ and the data-driven selection of the partition $\Gamma_{\max}$ by means of a decision tree yields a test whose power decreases only slightly with the dimension of the conditioning vector.

\subsection{Power study: Misspecification of the copulas in the lower trees}\label{secMisspec}
\begin{figure}[ht]
\begin{center}
\includegraphics[width=\textwidth]{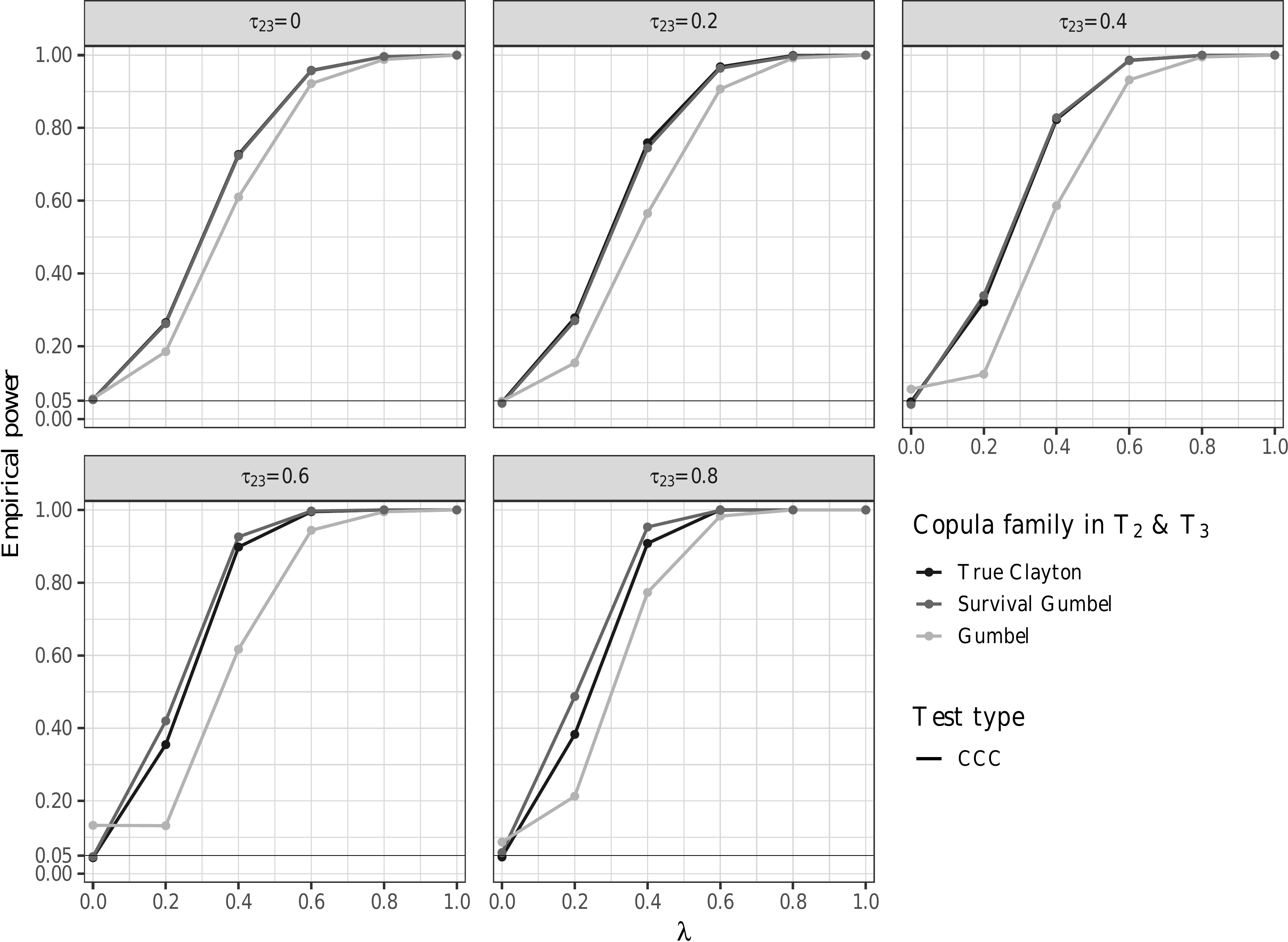}
\end{center}
\caption{
Empirical power of the CCC test $\Theta_\Myidx$ for the hypothesis \mbox{$H_{0}: (\pvc{U}_{1|23}, \pvc{U}_{4|23}) \perp U_{2:3}$} in \autoref{exSimStudy}.
Empirical power values are based on $1000$ samples of size $n = 1000$ and plotted against the parameter $\lambda$.
For $\lambda = 0$ the plots show the empirical size of the test.
Each plot corresponds to a specific value of Kendall's $\tau_{23}$ for the true copulas in the first and second tree.
The black lines show the results where the PPITs are estimated with the true Clayton copula family in the first and second tree.
The two grey lines correspond to the case where the PPITs are estimated with misspecified copula families (survival Gumbel and Gumbel, respectively).
}\label{figMisspec}
\end{figure}

The true family of the five copulas in the first and second tree $C_{12}, C_{23}, C_{34},\allowbreak C_{13\cs 2}, C_{24\cs 3}$  in \autoref{exSimStudy} is the Clayton copula.
To analyze the effect of misspecified copula families, we now vary the pairwise
value of Kendall's tau $\tau_{23} = \tau_{12} = \tau_{34} = \tau_{13} = 
\tau_{24}$ between $0$ and $0.8$ and estimate either survival Gumbel or Gumbel copulas for all five copulas in the lower trees.
The black lines in \autoref{figMisspec} show the results for correctly specified Clayton copulas in the lower trees as a benchmark.
Since the strength of the variation of $C_{14\ps 23}$ is more pronounced for higher values of $\tau_{23}$, the empirical power of the tests is also increasing
in $\tau_{23}$. 
The empirical size of the tests ($\lambda = 0$) is not influenced by $\tau_{23}$ and always close to the theoretical level of 5\%.

The dark grey line in \autoref{figMisspec} corresponds to a rather mild misspecification where we estimate survival Gumbel copulas in the first and second tree.
We see that the empirical size is still very close to the theoretical level of 5\% ($\lambda = 0$).
Moreover, the power of the test with misspecified survival Gumbel copulas is almost indistinguishable from the power of the test with correctly specified Clayton copulas.
If the degree of misspecification is severe and we fit Gumbel copulas (with upper tail dependence) to data generated from Clayton copulas (with lower tail dependence), differences in the empirical power of the CCC test become visible when comparing the light grey lines with the black lines in \autoref{figMisspec}.
In the majority of the considered scenarios the empirical power is now a little smaller.
In cases with high dependence, i.e., $\tau_{23} = 0.4, 0.6, 0.8$, the empirical size is increased.
This shows that the test might not control the size if the copula families in the lower trees are severely misspecified. 
Note that we misspecify five copula families and that the misspecification in the second tree might be even worse because the data in the edges of the second tree is no longer generated by Clayton copulas if the copulas in the first tree are misspecified.
Thus, the performance of the CCC test appears to be relatively robust w.r.t. such a severe misspecification.

\section{A hierarchical procedure for testing the \SA{} in vine copulas}\label{secSeqProcedure}
Up to now, we have tested single building blocks of vine copulas, i.e., the hypothesis \mbox{$H_0 \colon (\Ppitu, \Ppitv) \perp U_{\condset}$} for a fixed edge $\idxIterator \in \idxsett$.
To check the simplifying assumption for a $d$-dimensional R-vine copula with fixed structure $\mathcal{V}$, one has to check this constraint for $M = (\vineDim-1)(\vineDim-2)/2$ edges. 
In order to control the family-wise error rate $\alpha$, we apply the Bonferroni correction and test the set of hypotheses\footnote{
For each hypothesis corresponding to a fixed edge $\idxIterator \in \idxsett$ we use the same CCC test settings as before, i.e., decision trees with depth $J_{\max} = 2$, $\lambda_{\Myidx} = \frac{1}{\sqrt{\Myidx}}$ for the penalty and $\Gamma_0 = \Gamma_{\text{med}}$ (see equation~\eqref{gamma_med}) as base partition.}
\begin{align*}
\forall \idxIterator \in \idxsett: H_0: (\Ppitu, \Ppitv) \perp U_{\condset}.
\end{align*}
We use a hierarchical procedure and begin with testing the hypotheses in the second tree. 
We then only check the hypotheses for the next tree $T_{j+1}$ if the validity of the simplifying assumption could not be rejected for tree $T_j$, for $j= 2,\ldots, \vineDim-2$.
The hierarchical procedure is stopped whenever an individual hypothesis
$
H_0 \colon (\Ppitu, \Ppitv) \perp U_{\condset}
$
is rejected at a level of $\alpha/M$.
As a result, the hierarchical procedure detects critical building blocks of a vine copula model where 
\anuncop\ $\svc{C}_{h_e, i_e\cs D_e}$ does not seem to be an adequate model for the bivariate conditional copula $C_{h_e, i_e\cs D_e}$
of the bivariate conditional distribution $F_{h_e, i_e|D_e}$.
The procedure is also in line with the common sequential specification and estimation of simplified vine copulas and can be integrated in model selection algorithms as demonstrated in \citet{Kraus2017}.

\subsection{Simulation study}\label{secSimStudyMultCop}
In the following simulation study we test the simplifying assumption for the four-dimensional Clayton, Gaussian, Gumbel, and Frank copula with pair-wise values of Kendall's $\tau$ of $0.2, 0.4, 0.6, 0.8$.
Since each copula is exchangeable an arbitrary structure can be fixed.
We fit a D-vine copula and select the copula families of $\svc{C}_{12}$, $\svc{C}_{23}$, $\svc{C}_{34}$, $\svc{C}_{13; 2}$, $\svc{C}_{24; 3}$, $\svc{C}_{14; 23}$ using the AIC.
The hierarchical procedure is applied to test the simplifying assumption at a theoretical level of 5\%.
Thus, in tree $T_2$ two conditional copulas are tested with an individual level of 1.67\%.
If we do not reject the $H_0$ for both copulas in tree $T_2$, we continue in tree $T_3$ and test $H_{0}: (\pvc{U}_{1|23}, \pvc{U}_{4|23}) \perp U_{2:3}$ at an individual level of 1.67\%.

\begin{figure}[h]
\centering
\includegraphics[width=0.81\textwidth]{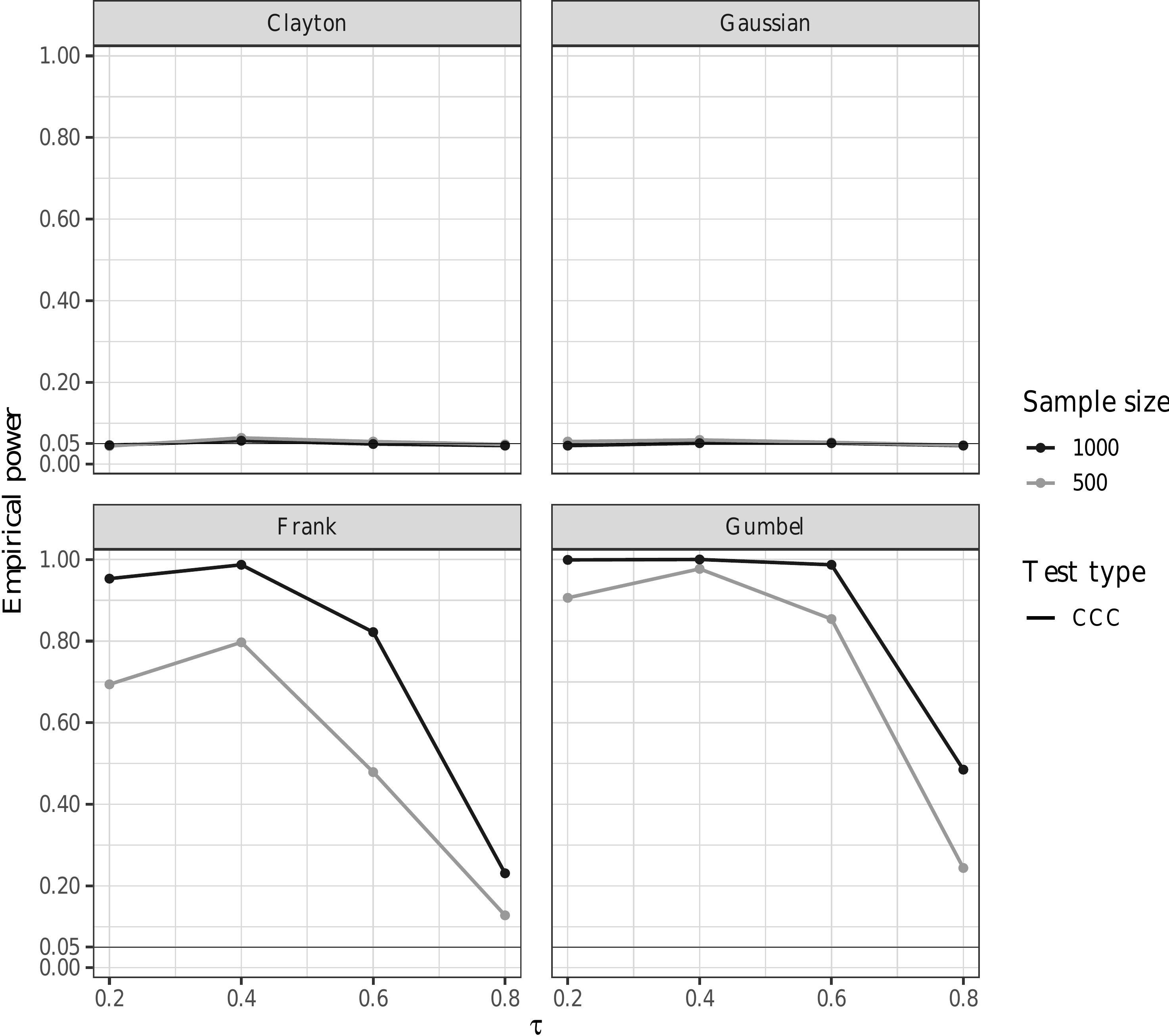}
\caption{
Empirical power of the hierarchical procedure with the CCC test $\Theta_\Myidx$ for the four-dimensional Clayton, Gaussian, Frank and Gumbel copula.
Vine copula models are estimated using the AIC for selecting the copula families.
Empirical power values are based on $1000$ samples and plotted against the pairwise value of Kendall's $\tau$.
Different sample sizes are shown color coded.
}\label{figMultivariateDistributions}
\end{figure}

\begin{figure}[h]
\begin{center}
\includegraphics[width=0.76\textwidth]{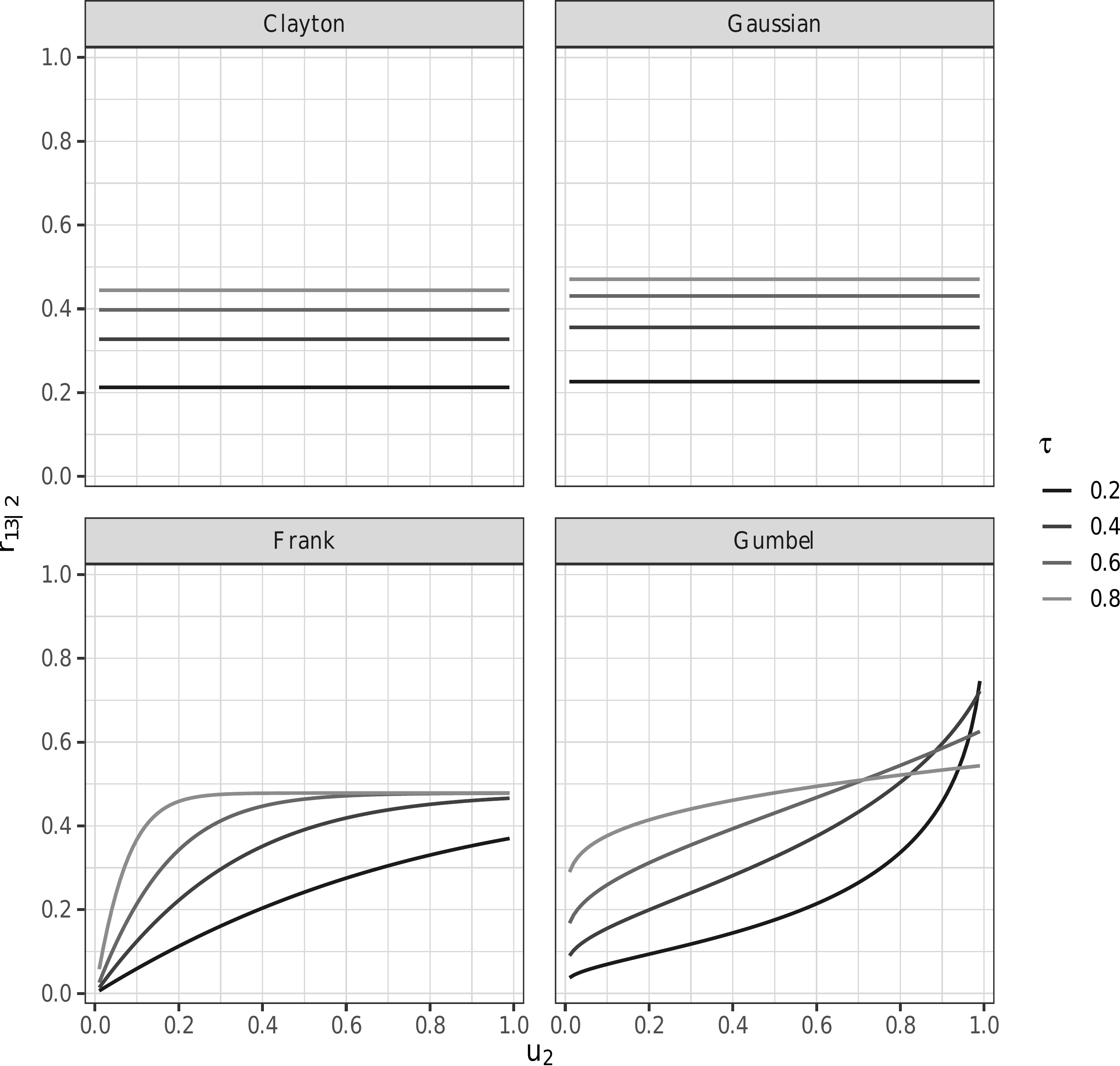}
\end{center}
\caption{
Conditional correlations $r_{13|2} = \Corr(U_{1|2}, U_{3|2} | U_{2})$ of $C_{13; 2}$ in tree $T_2$ of the Clayton, Gaussian, Frank and Gumbel copula.
}\label{figArchmCondCorr}
\end{figure}

The two upper panels in \autoref{figMultivariateDistributions} report the results for the four-dimensional Clayton and Gaussian copula for which the simplifying assumption is satisfied \citep{Stoeber2013}.
The empirical size of the hierarchical test procedure with the CCC test $\Theta_\Myidx$ is close to the theoretical level even under consideration of possibly misspecified copula families.
In the lower panels of \autoref{figMultivariateDistributions}, the empirical power for the four-dimensional Frank and Gumbel copula is plotted.
The Frank and Gumbel copula violate the simplifying assumption as long as $\tau \notin\{0,1\}$ \citep{Stoeber2013}.
Although the variation in the conditional copulas induced by the four-dimensional Frank and Gumbel copulas is rather mild (see \autoref{figArchmCondCorr}),
the CCC test often rejects the simplifying assumption. 
That the power has a minimum at $\tau = 0.8$ can be explained by the fact that both copulas satisfy the simplifying assumption for $\tau \to 1$
and the variation of the conditional correlations (see \autoref{figArchmCondCorr}) is less pronounced than for $\tau = 0.2, 0.4, 0.6$.

How the decision trees adapt to the variation of the conditional correlations can be seen by looking at the most frequently selected decision trees and the resulting $\Gamma_{\max}$ partitions.
For the Frank copula (with $n=1000$ and $\tau=0.2,0.4,0.6,0.8$) the most frequently selected decision tree in tree $T_2$ of the vine (selected in 21.28\% of all simulated cases) consists of a first split according to the 25\% quartile
and a second split at the 50\% quartile in the left part and at the 25\% quartile in the right part, i.e.,
\begin{align*}
\Gamma_{\max} = \Big\lbrace 
&\Lambda_{(0,l,l)} = [0, 0.125],\,
\Lambda_{(0,l,r)} = (0.125, 0.25],\,
\Lambda_{(0,r,l)} = (0.25, 0.4375], \\
&\Lambda_{(0,r,r)} = (0.4375, 1]
\Big\rbrace.
\end{align*}
Thus, the $\Gamma_{\max}$ partition is most granular for smaller values of the conditioning variable.
This is also the area where the variation of the Frank copula is most pronounced as can be seen from \autoref{figArchmCondCorr}.
In contrast, for the Gumbel copula (with $n=1000$ and $\tau=0.2,0.4,0.6,0.8$) the most common selected $\Gamma_{\max}$ partition in tree $T_2$  (selected in 18.51\% of all simulated cases) is 
\begin{align*}
\Gamma_{\max} = \Big\lbrace 
&\Lambda_{(0,l,l)} = [0, 0.375],\,
\Lambda_{(0,l,r)} = (0.375, 0.75],\,
\Lambda_{(0,r,l)} = (0.75, 0.875], \\
&\Lambda_{(0,r,r)} = (0.875, 1]
\Big\rbrace.
\end{align*}
Comparing $\Gamma_{\max}$ with \autoref{figArchmCondCorr}, which shows the conditional correlation for the Gumbel copula, we see that the decision tree adapts to the variation of the conditional correlation by selecting the most granular splits for higher values of the conditioning variable.

\begin{figure}[h]
\centering
\includegraphics[width=0.86\textwidth]{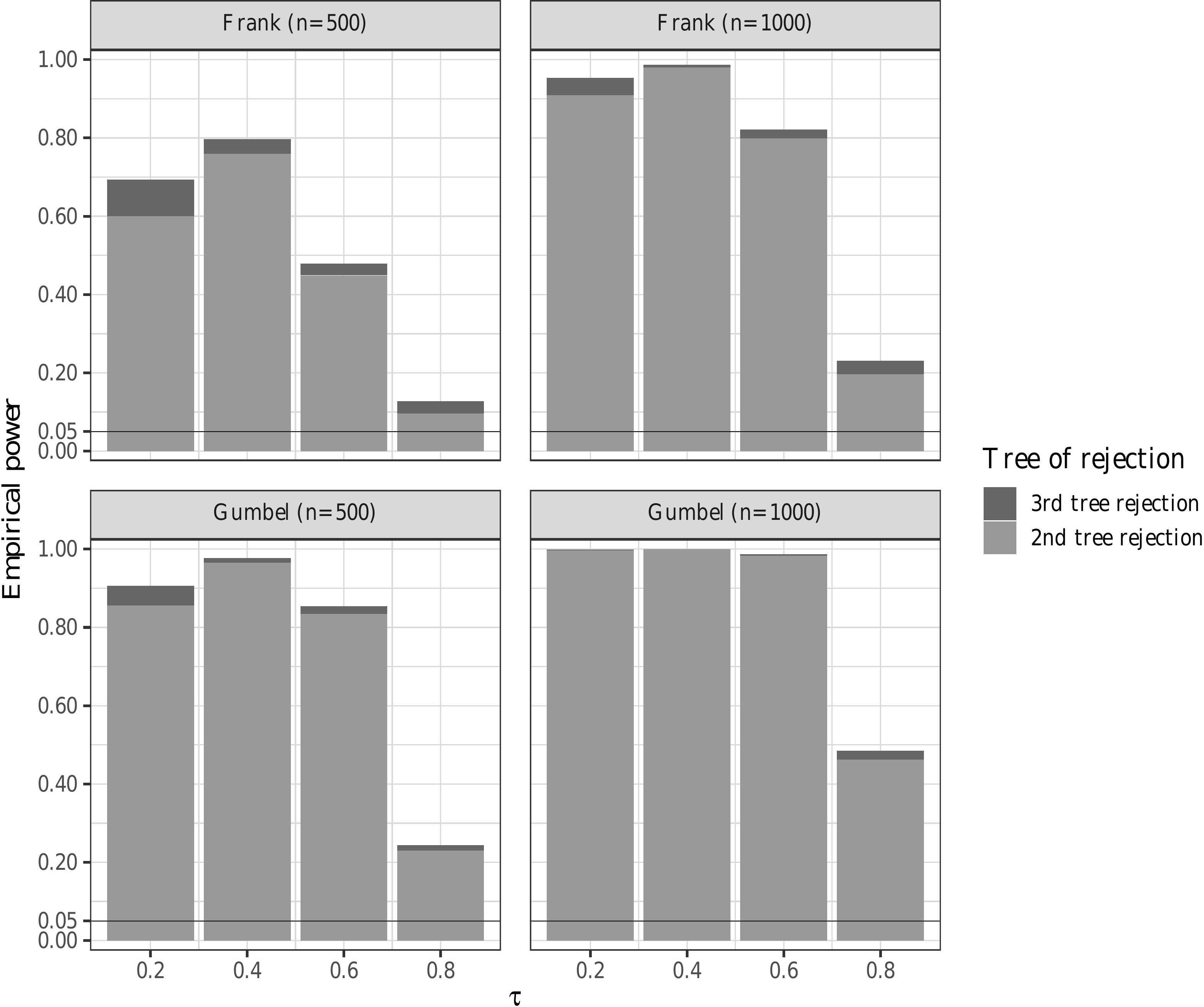}
\caption{
Empirical power of the hierarchical procedure with the CCC test $\Theta_\Myidx$ for the four-dimensional Frank and Gumbel copula.
The tree of rejections is shown color coded.
}\label{figTreeOfReject}
\end{figure}

Two hypotheses are tested in tree $T_2$ and, provided there is no rejection in tree $T_2$, one additional hypothesis is tested in tree $T_3$ when the hierarchical procedure for testing the simplifying assumption is applied to four-dimensional vine copulas.
In the simulation study, most of the rejections for the four-dimensional Frank and Gumbel copula happen in tree $T_2$.
In \autoref{figTreeOfReject} we decompose the empirical power of the hierarchical procedure with the CCC test $\Theta_\Myidx$ into second tree and third tree rejections.
Over all considered scenarios 94.09\% and 98.06\% of the rejections occur in the second tree $T_2$ for the Frank and Gumbel copula, respectively.

Finally, we revisit the up to twelve-dimensional \autoref{exSimStudy2} from \autoref{secPowerDim}.
In \autoref{secPowerDim} we only tested one hypothesis in the last tree and analyzed the effect of an increasing dimensionality of the conditioning set on the empirical power of the CCC test.
Now we apply the hierarchical procedure to test the simplifying assumption for the entire D-vine copula with fixed structure defined in \autoref{exSimStudy2} for $d=4, 6, 8, 10, 12$ and test the set of hypotheses
\begin{align*}
&\forall j=2, \ldots, d-1\, \forall i=1, \ldots, d-j:\\
&\qquad H_0: (\pvc{U}_{i|i+1:i+j-1}, \pvc{U}_{i+j|i+1:i+j-1}) \perp U_{i+1:i+j-1}.
\end{align*}
As mentioned before, the simplifying assumption for a $d$-dimensional vine copula is equivalent to $M=(d-1)(d-2)/2$ hypotheses.
This means that for \autoref{exSimStudy2} we might test $M=3, 10, 21, 36, 55$ hypotheses for $d=4, 6, 8, 10, 12$ but only the last hypothesis actually violates the simplifying assumption for $\lambda \neq 0$.

\begin{figure}[ht]
\centering
\includegraphics[width=\textwidth]{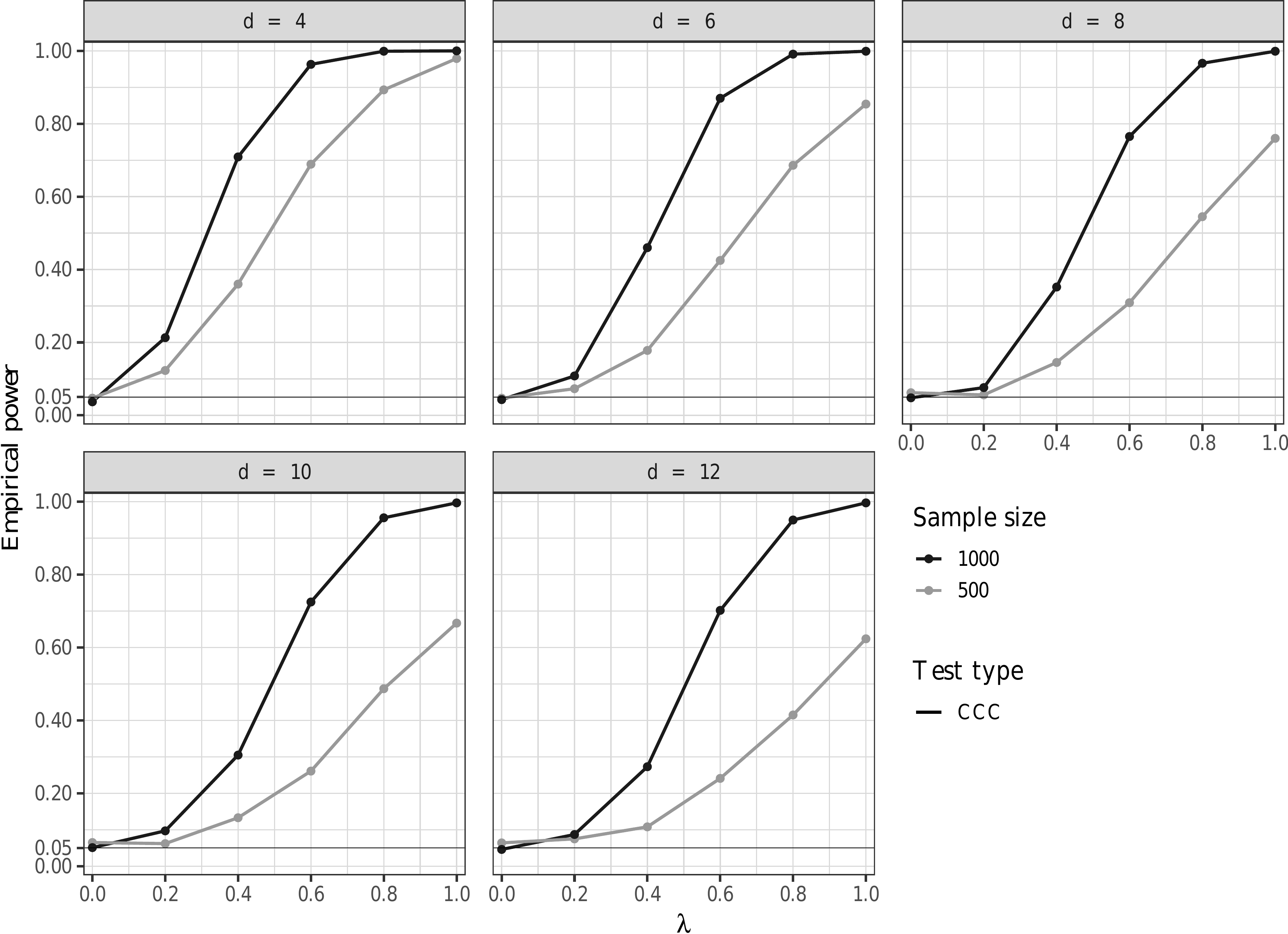}
\caption{
Empirical power of the hierarchical procedure with the CCC test $\Theta_n$ for the hypothesis that the simplifying assumption is satisfied for the vine copula defined in \autoref{exSimStudy2}.
Empirical power values are based on $1000$ samples and plotted against the parameter $\lambda$.
For $\lambda = 0$ the plots show the empirical size of the test.
Each plot corresponds to a specific dimension $d$ of the D-vine copula (\autoref{exSimStudy2}).
Different sample sizes are shown color coded.
}\label{figTestWholeVine}
\end{figure}

The empirical size and power of the hierarchical procedure with the CCC test $\Theta_\Myidx$ to test the simplifying assumption for data generated from \autoref{exSimStudy2} is shown in \autoref{figTestWholeVine}.
For $\lambda=0$ the simplifying assumption is satisfied and the empirical size is close to the theoretical level of 5\%.
Similar to \autoref{secPowerDim}, we observe that the empirical power of the hierarchical procedure with the CCC test $\Theta_\Myidx$ is slightly decreasing in the dimension $d$.
The effect is more pronounced for the smaller sample size $n=500$ as compared to $n=1000$.
Note that for $d=12$ only one of 55 building blocks of the vine copula violates the simplifying assumption.
This explains why for $n=500$ and $\lambda=1$ the empirical power of 62\% is lower than the empirical power of 94\% when only the last building block is tested (see \autoref{figPowerDimension}).
However, a sample size of $n=1000$ is sufficiently large so that the empirical power of the hierarchical procedure with the CCC test $\Theta_\Myidx$ is still almost 100\% in this scenario.

\subsection{Real data applications}\label{secRealData}
We now use the hierarchical procedure with the CCC test $\Theta_\Myidx$ to test the simplifying assumption for R-vine copulas fitted to four different real data sets.
The dimensionality of the data varies between $3$ and $10$ and the number of observations between $655$ and $5032$.
On the one hand, we consider the data set \texttt{uranium} \citep{Cook1986}
and obtain normalized ranks as pseudo-observations from the copula by means of the rescaled ecdf.
On the other hand, we consider three financial data sets  from the \textit{Kenneth R. French --  Data Library} (available under: \url{http://mba.tuck.dartmouth.edu/pages/faculty/ken.french/data_library.html}).
We apply ARMA($1$,$1$)-GARCH($1$,$1$)-filtering \citep{Engle1982,Bollerslev1986} with t-distributed innovations and apply the rescaled ecdf to the residuals to obtain pseudo-observations from the copulas.\footnote{Copula modeling for GARCH-filtered return data has been studied in \citep{Chen2006, Chan2009} and \citep{Patton2012} provides a review of copula models for economic time series.}

In order to test the simplifying assumption with the CCC test, the researcher has to specify the vine structure $\mathcal{V}$, which also determines the $H_0$, and the copula families for the edges. 
The selection of the vine structure $\mathcal{V}$ and copula families is commonly based on the data, e.g., by using the algorithm of \citet{Dissmann2013}. 
If model selection and statistical inference is done on the same data, statistical inference might no longer be valid (see \citet{Fithian2014} or \citet{Lee2016} for a discussion of post model selection inference).
Therefore, we randomly split the samples into two parts so that the partial vine copula model is selected on one half of the data and the other half of the data is used to apply the CCC test.
To obtain parametric models for the PVC we apply the standard R-vine model selection algorithm proposed by \citet{Dissmann2013} which is implemented in the \texttt{R}-package \texttt{VineCopula} \citep{Schepsmeier2017}.
The resulting vine structure $\mathcal{V}$ and copula families are then used to fit a model for the partial vine copula and conduct tests on the other half of the data. 

In \autoref{tableRealData} we provide information about all four data sets and report the results of the hierarchical test procedure with the CCC test $\Theta_\Myidx$.%
\footnote{Note that the CCC test $\Theta_\Myidx$ is not a consistent test because the conditional correlation can be constant if the simplifying assumption is false.
Moreover, if the conditional correlation is varying it may be possible that this variation is not detected by the proposed CCC test $\Theta_\Myidx$ with the decision tree based partitioning as described in \autoref{secDecTreeGroup}.}
For all cases where the validity of the simplifying assumption is rejected, we show the first tree in which we reject at least one null hypothesis of the form $H_0: (\Ppitu, \Ppitv)\indep U_{\condset}$ and stop the hierarchical test procedure.
Moreover, the smallest p-value of one hypothesis of the hierarchical procedure is also depicted.%
\begin{table}[htb]
{\scriptsize
\centering
\caption{Results for real data sets:
The simplifying assumption is tested by applying the hierarchical procedure with the CCC test $\Theta_\Myidx$.}\label{tableRealData}
\begin{tabular}{>{\RaggedRight\arraybackslash}p{2.5cm}@{\extracolsep{5pt}}>{\RaggedRight\arraybackslash}p{2.5cm}@{}>{\RaggedRight\arraybackslash}p{2.1cm}@{}>{\RaggedRight\arraybackslash}p{2.1cm}@{}>{\RaggedRight\arraybackslash}p{2.1cm}@{}}
\textbf{Name} &
\texttt{uranium} &
\texttt{FF3F} &
\texttt{FF5F} &
\texttt{Ind10}
\\ \hline \\[-3pt]
\textbf{Description} &
Uranium Exploration Data Set &
Fama/French 3 Factors &
Fama/French 5 Factors &
10 Industry Portfolios
\\ \\[-3pt]
\textbf{Source} &
\citet{Cook1986}, \texttt{R}-package \texttt{copula} & 
Kenneth R. French -- Data Library &
Kenneth R. French -- Data Library &
Kenneth R. French -- Data Library
\\ \\[-3pt]
\textbf{Variables} &
log concentration of \texttt{Uranium, Lithium, Cobalt, Potassium, Cesium, Scandium, Titanium} &
\texttt{SMB, HML, $R_m - R_f$} & 
\texttt{SMB, HML, RMW, CMA, $R_m - R_f$} & 
10 industry portfolios formed according to four-digit SIC codes.
\\ \\[-3pt]
\textbf{Period} &
--- &
02-Jan-2001 to \newline 31-Dec-2020 (daily) &
02-Jan-2001 to \newline 31-Dec-2020 (daily) &
02-Jan-2001 to \newline 31-Dec-2020 (daily)
\\ \\[-3pt]
\textbf{Dimension} &
7 \newline(21 pair-copulas) &
3 \newline(3 pair-copulas) &
5 \newline(10 pair-copulas) &
10 \newline(45 pair-copulas)
\\ \\[-3pt]
$\boldsymbol{n}$ & 
$655$ &
$5032$ &
$5032$ &
$5032$
\\[2pt] \hline \\[-3pt]
\textbf{Test decision of the hierarchical procedure} (at a 5\% significance level) &
The simplifying assumption can be \textbf{rejected} in tree $T_2$. &
The simplifying assumption \textbf{cannot} be rejected. &
The simplifying assumption \textbf{cannot} be rejected. &
The simplifying assumption \textbf{cannot} be rejected.
\\ \\[-3pt]
\textbf{Smallest p-value} [Bonferroni adjusted p-value]
of one hypothesis $H_0: (\Ppitu, \Ppitv) \perp U_{\condset}$
of the hierarchical procedure &
$<0.01$ \newline [$<0.01$] &
$0.20$ \newline [$0.20$] &
$0.21$ \newline [$1.00$] &
$0.10$ \newline [$1.00$]
\end{tabular}
}
\end{table}%

\begin{figure}[htb]
\centering
\includegraphics[width=\textwidth]{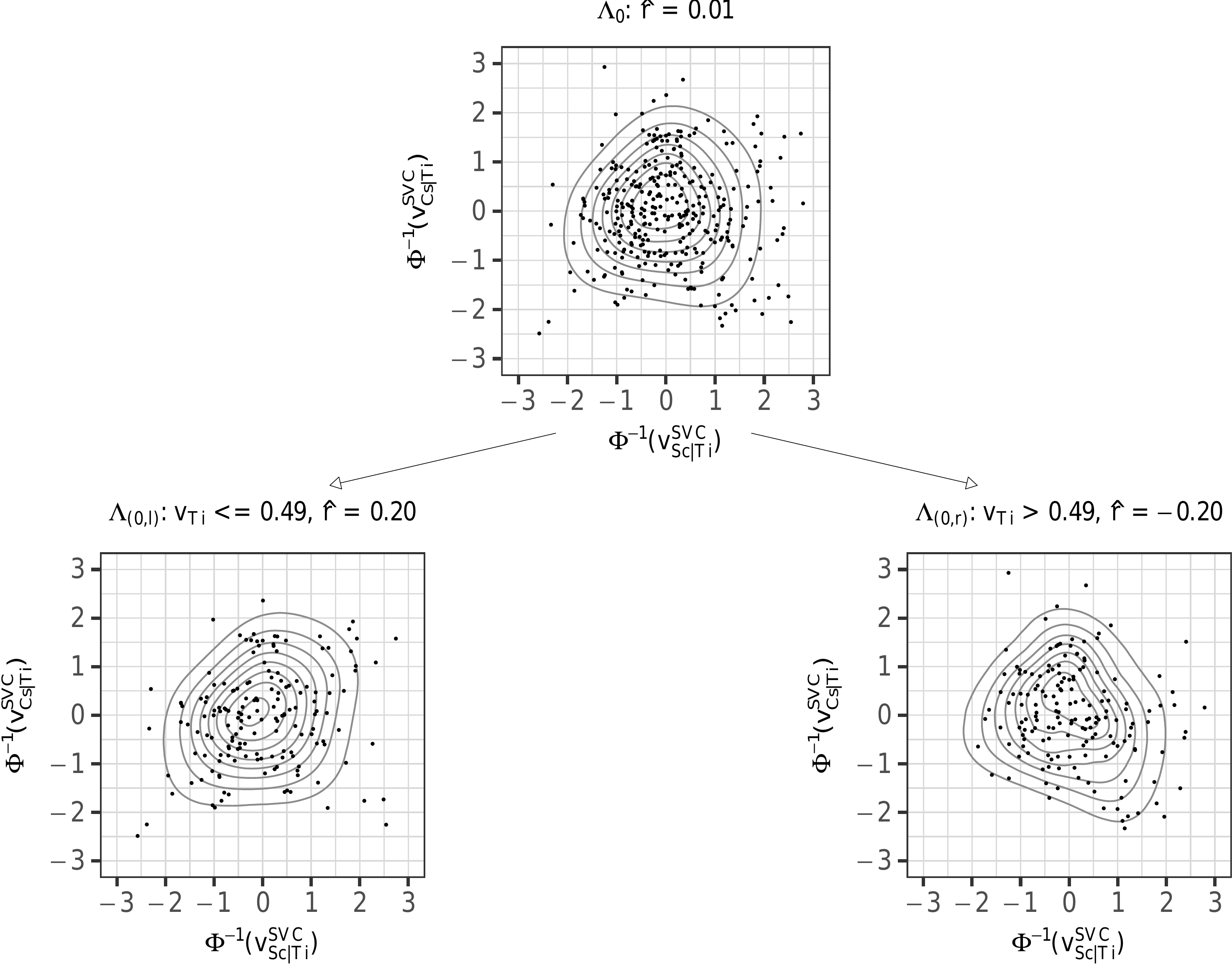}
\caption{
Scatterplots of pseudo-observations from the PPITs (see equation~\eqref{eq_pseudo_obs}) corresponding to the conditional copula of Scandium (\texttt{Sc}) and Cesium (\texttt{Cs}) given Titanium (\texttt{Ti}) for the \texttt{uranium} dat set.
The grouped scatterplots are formed according to the decision tree and show the data points that are used for the CCC test $\Theta_\Myidx$ to test the simplifying assumption.
The margins are transformed to be standard normal and the contours are derived from copula kernel density estimates.
}\label{figUraniumScatter}
\end{figure}

For the non-financial data example \texttt{uranium} we reject the simplifying assumption already in tree $T_2$.
The rejection is in line with the results reported by \citet{Gijbels2016} and \citet{Kraus2017}.
In \autoref{figUraniumScatter} we provide a visualization of the rejected building block for the \texttt{uranium} data set.
The grouped scatterplots are formed according to the decision tree used for the CCC test $\Theta_\Myidx$.
For easier visual inspection the margins are transformed to be standard normal and contours derived from copula kernel density estimates (\texttt{R}-package \texttt{kdecopula} \citep{kdecopula}) are shown.
The plots show that the conditional copula of Scandium (\texttt{Sc}) and Cesium (\texttt{Cs}) given Titanium (\texttt{Ti}) is varying with an estimated positive correlation of $0.20$ for values of Titanium being smaller than its median and a negative estimated correlation of $-0.20$ for larger values of Titanium.

For the three filtered financial returns the simplifying assumption cannot be rejected on the basis of the CCC test $\Theta_\Myidx$ (\autoref{tableRealData}).
This indicates that the possible violation of the \SA{} for the vine copulas selected by Di{\ss}mann's algorithm might be less severe for this kind of data as compared to \texttt{uranium}.
This is consistent with the findings of \citet{Kraus2017} who also use the CCC test and report that the simplifying assumption seems to be rather appropriate for filtered financial returns.

\section{Conclusion}\label{secConclusion}
In practical applications, a test for the simplifying assumption in high-dimensional vine copulas must be computationally feasible and tackle the curse of dimensionality.
The introduced hierarchical procedure with the CCC test $\Theta_n$ addresses these two issues.

The asymptotic distribution of the CCC test statistic is derived under the assumption of semi-parametrically estimated pseudo-observations from the partial probability integral transforms.
Since the test has a known asymptotic distribution and is based on the stepwise maximum likelihood estimator, it is computationally feasible also in high dimensions.
To prevent suffering from the curse of dimensionality, the CCC test utilizes a novel stochastic interpretation of the simplifying assumption based on the partial vine copula.
Moreover, we propose a discretization of the support of the conditioning vector into a finite number of subsets and incorporate a penalty in the test statistic.
A decision tree algorithm looks for the largest deviation from the simplifying assumption measured in terms of conditional correlations and also contributes to a computationally feasible test.

In a simulation study we provide a thorough analysis of the finite sample performance of the CCC test for various kinds of data generating processes.
The CCC test outperforms the vectorial independence test by a large margin if the conditional correlation is varying.
Even more important for high-dimensional applications, the simulation study demonstrates that the power of the test decreases only slightly with the dimension of the conditioning vector.
A moderate misspecification of the parametric copula families does not affect the power properties of the CCC test. 
We also investigate the performance of a hierarchical procedure that utilizes the CCC test to test the simplifying assumption.
An application to four real data sets demonstrates the usefulness of the test and indicates that the validity of the simplifying assumption should be checked individually for each data set.

The CCC test can also be utilized to improve the modeling of data with vine copulas.
\citet{Schellhase2017} make use of the CCC test to identify building blocks of vine copulas where the simplifying assumption does not seem to be adequate and the estimation of a conditional copula that is varying in its conditioning variables can improve the modeling.
Additionally, \citet{Kraus2017} use the CCC test to find alternative vine copula structures which might be more in line with the simplifying assumption.

\appendix
\section*{Appendix}

\renewcommand{\thesubsection}{A.\arabic{subsection}}
\renewcommand{\thesection}{A}

\subsection{Proof of \autoref{ProbStochInterpretation}}
That \ref{SI1} and \ref{SI4} are equivalent follows from the definition of the \SA{} in \autoref{testSA_simplidef} and the definition of the conditional and partial copula in \autoref{testSA_def_conditional_copula}.

For $\idxIterator \in \idxsett$, $(\Ppitu, \Ppitv)\perp U_{\condset}$
implies
$(\Ppitu, \Ppitv) \stackrel{\text{a.s.}}{=} (\Pitu, \Pitv)$
by Lemma 3.1 in \citet{Spanhel2016}. 
Thus,  
$(\Ppitu, \Ppitv, U_{\condset})$ and \linebreak$(\Pitu, \Pitv, U_{\condset})$ are equal in distribution and $(\Pitu, \Pitv)\perp U_{\condset}$.
This shows  \ref{SI4} $\Leftarrow$ \ref{SI3}.

To show \ref{SI4} $\Rightarrow$ \ref{SI3}, we use induction over the trees.
By the definition of the PPITs in the second tree $T_2$ we have that
\[
\forall \idxIterator \in E_{2}: \quad (\Ppitu, \Ppitv) \stackrel{\text{a.s.}}{=} (\Pitu, \Pitv).
\]
Because \ref{SI4} implies that 
\[
\forall \idxIterator \in E_{2}: \quad (\Pitu, \Pitv)\perp U_{\condset},
\]
it follows that
\[
\forall \idxIterator \in E_{2}: \quad (\Ppitu, \Ppitv)\perp U_{\condset}
\]
and the base case of the induction is proved.

We now assume the induction hypothesis that for tree $T_{j-1}$  
\[
\forall \bar{e} \in E_{j-1}: \quad
(\pvc{U}_{h_{\bar{e}}|D_{\bar{e}}}, \pvc{U}_{i_{\bar{e}}|D_{\bar{e}}}) \perp U_{D_{\bar{e}}}
\]
holds.
Note that Lemma 3.1 in \citet{Spanhel2016} then implies that
\begin{align}
\forall \bar{e} \in E_{j-1}: \quad
(\pvc{U}_{h_{\bar{e}}|D_{\bar{e}}}, \pvc{U}_{i_{\bar{e}}|D_{\bar{e}}}) \stackrel{\text{a.s.}}{=} (U_{h_{\bar{e}}|D_{\bar{e}}}, U_{i_{\bar{e}}|D_{\bar{e}}}). \label{indhyp}
\end{align}
By \ref{SI4}, we get for each $\idxIterator \in E_{j}$ in tree $T_j$ and $\indk=\indu,\indv$ that
\begin{align}
U_{\indk|\condset} &\stackrel{\text{a.s.}}{=} \partial_2 C_{l_{\bar{e}}, m_{\bar{e}} \ps D_{\bar{e}}}
(U_{l_{\bar{e}}|D_{\bar{e}}}, U_{m_{\bar{e}}|D_{\bar{e}}} \cs U_{D_{\bar{e}}}) \label{eq_cpit_ppit_1}
\\
&\stackrel{\text{a.s.}}{=} \partial_2\parsign{C}_{l_{\bar{e}}, m_{\bar{e}} \ps D_{\bar{e}}}
(U_{l_{\bar{e}}|D_{\bar{e}}}, U_{m_{\bar{e}}|D_{\bar{e}}}), \notag
\end{align}
where $\bar{e} \in E_{j-1}$ and
 $m_{\bar{e}} \in \condset$ are chosen such that
$(l_{\bar{e}}, m_{\bar{e}} \ps D_{\bar{e}}) = 
(l_{e}, m_{\bar{e}} \ps \condset \setminus m_{\bar{e}}) \in \mathcal{CV}$.\footnote{
By the proximity condition of the R-vine $\mathcal{V}$ (\autoref{testSA_rvinedef}) there exists a $m_{\bar{e}} \in \condset$ such that either
$(l_{e}, m_{\bar{e}} \ps \condset \setminus m_{\bar{e}}) \in \mathcal{CV}$
or $(m_{\bar{e}}, l_{e} \ps \condset \setminus m_{\bar{e}}) \in \mathcal{CV}$
with $\bar{e} \in E_{j-1} = N_{j}$
and w.l.o.g. we use the first case in \eqref{eq_cpit_ppit_1}.
}
Note that $\parsign{C}_{l_{\bar{e}}, m_{\bar{e}} \ps D_{\bar{e}}}$ is the distribution of $(U_{l_{\bar{e}}|D_{\bar{e}}}, U_{m_{\bar{e}}|D_{\bar{e}}})$
and $\pvc{C}_{l_{\bar{e}}, m_{\bar{e}} \ps D_{\bar{e}}}$ the distribution of $(\pvc{U}_{l_{\bar{e}}|D_{\bar{e}}}, \pvc{U}_{m_{\bar{e}}|D_{\bar{e}}})$.
By \eqref{indhyp} it follows that $\parsign{C}_{l_{\bar{e}}, m_{\bar{e}} \ps D_{\bar{e}}} = \pvc{C}_{l_{\bar{e}}, m_{\bar{e}} \ps D_{\bar{e}}}$.
Thus, 
\begin{align}
U_{\indk|\condset} &\stackrel{\text{a.s.}}{=}  \partial_2\pvc{C}_{l_{\bar{e}}, m_{\bar{e}} \ps D_{\bar{e}}}
(U_{l_{\bar{e}}|D_{\bar{e}}}, U_{m_{\bar{e}}|D_{\bar{e}}}) \nonumber 
\\
&\stackrel{\text{a.s.}}{=} \partial_2\pvc{C}_{l_{\bar{e}}, m_{\bar{e}} \ps D_{\bar{e}}}
(\pvc{U}_{l_{\bar{e}}|D_{\bar{e}}}, \pvc{U}_{m_{\bar{e}}|D_{\bar{e}}}) \label{eq_cpit_ppit_3} \\
&\stackrel{\text{a.s.}}{=} \pvc{U}_{\indk|\condset}, \label{eq_cpit_ppit_4}
\end{align}
where \eqref{eq_cpit_ppit_3}  follows from \eqref{indhyp} and \eqref{eq_cpit_ppit_4}
is the definition of $\pvc{U}_{\indk|\condset}$.
As a result, we have shown that 
\[
\forall \idxIterator \in E_{j}: \quad (\Ppitu, \Ppitv) \stackrel{\text{a.s.}}{=} (\Pitu, \Pitv).
\]
Finally, \ref{SI4} implies that 
\[
\forall \idxIterator \in E_{j}: \quad (\Ppitu, \Ppitv)\perp U_{\condset},
\]
which completes the induction to show \ref{SI4} $\Rightarrow$ \ref{SI3}.

\subsection{Proof of \autoref{Proposition2}}
\label{proof_proposition1}
We first prove the following lemma stating the asymptotic distribution of the test statistic $T_{\Myidx}^{\star}(\Gamma) = \Myidx (A \hat{R}_{\Gamma}^\star)^{T} (A\hat{\Sigma}_{R_{\Gamma}}^\star A^{T})^{-1} A \hat{R}_{\Gamma}^\star$ under $H_0: (\Ppitu,\Ppitv) \perp \CondVar$ and the assumption that observations from the PPITs are observable.
\begin{mylem}
\label{Proposition1}
Let $(U_{1:\vineDim}^{\myidx})_{\myidx = \oneToMyIdx}$ be $\Myidx$ independent copies of $U_{1:\vineDim} \sim C_{1:\vineDim}$ and
$\partElement_0 := \supp(\CondVar)$.
Consider a fixed R-vine structure $\mathcal{V}$
and let $\idxIterator \in \idxsett$ be a fixed edge.
Assume that the partition $\Gamma := \lbrace \partElement_1, \ldots, \partElement_\nGroups\rbrace$ satisfies the conditions stated in \autoref{Proposition2}.
Under \mbox{$H_0: (\Ppitu,\Ppitv) \perp \CondVar$} it holds that
\[
T_{\Myidx}^{\star}(\Gamma) \stackrel{d}{\rightarrow} \chi^2(\nGroups-1).
\]
\end{mylem}
\begin{myproof}
We first derive the asymptotic distribution of 
$\hat{R}_{\Gamma}^\star =\ (\hat{r}_1,\ldots,\hat{r}_L)$ under $H_0$ before showing that $T_\Myidx^\star(\Gamma)$ has an asymptotic chi-square distribution under $H_0$.
For this purpose, let $e_5 := (0,0,0,0,1)^T$,
$\otimes$ denote the Kronecker product,
$1_L$ be a $L \times 1$ column vector of ones and
$I_L$ be the $L \times L$ identity matrix,
so that $(I_L\otimes e_5)^T$ is a
$L \times 5L$ matrix that can be used to extract every fifth element from a $5L$-dimensional column vector. 
The correlations are then given by $\hat{R}_{\Gamma}^\star = (I_L\otimes e_5)^T\hat{\alpha}$, with $\hat{\alpha}$ being the unique solution of the estimating equation
\begin{align}
\frac{1}{\Myidx} \sum_{\myidx=1}^{\Myidx}
g_{\Gamma}(U_{1:\vineDim}^{\myidx}, \hat{\pi}, {\alpha})
\stackrel{!}{=} 0,
\label{esteq1}
\end{align}
where the estimating function $g_{\Gamma}$ will be stated in the following.

Define
\begin{align*}
g_{\pi}(U_{1:d}^k,\pi) = 
\bigg(
\pi_1-  \ivb{ U_{\condset}^{\myidx} \in \partElement_1},
\ldots,
\pi_L-  \ivb{ U_{\condset}^{\myidx} \in \partElement_\nGroups}
\bigg)^{T},
\end{align*}
where $\pi:= \pi_{1:L}\in \domainOfPi$.
The solution $\hat{\pi}$ of $\frac{1}{n}\sum_{k=1}^ng_{\pi}(U_{1:d}^k,\pi)\stackrel{!}{=}0$ denotes the random fraction of data corresponding to $\partElement_\iGroup$, i.e.,
\[
\hat{\pi}_{\iGroup} := 
\frac{1}{\Myidx} \sum_{\myidx=1}^{\Myidx} \ivb{ U_{\condset}^{\myidx} \in \partElement_\iGroup}, \quad l=1,\ldots,L.
\]

Define
\begin{align*}
h(U_{1:\vineDim}^{\myidx},\phi) &=
\left(
\begin{matrix}
\phi_{1} - \PpituIdx \\
\phi_{2} - \PpitvIdx \\
\phi_{3} - (\PpituIdx - \phi_{1})^2 \\
\phi_{4} - (\PpitvIdx - \phi_{2})^2 \\
\phi_{5} - (\PpituIdx - \phi_{1})(\PpitvIdx - \phi_{2})(\phi_{3} \phi_{4})^{-\frac{1}{2}} \\
\end{matrix}
\right),
\end{align*}
where $\phi:=\phi_{1:5} \in \mathbb{R}^5$.
For  $\partElement_\iGroup \in \Gamma = \lbrace \partElement_1, \ldots, \partElement_\nGroups \rbrace$ we set 
\[
f(U_{1:\vineDim}^{\myidx}, \pi_l,\phi) =
\pi_l^{-1} \ivb{ \CondVar^{\myidx} \in \partElement_\iGroup }
 h(U_{1:\vineDim}^{\myidx},\phi).
\]
The estimating function $g_{\Gamma}$ in \eqref{esteq1} is  given by 
\begin{align*}
g_{\Gamma}(U_{1:\vineDim}^{\myidx}, {\pi}, \alpha) &:= 
\bigg(  f(U_{1:\vineDim}^{\myidx}, \pi_1, \alpha_{1})^T, \ldots,  f(U_{1:\vineDim}^{\myidx}, \pi_\nGroups,
 \alpha_{\nGroups})^T \bigg)^{T}.
\end{align*}
Let $\pi_0$ be the unique solution of $\expec[g_{\pi}(U_{1:d}^k,\pi)]=0$,
$\phi_0$ be the unique solution of 
$\expec[h(U_{1:d}^k,\phi)]=0$, and 
$\alpha_{0} = (\phi_{0}^T, \ldots, \phi_{0}^T)^{T}$ so that
$\expec[g_{\Gamma}(U_{1:\vineDim}^{\myidx}, \pi, \alpha_0)]=0$ for all 
$\pi\in\domainOfPi$ under $H_0$ because for each $l$-th block element of $\expec[g_{\Gamma}(U_{1:\vineDim}^{\myidx}, \pi, \alpha_0)]$ it holds that
\begin{align*}
\notag
\big(\expec[ g_{\Gamma}(U_{1:\vineDim}^{\myidx}, \pi, \alpha_0)]\big)_l & := 
\expec[ f(U_{1:\vineDim}^{\myidx}, \pi_l, \phi_{0})]
=
\expec\left[
\pi_l^{-1} \ivb{ \CondVar^{\myidx} \in \partElement_{1}} h(U_{1:\vineDim}^{\myidx},\phi_0)
\right]
\\
& \stackrel{H_0}{=} \expec\left[\pi_l^{-1} \ivb{ \CondVar^{\myidx} \in \partElement_\iGroup }
\right]
\underbrace{\expec[h(U_{1:d}^k,\phi_0)]}_{= 0} = 0.
\end{align*}
Using the same steps it can be readily verified that $\expec[\partial_{\pi^T}g_{\Gamma}(U_{1:d}^k,\pi,\alpha_0)]=0$ for all $\pi\in\domainOfPi$ under $H_0$.
Thus, under $H_0$, the standard theory of estimating equations for two-step estimators, e.g., Theorem 6.1 in \citet{Newey1994},
yields that
\begin{align}
\label{mc1}
\sqrt{\Myidx}
(\hat{\alpha} - \alpha_0)
\stackrel{d}{\rightarrow} N_{5 L}(0, G_{\Gamma}^{-1} \Omega_{\Gamma} (G_{\Gamma}^{-1})^{T}),
\end{align}
where $G_{\Gamma}:= \expec\big[\partial_{\alpha^{T}}g_{\Gamma}(U_{1:\vineDim}^{\myidx}, \pi_0, \alpha_0)\big]$,
$\Omega_{\Gamma} := \Cov\big[g_\Gamma(U_{1:d}^k,\pi_0,\alpha_0)\big]$
and $N_{d}(\mu, \Sigma)$ denotes a $d$-dimensional normal distribution with mean vector $\mu$ and covariance matrix $\Sigma$.

If we now extract every fifth element from  $\hat{\alpha}$  using 
 $\hat{R}_{\Gamma}^\star =  (I_L\otimes e_5)^T \hat{\alpha}$ , we obtain the joint asymptotic distribution of the estimated  correlations under $H_0$ as
\begin{align*}
\sqrt{\Myidx}((\hat{r}_{1}, \ldots,\hat{r}_{\nGroups})^{T} - (r_{1}, \ldots,r_{\nGroups})^{T})=
\sqrt{\Myidx}
(\hat{R}_{\Gamma}^\star - R_{\Gamma})
\stackrel{d}{\rightarrow} N_{L}(0, \Sigma_{R_{\Gamma}}^\star),
\end{align*}
so that 
\begin{align*}
\Sigma_{R_{\Gamma}}^\star & =
(I_L\otimes e_5)^T G_{\Gamma}^{-1} \Omega_{\Gamma} (G_{\Gamma}^{-1})^{T} (I_L\otimes e_5).
\end{align*}

Under $H_0$  it holds that $r_{1} = \ldots = r_{\nGroups} = r = \Corr(\Ppitu, \Ppitv)$
and therefore it follows with the $(\nGroups-1) \times \nGroups$ first-order difference matrix $A$
and the continuous mapping theorem, that
\begin{align*}
&\sqrt{\Myidx}
A \hat{R}_{\Gamma}^\star
\stackrel{d}{\rightarrow} N_{L-1}(0, A \Sigma_{R_{\Gamma}}^\star A^{T}).
\end{align*}
To obtain the statistic of the CCC test when a sample from the PPITs is observable, the covariance matrix 
\[
\Sigma_{R_{\Gamma}}^\star = \Cov\big[(I_L\otimes e_5)^T G_{\Gamma}^{-1} 
g_{\Gamma}(U_{1:\vineDim}^{\myidx}, \pi_0, \alpha_0)\big]
\]
has to be consistently estimated, e.g., 
by
$\hat{\Sigma}_{R_{\Gamma}}^\star = 
\widehat{\Cov}[(I_L\otimes e_5)^T \hat{G}_{\Gamma}^{-1} g_{\Gamma}(U_{1:\vineDim}^{\myidx}, \hat{\pi}, \hat{\alpha})]$,
where $\widehat{\Cov}[X]$ denotes the sample covariance of the random vector $X$.
By applying once more the continuous mapping theorem and Slutsky's theorem, we get
\[
T_{\Myidx}^\star(\Gamma) = \Myidx (A \hat{R}_{\Gamma}^\star)^{T} (A\hat{\Sigma}_{R_{\Gamma}}^\star A^{T})^{-1} A \hat{R}_{\Gamma}^\star \stackrel{d}{\rightarrow} \chi^2(\nGroups-1)
\]
and \autoref{Proposition1} is proven. 
\end{myproof}

The remaining  part of the proof of \autoref{Proposition2} requires the definition of the pseudo stepwise maximum likelihood estimator of the vine copula parameters.
This estimator can be obtained as the solution of estimating equations (\citet{HobakHaff2013}, \citet{Spanhel2016c}, \citet{Tsukahara2005}).
By extending these estimating equations by the ones for the correlations defined in the proof of \autoref{Proposition1} we derive the asymptotic distribution of the CCC test when pseudo-observations from the PPITs  are estimated.
Consider a fixed R-vine structure $\mathcal{V}$ and let $\{\svc{C}_{1:d;\, \theta_{1:d-1}}\colon \theta_{1:d-1} \in \Upsilon\}$ be a parametric simplified vine copula such that $\exists \theta_{1:d-1;\, 0}\in \Upsilon$ so that $\svc{C}_{1:d;\, \theta_{1:d-1;\, 0}}= \pvc{C}_{1:d}$, where 
$\pvc{C}_{1:d}$ denotes the PVC of $(C_{1:d}, \mathcal{V})$.
The density of $\svc{C}_{1:d;\, \theta_{1:d-1;\, 0}}$ is given by
\[
\svc{c}_{1:\vineDim;\, \theta_{1:d-1}}(u_{1:\vineDim}) = \prod_{\idxIterator\in\idxset}
\svc{c}_{\indu, \indv\ps \condset;\, \theta_{j,e}}(\svc{u}_{\indu|\condset}(\theta_{1:j-1}),
\svc{u}_{\indv|\condset}(\theta_{1:j-1})),
\]
where $\svc{C}_{\indu, \indv\ps \condset;\, \theta_{j,e}}$ is a bivariate copula for each $\idxIterator\in\idxset$ with parameter $\theta_{j, e}$.
For $1\leq j \leq \vineDim-1$, the vector $\theta_{j}:=(\theta_{j,e})^{T}_{e \in E_j}$ collects the parameters in tree $T_j$ and the vector $\theta_{1:j} := (\theta_1, \ldots, \theta_j)^{T}$ collects all parameters up to and including tree $T_j$.
The individual stepwise pseudo score functions for the copulas in tree $T_j$ are given by
\begin{align*}
\partial_{\theta_j} \ell_{j}(V_{1:\vineDim}^k; \theta_{1:j}) :=
\partial_{\theta_j} \sum_{e \in E_{j}}
\ln\left(
\svc{c}_{\indu, \indv\ps \condset;\, \theta_{j,e}}(\svcIdx{V}_{\indu|\condset}(\theta_{1:j-1}),
\svcIdx{V}_{\indv|\condset}(\theta_{1:j-1}))
\right),
\end{align*}
$j=1 ,\ldots, \vineDim -1$.
Here, the pseudo-observations of the PPITs 
for $e \in E_{1}$, $\indk = \indu, \indv$ are defined by
\begin{align*}
\svcIdx{V}_{\indk|\condset}(\theta_{1:0}) &:= V_{\indk}^{\myidx}:= \frac{1}{\Myidx + 1} \sum_{m = 1}^{\Myidx} \ivb{ X_{\indk}^{m} \leq X_{\indk}^{\myidx}},
\end{align*}
and for $\idxIterator \in \idxsett$  as
\begin{align}
\svcIdx{V}_{\indk|\condset}(\theta_{1:j-1})
&:= \partial_2\svc{C}_{l_{\bar{e}}, m_{\bar{e}} \ps D_{\bar{e}};\, \theta_{j-1, \bar{e}}}(
\svcIdx{V}_{l_{\bar{e}}|D_{\bar{e}}}(\theta_{1:j-2}),
\svcIdx{V}_{m_{\bar{e}}|D_{\bar{e}}}(\theta_{1:j-2})
), \label{eq_pseudo_obs}
\end{align}
where $\indk = \indu, \indv$ and
$(l_{\bar{e}}, m_{\bar{e}} \ps D_{\bar{e}}) = 
(l_{e}, m_{\bar{e}} \ps \condset \setminus m_{\bar{e}}) \in \mathcal{CV}$
is selected as in \autoref{testSA_svc}.

Set $\theta := \theta_{1:\vineDim-1}$ and
define the estimating function
\begin{align*}
g_{S}(V_{1:\vineDim}^{\myidx}, \theta)   =
(\partial_{\theta_1} \ell_{1}(V_{1:\vineDim}^{\myidx}; \theta_{1})^{T}, \ldots,
\partial_{\theta_{d-1}} \ell_{\vineDim-1}(V_{1:\vineDim}^{\myidx}; \theta_{1:\vineDim-1})^{T})^{T},
\end{align*}
so that the solution $\hat{\theta}$ of 
$\frac{1}{n}\sum_{k=1}^ng_S(V_{1:d}^k,\theta)\stackrel{!}{=}0$ is the pseudo stepwise maximum likelihood estimator. 

Moreover, $g_{\Gamma}(V_{1:\vineDim}^{\myidx}, {\pi},{\theta}, \alpha)$ denotes the estimating function of the correlations when pseudo-observations from the PPITs are used,
i.e., 
\begin{align*}
g_{\Gamma}(V_{1:\vineDim}^{\myidx}, {\pi}, \theta, \alpha) &:= 
\bigg(  f(V_{1:\vineDim}^{\myidx}, \pi_1, \theta,\alpha_{1})^T, \ldots,  f(V_{1:\vineDim}^{\myidx}, \pi_\nGroups, \theta,
 \alpha_{\nGroups})^T \bigg)^{T},
\end{align*}
where
\begin{align*}
f(V_{1:\vineDim}^{\myidx}, \pi_l,\theta,\phi) =
\pi_l^{-1} \ivb{ \CondVar^{\myidx} \in \partElement_\iGroup }
\begin{pmatrix}
\phi_{1} - \svcIdx{V}_{\indu|\condset}(\theta_{1:j-1}) \\
\phi_{2} - \svcIdx{V}_{\indv|\condset}(\theta_{1:j-1}) \\
\phi_{3} - (\svcIdx{V}_{\indu|\condset}(\theta_{1:j-1}) - \phi_{1})^2 \\
\phi_{4} - (\svcIdx{V}_{\indv|\condset}(\theta_{1:j-1}) - \phi_{2})^2 \\
\phi_{5} - \frac{(\svcIdx{V}_{\indu|\condset}(\theta_{1:j-1}) - \phi_{1})(\svcIdx{V}_{\indv|\condset}(\theta_{1:j-1}) - \phi_{2})}{(\phi_{3} \phi_{4})^{\frac{1}{2}}} 
\end{pmatrix}.
\end{align*}

Let $\beta := (\theta^T, \alpha^T)^{T}$ 
so that the   estimating function of the vine copula parameters and the correlations is given by
\[
g(V_{1:\vineDim}^{\myidx}, {\pi}, \beta) = 
\left(\begin{matrix}
g_{S}(V_{1:\vineDim}^{\myidx}, \theta) \\
g_{\Gamma}(V_{1:\vineDim}^{\myidx}, {\pi}, \theta, \alpha) \\
\end{matrix}
\right).
\]
The rank approximate estimator $\hat{\beta}$ is then given as the solution of \linebreak$\frac{1}{n}\sum_{k=1}^ng(V_{1:d}^{\myidx}, \hat{\pi},\beta)\stackrel{!}{=}0$ where $\hat{\pi}$ is given as in the proof of \autoref{Proposition1}.
To derive the asymptotic distribution of $\hat{\beta}$, 
introduce
\begin{align*}
W_i &:= \int \partial_{u_i} g(u_{1:\vineDim}, \pi_0,\beta_0) \ivb{ U_i \leq u_i} \text{d}C_{1:\vineDim}(u_{1:\vineDim}),
&i=1,\ldots,\vineDim, 
\\
\tilde{\Omega} &:= \Cov\left(g(U_{1:\vineDim}^k, \pi_0,\beta_0) + \sum_{i=1}^{\vineDim} W_i \right) , 
\end{align*}
where $\beta_0 := (\theta_{1:d-1;\, 0}^T, \alpha_0^T)^{T}$ is the unique solution of $\mathbb{E}[g(U_{1:\vineDim}^{\myidx}, \pi,\beta_{0})] = 0$ for all $\pi\in\domainOfPi$ under $H_0$.
By the same reasoning as in the proof of \autoref{Proposition1} and because $g_S(U_{1:d}^k, \theta)$ does not depend on $\pi$ it follows that $\expec\big[\partial_{\pi^T} g(U_{1:d}^k,\pi,\beta_0)\big]= 0$ for all $\pi\in\domainOfPi$ under $H_0$.
Thus, provided the regularity conditions in Theorem 1 in
\citet{HobakHaff2013} are satisfied, it follows that
\[
\sqrt{\Myidx}
(\hat{\beta} - \beta_0)
\stackrel{d}{\rightarrow} N_{n_{\theta} + 5L}(0, \tilde{\Sigma}),
\]
where $\tilde{\Sigma} = G^{-1} {\tilde\Omega} (G^{-1})^{T}$ with 
$G := - \mathbb{E}[\partial_{\beta^T} g(U_{1:\vineDim}^{\Myidx}, \pi_0,\beta_0)]$
and $n_\theta$ is the number of vine copula parameters, i.e., the length of the vector $\theta$.

To extract the estimated correlations $\hat{R}_\Gamma$ from $\hat{\beta}$ and to obtain the corresponding asymptotic covariance matrix, we can exploit the block-structure of $G$ as follows
\begin{align*}
G &=
\begin{pmatrix}  \expec[\partial_{\theta^T} g_{S}(U_{1:\vineDim}^{\myidx}, \theta_0)] &
0 \\
\expec[\partial_{\theta^T}g_{\Gamma}(U_{1:\vineDim}^{k}, \pi_0,\theta_0,\alpha_0)]
& \expec[\partial_{\alpha^T}g_{\Gamma}(U_{1:\vineDim}^{k}, \pi_0,\theta_0,\alpha_0)]
\end{pmatrix} \\
&= \begin{pmatrix}  \expec[\partial_{\theta^T} g_{S}(U_{1:\vineDim}^{\myidx}, \theta_0)] &
0 \\
\expec[\partial_{\theta^T}g_{\Gamma}(U_{1:\vineDim}^{k}, \pi_0,\theta_0,\alpha_0)]
& G_\Gamma
\end{pmatrix}.
\end{align*}
Denote the $n_\theta \times L$ matrix consisting of zeros by $0_{n_\theta \times L}$ and define
\linebreak$\delta := (0_{n_\theta \times L}^T, (I_L\otimes e_5)^T)^T$ so that $\hat{R}_\Gamma = \delta^T \hat{\beta}$.
The asymptotic covariance matrix of $\sqrt{\Myidx} (\hat{R}_{\Gamma} - R_{\Gamma})$ is then 
\begin{align*}
{\Sigma}_{R_{\Gamma}} &= 
 \Cov\big[\delta^T G^{-1} \big( g(U_{1:\vineDim}, \pi_0,\beta_0) + \sum_{i=1}^{\vineDim} W_i \big) \big]
\\
&= \Sigma_{R_{\Gamma}}^\star + 
\big( \Cov\big[\delta^T G^{-1}g(U_{1:d},\pi_0,\beta_0)\big] - \Sigma_{R_{\Gamma}}^\star \big)
\\
& \quad +  \Big( \Cov\big[\delta^T G^{-1}\big( g(U_{1:\vineDim}, \pi_0,\beta_0) + \sum_{i=1}^{\vineDim} W_i \big] \big)
- \Cov\big[ \delta^T G^{-1}g(U_{1:d},\pi_0,\beta_0) \big]\Big)
\\
& = \Sigma_{R_{\Gamma}}^\star + \Sigma_{\text{\tiny PVC}} + \Sigma_r.
\end{align*} 
Thus, under $H_0$ it follows that
\begin{align*}
\sqrt{\Myidx}
(\hat{R}_{\Gamma} - R_{\Gamma})
\stackrel{d}{\rightarrow} N_{L}(0, {\Sigma}_{R_{\Gamma}}).
\end{align*}
With the same arguments as in the proof of \autoref{Proposition1} this implies under $H_0$
\[
T_{\Myidx}(\Gamma) = \Myidx (A \hat{R}_{\Gamma})^{T} (A\hat{\Sigma}_{R_{\Gamma}} A^{T})^{-1} A \hat{R}_{\Gamma} \stackrel{d}{\rightarrow} \chi^2(\nGroups-1),
\]
where
\begin{align}
\hat{\Sigma}_{R_{\Gamma}} = 
\widehat{\Cov}\left(\delta^T \hat{G}^{-1}(g(U_{1:\vineDim}^{\myidx}, \hat{\pi}, \hat{\beta}) + \sum_{i=1}^{\vineDim} \hat{W}_i) \right)
\label{eq_sigma_hat}
\end{align}
is a consistent estimator of $\Sigma_{R_{\Gamma}}$.\footnote{
See \citet{genest1995} for a consistent estimator of $W_i = \int \partial_{u_i} g(u_{1:\vineDim}, \pi_0,\beta_0) \ivb{ U_i \leq u_i} \text{d}C_{1:\vineDim}(u_{1:\vineDim})$.}

\subsection{Proof of \autoref{Proposition3}}\label{proof_proposition3}
To obtain the asymptotic distribution of the test statistic $\Theta_{\Myidx}$, we need the following lemmas.
\begin{mylem}
\label{Lemma1}
Let $Y \sim F_Y$, where $F_Y$ is the cdf of a continuous probability distribution.
Additionally, let $\lambda_{\Myidx}$ be a penalty function
satisfying the conditions stated in \autoref{Proposition3}.
If $Y_\Myidx \stackrel{d}{\rightarrow} Y$
it holds that  $Y_\Myidx - \Myidx\lambda_{\Myidx}
\stackrel{p}{\to} -\infty$, i.e.,
\[
\forall \alpha \in \mathbb{R}: \lim_{\Myidx\rightarrow \infty} \mathbb{P}(Y_\Myidx - \Myidx \lambda_{\Myidx} \leq \alpha) = 1.
\]
\end{mylem}
\begin{myproof}
Let $\alpha \in \mathbb{R}$. 
Since  $\Myidx \lambda_{\Myidx} \rightarrow \infty$ it holds that
\begin{align}
\forall \varepsilon>0 \; \exists \Myidx_1\in\mathbb{N} \; \forall \Myidx\geq \Myidx_1: F_{Y}(\alpha + \Myidx\lambda_{\Myidx}) > 1-\frac{\varepsilon}{2}.\label{eq1InLemmaProof}
\end{align}
By assumption $Y_\Myidx$ converges in distribution to $Y\sim F_Y$, therefore
\begin{align}
\forall \varepsilon>0 \; \forall \Myidx_1\in\mathbb{N} \; \exists \Myidx_2\in\mathbb{N}  \; \forall \Myidx\geq \Myidx_2 : \left| F_{Y_\Myidx}(\alpha + \Myidx_1 \lambda_{\Myidx_1} ) - F_{Y}(\alpha + \Myidx_1 \lambda_{\Myidx_1}) \right| < \frac{\varepsilon}{2}.\label{eq2InLemmaProof}
\end{align}
Moreover, $\exists n_3\in\mathbb{N}\ \forall n \geq n_3\colon n\lambda_{\Myidx}\geq n_1\lambda_{\Myidx_1}$. 
Thus, $\forall \varepsilon>0 \; \forall \Myidx \geq \text{max}(\Myidx_1,\Myidx_2,\Myidx_3)$ it holds that
\begin{align*}
\mathbb{P}(Y_\Myidx \leq \alpha + \Myidx\lambda_{\Myidx})
&\geq \mathbb{P}(Y_\Myidx \leq \alpha + \Myidx_1 \lambda_{\Myidx_1})
= F_{Y_\Myidx}(\alpha + \Myidx_1 \lambda_{\Myidx_1}) \\
&= F_{Y_\Myidx}(\alpha + \Myidx_1 \lambda_{\Myidx_1}) - F_{Y}(\alpha + \Myidx_1 \lambda_{\Myidx_1}) + F_{Y}(\alpha + \Myidx_1 \lambda_{\Myidx_1})\\
&\stackrel{\eqref{eq2InLemmaProof}}{>} F_{Y}(\alpha + \Myidx_1 \lambda_{\Myidx_1}) - \frac{\varepsilon}{2}
\stackrel{\eqref{eq1InLemmaProof}}{>} 1 - \varepsilon.
\end{align*}
Thus, 
\[
\lim_{\Myidx\rightarrow \infty} \mathbb{P}(Y_\Myidx - \Myidx \lambda_{\Myidx} \leq \alpha) = 1.
\]
\vspace{0pt} 
\end{myproof}

In the following \autoref{Lemma2} the asymptotic behavior of
$\delta_\Myidx := \text{max}\big\lbrace Y_{0}^{\Myidx} + \Myidx \lambda_{\Myidx}, Y_{1}^{\Myidx},\linebreak Y_{2}^{\Myidx}, \ldots, Y_{\nPartitions}^{\Myidx}\big\rbrace - \Myidx \lambda_{\Myidx}$
is analyzed.
\begin{mylem}
\label{Lemma2}
Let $(Y_{0}^{\myidx})_{\myidx=1:\Myidx}, (Y_{1}^{\myidx})_{\myidx=1:\Myidx}, \ldots, (Y_{\nPartitions}^{\myidx})_{\myidx=1:\Myidx}$ be $(\nPartitions+1)$ sequences of 
random variables and
$Y_{\iPartition} \sim F_{Y_{\iPartition}}$, $0\leq \iPartition \leq \nPartitions$, random variables with continuous cumulative distribution functions.
Further let $\lambda_{\Myidx}: \mathbb{N} \rightarrow \mathbb{R}^{+}$ be a penalty function satisfying the conditions stated in \autoref{Proposition3}.\\
Define $\delta_\Myidx := \max \left\lbrace Y_{0}^{\Myidx} + \Myidx \lambda_{\Myidx}, Y_{1}^{\Myidx}, Y_{2}^{\Myidx}, \ldots, Y_{\nPartitions}^{\Myidx}\right\rbrace - \Myidx \lambda_{\Myidx}$.
\begin{enumerate}[label=(\roman*)]
\item \label{pp1}
If $Y_{\iPartition}^{\Myidx} \stackrel{d}{\rightarrow} Y_{\iPartition}$ for each $0\leq \iPartition \leq \nPartitions$, it holds that
$\delta_\Myidx \stackrel{d}{\rightarrow} Y_{0}$.
\item \label{pp2}
If there is an $m^{\star}\in \{0,\ldots,M\}$ such that 
$\plim{\Myidx} \frac{1}{\Myidx} Y_{m^{\star}}^\Myidx = y_{m^{\star}} > 0$ then $\delta_\Myidx\stackrel{p}{\to} \infty$.
\end{enumerate}
\end{mylem}
\begin{myproof}
Proof of \ref{pp1}.
Let $\alpha \in \mathbb{R}$, then
\begin{align*}
\mathbb{P}(\delta_\Myidx \leq \alpha)
&= \mathbb{P}\left( \text{max} \left\lbrace Y_{0}^{\Myidx} + \Myidx \lambda_{\Myidx}, Y_{1}^{\Myidx}, \ldots, Y_{\nPartitions}^{\Myidx} \right\rbrace  - \Myidx\lambda_{\Myidx} \leq \alpha \right) \\
&= \mathbb{P}\left( \text{max}\left\lbrace Y_{0}^{\Myidx}, Y_{1}^{\Myidx} - \Myidx\lambda_{\Myidx}, \ldots, Y_{\nPartitions}^{\Myidx} - \Myidx \lambda_{\Myidx}\right\rbrace  \leq \alpha \right) \\
&= \mathbb{P}\left(Y_{0}^{\Myidx} \leq \alpha, Y_{1}^{\Myidx} - \Myidx\lambda_{\Myidx} \leq \alpha, \ldots, Y_{\nPartitions}^{\Myidx} - \Myidx \lambda_{\Myidx} \leq \alpha \right).
\end{align*}
Using the Fr\'{e}chet-Hoeffding inequalities \citep{Frechet1951,Hoeffding1940} we have
\begin{align*}
&\mathbb{P}(\delta_\Myidx \leq \alpha) \geq
\text{max} \bigg\lbrace 0, \mathbb{P}\left(Y_{0}^{\Myidx} \leq \alpha\right) + \sum_{\iPartition=1}^{\nPartitions} \mathbb{P}\left(Y_{\iPartition}^{\Myidx} - \Myidx \lambda_{\Myidx} \leq \alpha\right) - \nPartitions \bigg\rbrace
\intertext{and}
&\mathbb{P}(\delta_\Myidx \leq \alpha) \leq
\text{min} \bigg\lbrace \mathbb{P}\left(Y_{0}^{\Myidx} \leq \alpha\right), \underset{1 \leq \iPartition \leq \nPartitions}{\text{min}} \Big\lbrace \mathbb{P}\left(Y_{\iPartition}^{\Myidx} - \Myidx \lambda_{\Myidx} \leq \alpha\right) \Big\rbrace \bigg\rbrace.
\end{align*}
Due to the continuity of the minimum and maximum as well as \autoref{Lemma1} it follows that
\begin{align}
\notag
\lim_{\Myidx\rightarrow \infty} \mathbb{P}(\delta_\Myidx \leq \alpha) &\geq
\lim_{\Myidx\rightarrow \infty} \text{max} \bigg\lbrace 0, \mathbb{P}\left(Y_{0}^{\Myidx} \leq \alpha\right) + \sum_{\iPartition=1}^{\nPartitions} \mathbb{P}\left(Y_{\iPartition}^{\Myidx} - \Myidx \lambda_{\Myidx} \leq \alpha\right) - \nPartitions \bigg\rbrace \\
\notag
&= \text{max} \bigg\lbrace 0, \lim_{\Myidx\rightarrow \infty} \mathbb{P}\left(Y_{0}^{\Myidx} \leq \alpha\right) + \sum_{\iPartition=1}^{\nPartitions} \lim_{\Myidx\rightarrow \infty} \mathbb{P}\left(Y_{\iPartition}^{\Myidx} - \Myidx \lambda_{\Myidx} \leq \alpha\right) - \nPartitions \bigg\rbrace \\
\notag
&= \text{max} \bigg\lbrace 0, F_{Y_{0}}(\alpha) + \sum_{\iPartition=1}^{\nPartitions} 1 - \nPartitions \bigg\rbrace
= F_{Y_{0}}(\alpha)
\intertext{and}
\notag
\lim_{\Myidx\rightarrow \infty} \mathbb{P}(\delta_\Myidx \leq \alpha) &\leq
\lim_{\Myidx\rightarrow \infty} \text{min} \bigg\lbrace \mathbb{P}\left(Y_{0}^{\Myidx} \leq \alpha\right), \underset{1 \leq \iPartition \leq \nPartitions}{\text{min}} \Big\lbrace \mathbb{P}\left(Y_{\iPartition}^{\Myidx} - \Myidx \lambda_{\Myidx} \leq \alpha\right) \Big\rbrace \bigg\rbrace \\
&= \text{min} \bigg\lbrace \lim_{\Myidx\rightarrow \infty} \mathbb{P}\left(Y_{0}^{\Myidx} \leq \alpha\right), \underset{1 \leq \iPartition \leq \nPartitions}{\text{min}} \Big\lbrace \lim_{\Myidx\rightarrow \infty} \mathbb{P}\left(Y_{\iPartition}^{\Myidx} - \Myidx \lambda_{\Myidx} \leq \alpha\right) \Big\rbrace \bigg\rbrace \label{refh}
\\
\notag
&= \text{min} \bigg\lbrace F_{Y_{0}}(\alpha), \underset{1 \leq \iPartition \leq \nPartitions}{\text{min}} \big\lbrace 1 \big\rbrace \bigg\rbrace
= F_{Y_{0}}(\alpha).
\end{align}
Thus,
\[
\delta_\Myidx \stackrel{d}{\rightarrow} Y_{0}.
\]

Proof of \ref{pp2}.
For $m=0,\ldots,M$, define $Z_m^n := Y_m^\Myidx - \Myidx\lambda_{\Myidx}\ivb{m\neq 0}$.
Because $\lambda_{\Myidx}\to 0$ it follows that 
$\plim{\Myidx} \frac{1}{\Myidx} Z_{m^{\star}}^n = y_{m^{\star}}$.
Note that $\mathbb{P}(Z_{m^{\star}}^\Myidx \leq \alpha) \stackrel{n\to\infty}{\to}0$ for all $\alpha\in \R$ because $ \frac{1}{\Myidx} Z_{m^{\star}}^n \stackrel{p}{\to} y_{m^{\star}} > 0$.

Thus, the Fr\'{e}chet-Hoeffding upper bound implies that for any $\alpha \in \R$,
\begin{align*}
\lim_{\Myidx\rightarrow \infty} \mathbb{P}(\delta_\Myidx \leq \alpha) &\leq
\lim_{\Myidx\rightarrow \infty} \text{min} \Big\lbrace 
\mathbb{P}(Z_{m^{\star}}^n \leq \alpha), \underset{m\in\{0,\ldots,M\}\backslash m^{\star}}{\text{min}} \Big\lbrace \mathbb{P}\left(Z_{m}^n \leq \alpha\right) \Big\rbrace \Big\rbrace \\
& = \text{min} \Big\lbrace 
\lim_{\Myidx\rightarrow \infty} \mathbb{P}(Z_{m^{\star}}^n \leq \alpha), \underset{m\in\{0,\ldots,M\}\backslash m^{\star}}{\text{min}} \Big\lbrace \lim_{\Myidx\rightarrow \infty} \mathbb{P}\left(Z_{m}^n \leq \alpha\right) \Big\rbrace \Big\rbrace \\
& = \text{min} \Big\lbrace 
0, \underbrace{\underset{m\in\{0,\ldots,M\}\backslash m^{\star}}{\text{min}} \Big\lbrace \lim_{\Myidx\rightarrow \infty} \mathbb{P}\left(Z_{m}^n \leq \alpha\right) \Big\rbrace}_{\in [0,1]} \Big\rbrace = 0,
\end{align*}
and the proof is complete.
\end{myproof}

Using \autoref{Proposition2} and setting $Y_m^n=T_n(\Gamma_m)$ in \autoref{Lemma2} \ref{pp1} it follows that the statistic $\Theta_\Myidx$ converges under $H_0$  to a $\chi^2(\nGroups_0-1)$ distribution.

Now assume that the correlations conditional on the subsets of the partition $\Gamma_{m^\star}$ are not identical, i.e.,  $\plim{\Myidx} \frac{1}{\Myidx} T_\Myidx(\Gamma_{m^\star}) = c>0$ for some $m^{\star}\in\{0,\ldots,M\}$.
Setting $Y_m^n=T_\Myidx(\Gamma_m)$ in \autoref{Lemma2} \ref{pp2} shows that the test statistic $\Theta_{\Myidx}$ converges in probability to infinity.

\subsection{The decision tree: Algorithmic details}\label{appDecisionTreeAlgorithm}
\begin{figure}[ht]
\centering
\includegraphics[width=\textwidth]{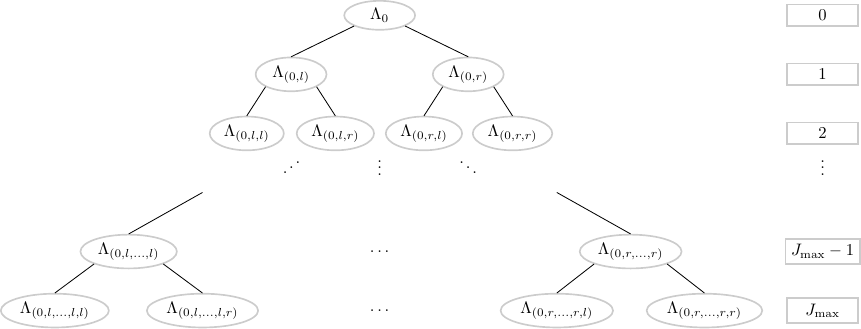}
\caption{Partitioning of the support $\partElement_{0} = \text{supp}(\CondVar)$ of the random vector $\CondVar$ into disjoint subsets $\partElement_{\gamma_{0:J}}$, where $\gamma_{0:J}:=(\gamma_{0},\gamma_{1},\ldots,\gamma_{J}) \in \lbrace 0 \rbrace \times \lbrace l,r \rbrace^{J}$, $0\leq J \leq J_{\text{max}}$, 
using a decision tree algorithm with maximum depth $J_{\text{max}}$.}\label{FigDecTreeGeneral}
\end{figure}

Every leaf in the decision tree represents a subset of the support $\partElement_{0}$ of the random vector $\CondVar$.
The maximum depth of the decision tree is denoted by $J_{\text{max}}$ and every leaf is assigned to a level $J$ in the decision tree ($0\leq J \leq J_{\text{max}}$).
The level of a leaf refers to the number of splits which have already been used to arrive at the leaf, starting from the root leaf $\Lambda_0$ (see \autoref{FigDecTreeGeneral}).

A leaf is denoted by $\partElement_{\gamma_{0:J}}$, where the $(J+1)$-dimensional vector $\gamma_{0:J}:=(\gamma_{0},\gamma_{1},\ldots,\gamma_{J}) \in \lbrace 0 \rbrace \times \lbrace l,r \rbrace^{J}$ is the unique identifier for a leaf in the $J$-th level of the decision tree.
That is, the two leaves in the $(J+1)$-th level of the decision tree being connected via edges to the leaf $\gamma_{0:J}=(\gamma_{0},\gamma_{1},\ldots,\gamma_{J})$ in the $J$-th level are  identified by $\gamma_{0:J+1} := (\gamma_{0:J},k):=(\gamma_{0},\gamma_{1},\ldots,\gamma_{J},k)$ with $k\in\lbrace l,r \rbrace$.
The subsets assigned to the leafs in the $(J+1)$-th level by a binary split are given by
\[
\partElement_{(\gamma_{0:J},k)} \subset\partElement_{\gamma_{0:J}},
\qquad J \geq 0, \qquad k\in\lbrace l,r \rbrace,
\qquad \partElement_{(\gamma_{0:J},l)} \cap \partElement_{(\gamma_{0:J},r)} = \emptyset.
\]
Every split is chosen out of a finite number $\mathcal{M}$ of possible splits.
A possible split $S_{\gamma_{0:J}}^{m}$ in the leaf $\gamma_{0:J}$ is defined as a pair of disjoint subsets of $\partElement_{\gamma_{0:J}}$, i.e., $S_{\gamma_{0:J}}^{m} := (\partElement_{(\gamma_{0:J},l)}^{m},\partElement_{(\gamma_{0:J},r)}^{m}) \subset \partElement_{\gamma_{0:J}} \times \partElement_{\gamma_{0:J}}$ with $\partElement_{(\gamma_{0:J},l)}^{m} \cap \partElement_{(\gamma_{0:J},r)}^{m} = \emptyset$.
From these possible splits, the split is selected that maximizes 
the statistic of the CCC test.
Meaning that every split is defined as
\[
S_{\gamma_{0:J}} = (\partElement_{(\gamma_{0:J},l)},\partElement_{(\gamma_{0:J},r)}) = \underset{
S_{\gamma_{0:J}}^{m} \in \lbrace S_{\gamma_{0:J}}^{1}, \ldots, S_{\gamma_{0:J}}^{\mathcal{M}} \rbrace}{\text{argmax}} 
T_\Myidx \left( S_{\gamma_{0:J}}^{m} \right).
\]
Thus, the subsets that are transferred to leaf $(\gamma_{0:J},k)$, $k \in \lbrace l,r \rbrace$, after using the optimal split $S_{\gamma_{0:J}}$, are given by
$\partElement_{(\gamma_{0:J},l)}$ and $\partElement_{(\gamma_{0:J},r)}$.
In the last level $J_{\text{max}}$ we obtain a final partition of the support $\partElement_0$ into $2^{J_{\text{max}}}$ disjoint subsets given by $\Gamma_{\max} := \lbrace \partElement_{(0,l,\ldots,l,l)}, \partElement_{(0,l,\ldots,l,r)}, \ldots, \partElement_{(0,r,\ldots,r,l)}, \partElement_{(0,r,\ldots,r,r)} \rbrace$.
For the final partition we compute the value of the test statistic 
\[
\Theta_\Myidx = \text{max} \lbrace T_{\Myidx}(\Gamma_0) + \Myidx \lambda_{\Myidx}, T_{\Myidx}(\Gamma_{\max}) \rbrace - \Myidx \lambda_{\Myidx}.
\]

In all simulations in \autoref{secSimStudy} and the real data applications in \autoref{secRealData}, we choose $\lambda_{\Myidx} = \frac{1}{\sqrt{\Myidx}}$ for the penalty and $\Gamma_0 = \Gamma_{\text{med}}$ as base partition.%
\footnote{The partition $\Gamma_{\text{med}}$ is defined in equation~\eqref{gamma_med} in \autoref{secNaiveGroup}.}
Further tuning parameters of the  decision tree are the maximum depth $J_{\text{max}}$ of the tree and the set of possible splits 
$\lbrace S_{\gamma_{0:J}}^{1},\ldots,S_{\gamma_{0:J}}^{\mathcal{M}} \rbrace$.
To keep the test computationally feasible and because it performs well in simulations, we consider a maximum depth of $J_{\text{max}} = 2$ and the number of possible splits in each leaf $\gamma_{0:J}$ is restricted to be at most $\mathcal{M} = 3(\Mydim+\ivb{ j\geq 3})$ in tree $T_j$.
The considered splits are as follows:
To obtain the sets $\Lambda_{(0,l)}$ and $\Lambda_{(0,r)}$ for the two leaves in level 1, we consider the empirical 25\%, 50\% and 75\% quartiles for each conditioning variable $U_{k}$, $k \in \condset$.
If $j \geq 3$, we additionally take the empirical 25\%, 50\% and 75\% quartiles of the mean aggregated conditioning variables into account, resulting in $3 \cdot j$ possible splits.
A formal definition of the set of possible splits is given in  \appref{appFormalDefinitionOfSplits}.
The sets $\{\Lambda_{(0,l,l)}, \Lambda_{(0,l,r)}\}$ and
$\{\Lambda_{(0,r,l)}, \Lambda_{(0,r,r)}\}$ for the four leaves in level 2
are obtained in the same fashion except that we now condition on $U_{\condset}\in \Lambda_{(0,l)}$ or  $U_{\condset}\in \Lambda_{(0,r)}$, respectively. 
Furthermore, we use several restrictions in the decision tree algorithm to guarantee that the final data sets do not become too small.%
\footnote{
A decision tree with two or three splits is only applied if we have a certain amount of data.
This is implemented by introducing a tuning parameter which controls the minimum sample size per leaf in the decision tree (the default value is 100 observations).
As a result we do not always use the 25\%, 50\% and 75\% quartiles as thresholds but depending on the available sample size we may only use the 50\% quartile or even don't apply any additional split at all.}

\subsection{Formal definition of the set of possible splits for the decision tree}\label{appFormalDefinitionOfSplits}
If $Q_{q}[(X_{i})_{i\in\mathcal{I}}]$ denotes the empirical $q$-quantile of the vector $(X_{i})_{i\in\mathcal{I}}$, the set of possible splits in the leaf $\gamma_{0:J}$, for $J=0,1$
and $e \in E_j$, $j\geq 3$ with $(j-1)$-dimensional conditioning set $D_e:=D_e^{1:j-1}$ is given by
\begin{align*}
\mathcal{S}_{\gamma_{0:J}}^{1:3 j} &:=
\left\lbrace S_{\gamma_{0:J}}^{1},\ldots,S_{\gamma_{0:J}}^{3 j}\right\rbrace \\
&= \left\lbrace(\partElement_{(\gamma_{0:J},l)}^{1},\partElement_{(\gamma_{0:J},r)}^{1}),\ldots,(\partElement_{(\gamma_{0:J},l)}^{3 j},\partElement_{(\gamma_{0:J},r)}^{3 j})\right\rbrace,
\end{align*}
with
{\footnotesize%
\begin{align*}
\partElement_{(\gamma_{0:J},l)}^{m} &:= \begin{cases}
\left\lbrace \condVar \in \partElement_{\gamma_{0:J}}: u_{D_e^{m}} \leq  Q_{0.25}
[(V_{D_e^{m}}^{\myidx})_{\myidx\in\mathcal{I}_{\gamma_{0:J}}}] \right\rbrace
&,1 \leq m\leq \Mydim \\
\left\lbrace \condVar \in \partElement_{\gamma_{0:J}}: u_{D_e^{m_1}} \leq  Q_{0.5}
[(V_{D_e^{m_1}}^{\myidx})_{\myidx\in\mathcal{I}_{\gamma_{0:J}}}] \right\rbrace
&,j \leq m\leq 2j - 2 \\
\left\lbrace \condVar \in \partElement_{\gamma_{0:J}}: u_{D_e^{m_2}} \leq  Q_{0.75}
[(V_{D_e^{m_2}}^{\myidx})_{\myidx\in\mathcal{I}_{\gamma_{0:J}}}] \right\rbrace
&,2j - 1 \leq m\leq 3j - 3 \\
\left\lbrace \condVar \in \partElement_{\gamma_{0:J}}: g(\condVar) \leq  Q_{0.25 m_3}
[(g(V_{\condset}^{\myidx}))_{\myidx\in\mathcal{I}_{\gamma_{0:J}}}] \right\rbrace
&,3j - 2 \leq m \leq 3j
\end{cases}
\end{align*}}%
and $\partElement_{(\gamma_{0:J},r)}^{m} = \partElement_{\gamma_{0:J}} \setminus \partElement_{(\gamma_{0:J},l)}^{m}$.
We further used the notation $m_\ell := m-\ell(\Mydim)$ for $1\leq \ell \leq 3$ and the index set $\mathcal{I}_{\gamma_{0:J}}$ is defined as $\mathcal{I}_{\gamma_{0:J}} := \lbrace \myidx \in \lbrace 1,\ldots,\Myidx\rbrace: V_{\condset}^{\myidx}
\in\partElement_{\gamma_{0:J}}\rbrace$.

\subsection{Choosing the penalty function: A finite sample analysis}\label{appPenalty}
To apply the test based on the statistic $\Theta_\Myidx$, a penalty function $\lambda_{\Myidx}$ has to be specified and
any choice satisfying the conditions stated in \autoref{Proposition3} results in an asymptotically valid test.
However, the size and power for finite sample sizes depends on the chosen penalty function  $\lambda_{\Myidx}$.
In the following, the choice of the penalty function in finite samples will be analyzed in a simulation study under $H_0$, i.e., with a focus on the empirical size.

In all simulations in \autoref{secSimStudy} and the real data applications in \autoref{secRealData}, we choose $\lambda_{\Myidx} = \frac{1}{\sqrt{\Myidx}}$ for the penalty and $\Gamma_0 = \Gamma_{\text{med}}$ as base partition.%
\footnote{The partition $\Gamma_{\text{med}}$ is defined in equation~\eqref{gamma_med} in \autoref{secNaiveGroup}.}
As mentioned in \autoref{secPowerGain}, the test statistic $T_\Myidx(\Gamma_0)$ (the CCC test with fixed partition $\Gamma_0$) is related to the test statistic $\Theta_\Myidx$ (the CCC test with a decision tree selected partition) in the following way.
For $\Theta_\Myidx = \text{max} \lbrace T_{\Myidx}(\Gamma_0), T_{\Myidx}(\Gamma_{\max}) - \Myidx\lambda_{\Myidx} \rbrace$,
with $\Gamma_{\max} := \text{argmax}_{\Gamma_m \in \lbrace \Gamma_1, \ldots, \Gamma_M\rbrace} T_\Myidx(\Gamma_m)$, it holds
\[
T_\Myidx(\Gamma_0) \leq \Theta_\Myidx,
\]
meaning that if we reject based on $T_\Myidx(\Gamma_0)$, we also reject based on $\Theta_\Myidx$.
It follows that the empirical size of $\Theta_\Myidx$ is bounded from below by the empirical size of $T_\Myidx(\Gamma_0)$ when both tests are applied to the same collection of data sets in a monte carlo simulation to compute the empirical size.

We now derive a condition on $\lambda_{\Myidx}$ such that $\Theta_\Myidx$ and $T_\Myidx(\Gamma_0)$ result in equivalent test decisions.
This means that the test statistic $\Theta_\Myidx$ is analyzed relative to $T_\Myidx(\Gamma_0)$.\footnote{
An extensive simulation study of the finite sample performance of the proposed test $\Theta_\Myidx$ is presented in \autoref{secSimStudy} where the empirical size relative to the theoretical level of the test is studied.}
Let $\tau := F_{\chi^2(L_0-1)}^{-1}(1-\alpha)$, where $L_0$ is the cardinality of $\Gamma_0$, i.e., the number of subsets forming the partition $\Gamma_0$.
If the penalty function $\lambda_{\Myidx}$ satisfies
\begin{align}
\lambda_{\Myidx} > \frac{1}{\Myidx} \Big( T_{\Myidx}(\Gamma_{\max}) - \tau \Big) =: b_\Myidx,
\label{eqBoundPenalty}
\end{align}
it follows that
$\Theta_\Myidx = \text{max} \lbrace T_{\Myidx}(\Gamma_0), T_{\Myidx}(\Gamma_{\max}) - \Myidx\lambda_{\Myidx} \rbrace
\leq \text{max} \lbrace T_{\Myidx}(\Gamma_0), \tau \rbrace$.
Therefore, if we cannot reject at a $\alpha$-level based on $T_\Myidx(\Gamma_0)$ and if $\lambda_{\Myidx}$ satisfies \eqref{eqBoundPenalty}, we also cannot reject based on $\Theta_\Myidx$, i.e., if $\lambda_{\Myidx} > b_\Myidx$ it holds
\begin{align*}
T_{\Myidx}(\Gamma_0) < \tau
\quad \Rightarrow \quad
\Theta_\Myidx < \tau.
\end{align*}
As a result, if $\lambda_{\Myidx} > b_\Myidx$, both tests result in the same $\alpha$-level test decisions, i.e.,
\begin{align*}
T_{\Myidx}(\Gamma_0) \geq \tau
\quad \Leftrightarrow \quad
\Theta_\Myidx \geq \tau.
\end{align*}

Note that $T_\Myidx (\Gamma)$ converges in distribution to a $\chi^2$-distribution under $H_0$ and by Slutsky's theorem it follows that $b_n \stackrel{p}{\rightarrow} 0$.
Therefore, the lower bound $b_n$ is bounded in probability, i.e.,
\begin{align*}
\forall \varepsilon >0\; \exists B>0\; \forall n \geq 1:\; \mathbb{P}(|b_n| \geq B) < \varepsilon.
\end{align*}
Meaning that for any $\varepsilon >0$, we can choose $\lambda_{\Myidx}$ such that $\mathbb{P}(|b_n| \geq \lambda_{\Myidx}) < \varepsilon$, which restricts the probability of different test decisions (i.e., rejecting the $H_0$ with $\Theta_\Myidx$ but not rejecting the $H_0$ with $T_\Myidx(\Gamma_0)$) at a $\alpha$-level to $\varepsilon$ because
\begin{align*}
\mathbb{P}(\Theta_n \geq \tau, T_n(\Gamma_0) < \tau)
&= \mathbb{P}(\text{max} \lbrace T_{\Myidx}(\Gamma_0), T_{\Myidx}(\Gamma_{\max}) - \Myidx\lambda_{\Myidx} \rbrace \geq \tau, T_n(\Gamma_0) < \tau) \\
&= \mathbb{P}(T_{\Myidx}(\Gamma_{\max}) - \Myidx\lambda_{\Myidx} \geq \tau, T_n(\Gamma_0) < \tau) \\
&\leq \mathbb{P}(T_{\Myidx}(\Gamma_{\max}) - \Myidx\lambda_{\Myidx} \geq \tau)
= \mathbb{P}(b_n \geq \lambda_{\Myidx}) < \varepsilon.
\end{align*}
This implies that for any $\varepsilon >0$, we can choose $\lambda_{\Myidx}$ such that $\mathbb{P}(|b_n| \geq \lambda_{\Myidx}) < \varepsilon$ and therefore
\begin{align*}
&\mathbb{P}\Big( T_{\Myidx}(\Gamma_0) \geq \tau
\quad \Leftrightarrow \quad
\Theta_\Myidx \geq \tau \Big) \\
&\qquad = \P(T_{\Myidx}(\Gamma_0) \geq \tau, \Theta_\Myidx \geq \tau)
+ \P(T_{\Myidx}(\Gamma_0) < \tau, \Theta_\Myidx < \tau) \\
&\qquad = 1 - \underbrace{\P(T_{\Myidx}(\Gamma_0) \geq \tau, \Theta_\Myidx < \tau)}_{=0}
- \P(T_{\Myidx}(\Gamma_0) < \tau, \Theta_\Myidx \geq \tau)
\geq 1 - \varepsilon.
\end{align*}

In practical applications, we are interested in the finite sample distribution of the lower bound $b_n$ of the penalty function $\lambda_{\Myidx}$.
Using resampling techniques, we can determine this lower bound for $\lambda_{\Myidx}$ under $H_0$.

\begin{figure}[ht]
\begin{center}
\includegraphics[width=1\textwidth]{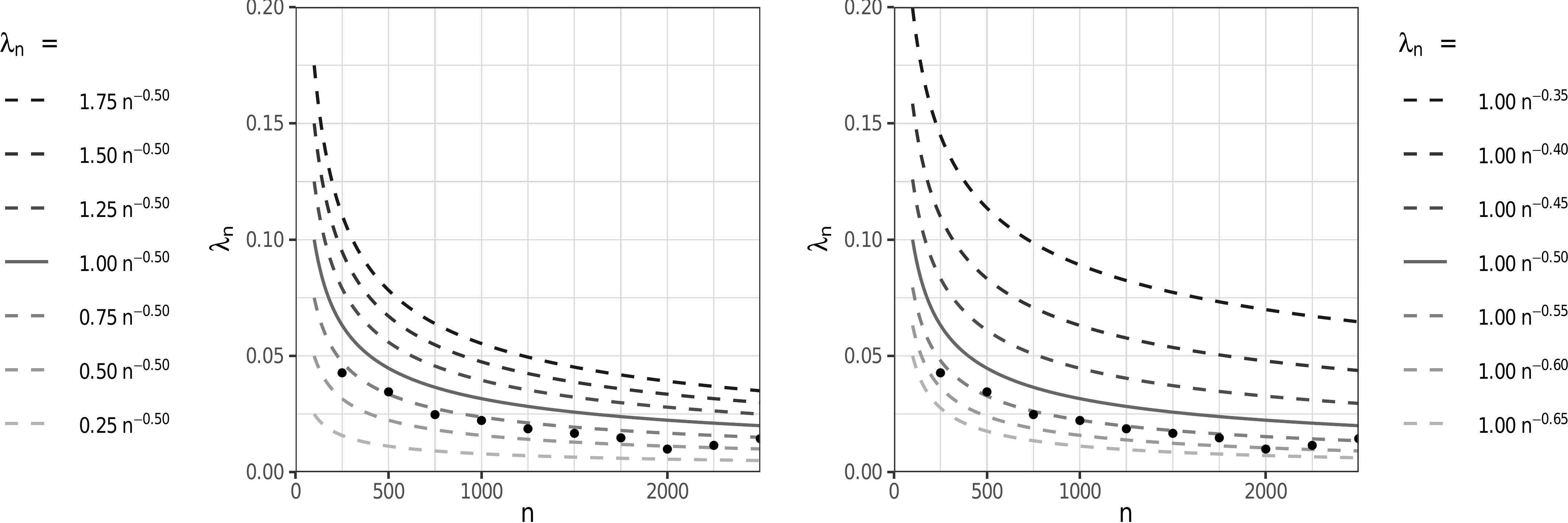}
\end{center}
\caption{
The dots correspond to the maximum of the lower bound for the penalty function in $1000$ simulated samples each consisting of $\Myidx$ observations.
The data generating process is the four-dimensional D-vine copula defined in \autoref{exSimStudy}.
The lines correspond to different choices of the penalty function $\lambda_{\Myidx} = c \Myidx^{-\beta}$.
The solid line corresponds to the default penalty function $\lambda_{\Myidx} = \frac{1}{\sqrt{\Myidx}}$ used in all simulations and applications.
On the left hand side the dashed lines correspond to different choices of the level $c$ of the penalty function $\lambda_{\Myidx}$ and on the right hand side to different choices for the power $\beta$ of the penalty function.
}\label{figPenalty}
\end{figure}

To illustrate how one can use resampling techniques to determine the parameters $c$ and $\beta$ of the penalty function $\lambda_{\Myidx} = c \Myidx^{-\beta}$, we again consider the data generating process given in \autoref{exSimStudy}.
For $\lambda = 0$, the null hypothesis $H_{0}: (\pvc{U}_{1|23}, \pvc{U}_{4|23}) \perp U_{2:3}$ is true.
For each considered sample size $\Myidx$ we generate $1000$ random samples of size $\Myidx$ from the four-dimensional D-vine copula and compute for each sample the lower bound
$b_n$
of the penalty function $\lambda_{\Myidx}$.
In \autoref{figPenalty}, the maximum of all $1000$ lower bounds in the different samples is plotted for different sample sizes $\Myidx$ as dots.
By taking the maximum over all resampled lower bounds we identify a lower bound for the penalty which would guarantee that in every of the $1000$ samples the asymptotic $\alpha$-level test decisions are the same, i.e.,
$T_{\Myidx}(\Gamma_0) \geq \tau
\quad \Leftrightarrow \quad
\Theta_\Myidx \geq \tau$.
The level of the test is chosen to be 5\% and
the lines correspond to different choices of the penalty function $\lambda_{\Myidx} = c \Myidx^{-\beta}$.
The level $c$ of the penalty function is varied for a fixed power of $\beta = 0.5$ in the plot on the left hand side of \autoref{figPenalty} and the power $\beta$ of the penalty function is varied for a fixed level of $c= 1$ on the right hand side.
The solid line corresponds to the penalty function $\lambda_{\Myidx} = \frac{1}{\sqrt{\Myidx}}$ which we use in all the simulations and applications.
One can see that the choice of the penalty function is reasonable in comparison to the lower bounds obtained via resampling techniques for all sample sizes between $250$ and $2500$ observations, as the penalty is for all sample sizes considerably larger than the lower bounds.

\subsection{An analysis of the typical partitions selected by the decision trees}\label{appTypicalParts}
\begin{figure}[ht]
\centering
\includegraphics[width=\textwidth]{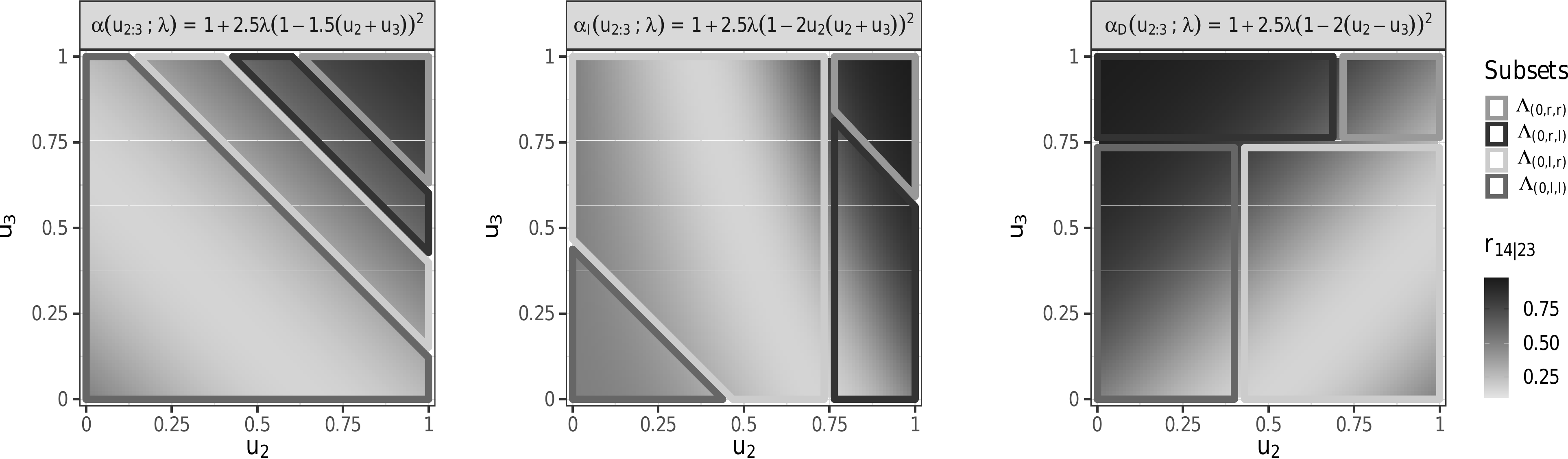}
\caption{
Visualization of the most frequently selected $\Gamma_{\max}:= \lbrace \partElement_{(0,l,l)}, \partElement_{(0,l,r)}, \partElement_{(0,r,l)},\break \partElement_{(0,r,r)} \rbrace$ partitions with the decision trees in the simulation study for $\lambda=1$ and $1000$ samples of size $n=1000$.
The three panels correspond to the parameter functions $\alpha(\cdot)$, $\alpha_{I}(\cdot)$ and $\alpha_{D}(\cdot)$, respectively.
The shaded background shows the conditional correlation $r_{14|23}$ of $C_{14\cs 23}$ as a function of $U_2$ and $U_3$.
}\label{figMostFrequentPartitions}
\end{figure}

The following analysis complements the simulation study in \autoref{secPowerGain} by an analysis of the partitions typically selected by the decision trees.
To investigate what kind of partitions are selected by the decision trees and how they adapt to the different variations in the conditional correlation determined by $\alpha(\cdot), \alpha_I(\cdot),$ and $\alpha_D(\cdot)$ (see \eqref{eqParFunctional}, \eqref{eqParFunctionalAlternative1} and \eqref{eqParFunctionalAlternative2}), we focus on the case $\lambda=1$ and a sample size of $n=1000$.
If the variation of the copula parameter is given by $\alpha(\cdot)$, the first split is most often done with respect to the 75\% quartile of $\bar{U}_{2:3}:= \frac{1}{2} (U_2 + U_3)$ (in 93.6\% of the simulated samples).
The most frequently selected $\Gamma_{\max}$ partition (in 40.6\% of the samples) consists of two more splits with respect to the quartiles of $\bar{U}_{2:3}$ in the next level.
This partition is visualized in the left panel of \autoref{figMostFrequentPartitions}.
The adaption of the decision tree based partition to the variation in the conditional correlation becomes evident when looking at the shaded background which shows the conditional correlation $r_{14|23}$ of $C_{14\cs 23}$ as a function of $U_2$ and $U_3$.

The middle panel in \autoref{figMostFrequentPartitions} shows the most frequently selected $\Gamma_{\max}$ partition (in 14.8\% of the samples) if the variation of the copula parameter is given by $\alpha_I(\cdot)$ which contains an interaction effect between the two conditioning variables.
In this case, the first split is most often done with respect to the 75\% quartile of $U_2$ (in 61.7\% of the samples).
The decision tree adapts to this variation in the conditional correlation, shown as shaded background, by choosing splits with respect to the quartiles of $\bar{U}_{2:3}$ in the second level.

The right panel in \autoref{figMostFrequentPartitions} visualizes the most frequently selected $\Gamma_{\max}$ partition (in 22.7\% of the samples) if the variation of the copula parameter is given by $\alpha_D(\cdot)$.
Here, the two conditioning variables have an opposite sign in the definition of $\alpha_D(\cdot)$. 
In this case, the first split is most often done with respect to the 75\% quartile of $U_3$ (in 52.8\% of the samples).
Once again, the decision tree adapts to the variation in the conditional correlation, shown as shaded background, by choosing splits with respect to the quartiles of the other variable $U_{2}$ in the second level.

\begin{acks}[Acknowledgments]
We are very grateful for the helpful comments of two anonymous reviewers, an associate editor and the editor.
Malte S. Kurz acknowledges funding by the Deutsche Forschungsgemeinschaft (DFG, German Research Foundation) -- Project Number 431701914.
\end{acks}

\bibliographystyle{imsart-nameyear}
\bibliography{Bib}

\begin{thebibliography}{46}

\bibitem[\protect\citeauthoryear{Aas et~al.}{2009}]{Aas2009}
\begin{barticle}[author]
\bauthor{\bsnm{Aas},~\bfnm{K.}\binits{K.}},
  \bauthor{\bsnm{Czado},~\bfnm{C.}\binits{C.}},
  \bauthor{\bsnm{Frigessi},~\bfnm{A.}\binits{A.}} \AND
  \bauthor{\bsnm{Bakken},~\bfnm{H.}\binits{H.}}
(\byear{2009}).
\btitle{Pair-copula constructions of multiple dependence}.
\bjournal{Insurance: Mathematics and Economics}
\bvolume{44}
\bpages{182--198}.
\end{barticle}
\endbibitem

\bibitem[\protect\citeauthoryear{Acar, Craiu and Yao}{2013}]{Acar2013}
\begin{barticle}[author]
\bauthor{\bsnm{Acar},~\bfnm{E.~F.}\binits{E.~F.}},
  \bauthor{\bsnm{Craiu},~\bfnm{R.~V.}\binits{R.~V.}} \AND
  \bauthor{\bsnm{Yao},~\bfnm{F.}\binits{F.}}
(\byear{2013}).
\btitle{Statistical testing of covariate effects in conditional copula models}.
\bjournal{Electronic Journal of Statistics}
\bvolume{7}
\bpages{2822--2850}.
\end{barticle}
\endbibitem

\bibitem[\protect\citeauthoryear{Bedford and Cooke}{2001}]{Bedford2001}
\begin{barticle}[author]
\bauthor{\bsnm{Bedford},~\bfnm{T.}\binits{T.}} \AND
  \bauthor{\bsnm{Cooke},~\bfnm{R.~M.}\binits{R.~M.}}
(\byear{2001}).
\btitle{Probability density decomposition for conditionally dependent random
  variables modeled by vines}.
\bjournal{Annals of Mathematics and Artificial Intelligence}
\bvolume{32}
\bpages{245--268}.
\end{barticle}
\endbibitem

\bibitem[\protect\citeauthoryear{Bedford and Cooke}{2002}]{Bedford2002}
\begin{barticle}[author]
\bauthor{\bsnm{Bedford},~\bfnm{T.}\binits{T.}} \AND
  \bauthor{\bsnm{Cooke},~\bfnm{R.~M.}\binits{R.~M.}}
(\byear{2002}).
\btitle{Vines -- A new graphical model for dependent random variables}.
\bjournal{The Annals of Statistics}
\bvolume{30}
\bpages{1031--1068}.
\end{barticle}
\endbibitem

\bibitem[\protect\citeauthoryear{Bergsma}{2004}]{Bergsma2004}
\begin{bunpublished}[author]
\bauthor{\bsnm{Bergsma},~\bfnm{W.~P.}\binits{W.~P.}}
(\byear{2004}).
\btitle{Testing conditional independence for continuous random variables}.
\bnote{URL: \url{https://www.eurandom.tue.nl/reports/2004/048-report.pdf}}.
\end{bunpublished}
\endbibitem

\bibitem[\protect\citeauthoryear{Bollerslev}{1986}]{Bollerslev1986}
\begin{barticle}[author]
\bauthor{\bsnm{Bollerslev},~\bfnm{Tim}\binits{T.}}
(\byear{1986}).
\btitle{Generalized autoregressive conditional heteroskedasticity}.
\bjournal{Journal of Econometrics}
\bvolume{31}
\bpages{307--327}.
\end{barticle}
\endbibitem

\bibitem[\protect\citeauthoryear{Chan et~al.}{2009}]{Chan2009}
\begin{barticle}[author]
\bauthor{\bsnm{Chan},~\bfnm{Ngai-Hang}\binits{N.-H.}},
  \bauthor{\bsnm{Chen},~\bfnm{Jian}\binits{J.}},
  \bauthor{\bsnm{Chen},~\bfnm{Xiaohong}\binits{X.}},
  \bauthor{\bsnm{Fan},~\bfnm{Yanqin}\binits{Y.}} \AND
  \bauthor{\bsnm{Peng},~\bfnm{Liang}\binits{L.}}
(\byear{2009}).
\btitle{Statistical Inference for Multivariate Residual Copula of {GARCH}
  Models}.
\bjournal{Statistica Sinica}
\bvolume{19}
\bpages{53--70}.
\end{barticle}
\endbibitem

\bibitem[\protect\citeauthoryear{Chen and Fan}{2006}]{Chen2006}
\begin{barticle}[author]
\bauthor{\bsnm{Chen},~\bfnm{Xiaohong}\binits{X.}} \AND
  \bauthor{\bsnm{Fan},~\bfnm{Yanqin}\binits{Y.}}
(\byear{2006}).
\btitle{Estimation of copula-based semiparametric time series models}.
\bjournal{Journal of Econometrics}
\bvolume{130}
\bpages{307--335}.
\end{barticle}
\endbibitem

\bibitem[\protect\citeauthoryear{Cook and Johnson}{1986}]{Cook1986}
\begin{barticle}[author]
\bauthor{\bsnm{Cook},~\bfnm{R.~Dennis}\binits{R.~D.}} \AND
  \bauthor{\bsnm{Johnson},~\bfnm{Mark~E.}\binits{M.~E.}}
(\byear{1986}).
\btitle{Generalized Burr-Pareto-Logistic Distributions With Applications to a
  Uranium Exploration Data Set}.
\bjournal{Technometrics}
\bvolume{28}
\bpages{123--131}.
\end{barticle}
\endbibitem

\bibitem[\protect\citeauthoryear{{Derumigny} and
  {Fermanian}}{2017}]{Derumigny2016}
\begin{barticle}[author]
\bauthor{\bsnm{{Derumigny}},~\bfnm{A.}\binits{A.}} \AND
  \bauthor{\bsnm{{Fermanian}},~\bfnm{J.~D.}\binits{J.~D.}}
(\byear{2017}).
\btitle{About tests of the ``simplifying'' assumption for conditional copulas}.
\bjournal{Dependence Modeling}
\bvolume{5}
\bpages{154--197}.
\end{barticle}
\endbibitem

\bibitem[\protect\citeauthoryear{Di{\ss}mann et~al.}{2013}]{Dissmann2013}
\begin{barticle}[author]
\bauthor{\bsnm{Di{\ss}mann},~\bfnm{J.}\binits{J.}},
  \bauthor{\bsnm{Brechmann},~\bfnm{E.~C.}\binits{E.~C.}},
  \bauthor{\bsnm{Czado},~\bfnm{C.}\binits{C.}} \AND
  \bauthor{\bsnm{Kurowicka},~\bfnm{D.}\binits{D.}}
(\byear{2013}).
\btitle{Selecting and estimating regular vine copulae and application to
  financial returns}.
\bjournal{Computational Statistics \& Data Analysis}
\bvolume{59}
\bpages{52--69}.
\end{barticle}
\endbibitem

\bibitem[\protect\citeauthoryear{Engle}{1982}]{Engle1982}
\begin{barticle}[author]
\bauthor{\bsnm{Engle},~\bfnm{Robert~F.}\binits{R.~F.}}
(\byear{1982}).
\btitle{Autoregressive Conditional Heteroscedasticity with Estimates of the
  Variance of United Kingdom Inflation}.
\bjournal{Econometrica}
\bvolume{50}
\bpages{987--1007}.
\end{barticle}
\endbibitem

\bibitem[\protect\citeauthoryear{{Fithian}, {Sun} and
  {Taylor}}{2014}]{Fithian2014}
\begin{bunpublished}[author]
\bauthor{\bsnm{{Fithian}},~\bfnm{William}\binits{W.}},
  \bauthor{\bsnm{{Sun}},~\bfnm{Dennis}\binits{D.}} \AND
  \bauthor{\bsnm{{Taylor}},~\bfnm{Jonathan}\binits{J.}}
(\byear{2014}).
\btitle{{Optimal Inference After Model Selection}}.
\bnote{ArXiv e-prints
  \href{http://arxiv.org/abs/1410.2597}{\nolinkurl{arXiv:1410.2597}}}.
\end{bunpublished}
\endbibitem

\bibitem[\protect\citeauthoryear{Fr\'{e}chet}{1951}]{Frechet1951}
\begin{barticle}[author]
\bauthor{\bsnm{Fr\'{e}chet},~\bfnm{M.}\binits{M.}}
(\byear{1951}).
\btitle{Sur les tableaux de corr\'{e}lation dont les marges sont donn\'{e}es}.
\bjournal{Annales de l'Universit\'{e} de Lyon, Section A, Series 3}
\bvolume{14}
\bpages{53--77}.
\end{barticle}
\endbibitem

\bibitem[\protect\citeauthoryear{Genest, Ghoudi and Rivest}{1995}]{genest1995}
\begin{barticle}[author]
\bauthor{\bsnm{Genest},~\bfnm{C.}\binits{C.}},
  \bauthor{\bsnm{Ghoudi},~\bfnm{K.}\binits{K.}} \AND
  \bauthor{\bsnm{Rivest},~\bfnm{L.~P.}\binits{L.~P.}}
(\byear{1995}).
\btitle{A semiparametric estimation procedure of dependence parameters in
  multivariate families of distributions}.
\bjournal{Biometrika}
\bvolume{82}
\bpages{543--552}.
\end{barticle}
\endbibitem

\bibitem[\protect\citeauthoryear{Gijbels, Omelka and
  Veraverbeke}{2015}]{Gijbels2015b}
\begin{barticle}[author]
\bauthor{\bsnm{Gijbels},~\bfnm{Ir{\`e}ne}\binits{I.}},
  \bauthor{\bsnm{Omelka},~\bfnm{Marek}\binits{M.}} \AND
  \bauthor{\bsnm{Veraverbeke},~\bfnm{No{\"e}l}\binits{N.}}
(\byear{2015}).
\btitle{Partial and average copulas and association measures}.
\bjournal{Electronic Journal of Statistics}
\bvolume{9}
\bpages{2420--2474}.
\end{barticle}
\endbibitem

\bibitem[\protect\citeauthoryear{Gijbels, Omelka and
  Veraverbeke}{2017}]{Gijbels2016}
\begin{barticle}[author]
\bauthor{\bsnm{Gijbels},~\bfnm{Ir{\`e}ne}\binits{I.}},
  \bauthor{\bsnm{Omelka},~\bfnm{Marek}\binits{M.}} \AND
  \bauthor{\bsnm{Veraverbeke},~\bfnm{No{\"e}l}\binits{N.}}
(\byear{2017}).
\btitle{Nonparametric testing for no covariate effects in conditional copulas}.
\bjournal{Statistics}
\bvolume{51}
\bpages{475--509}.
\end{barticle}
\endbibitem

\bibitem[\protect\citeauthoryear{Gijbels et~al.}{2017}]{Gijbels2017}
\begin{barticle}[author]
\bauthor{\bsnm{Gijbels},~\bfnm{Ir{\`e}ne}\binits{I.}},
  \bauthor{\bsnm{Omelka},~\bfnm{Marek}\binits{M.}},
  \bauthor{\bsnm{Pe{\v{s}}ta},~\bfnm{Michal}\binits{M.}} \AND
  \bauthor{\bsnm{Veraverbeke},~\bfnm{No{\"e}l}\binits{N.}}
(\byear{2017}).
\btitle{Score tests for covariate effects in conditional copulas}.
\bjournal{Journal of Multivariate Analysis}
\bvolume{159}
\bpages{111--133}.
\end{barticle}
\endbibitem

\bibitem[\protect\citeauthoryear{Grothe and Nicklas}{2013}]{Grothe2013}
\begin{barticle}[author]
\bauthor{\bsnm{Grothe},~\bfnm{Oliver}\binits{O.}} \AND
  \bauthor{\bsnm{Nicklas},~\bfnm{Stephan}\binits{S.}}
(\byear{2013}).
\btitle{{V}ine constructions of {L}{\'e}vy copulas}.
\bjournal{Journal of Multivariate Analysis}
\bvolume{119}
\bpages{1--15}.
\end{barticle}
\endbibitem

\bibitem[\protect\citeauthoryear{{Hob{\ae}k Haff}}{2013}]{HobakHaff2013}
\begin{barticle}[author]
\bauthor{\bsnm{{Hob{\ae}k Haff}},~\bfnm{I.}\binits{I.}}
(\byear{2013}).
\btitle{Parameter estimation for pair-copula constructions}.
\bjournal{Bernoulli}
\bvolume{19}
\bpages{462--491}.
\end{barticle}
\endbibitem

\bibitem[\protect\citeauthoryear{{Hob{\ae}k Haff}, Aas and
  Frigessi}{2010}]{HobakHaff2010}
\begin{barticle}[author]
\bauthor{\bsnm{{Hob{\ae}k Haff}},~\bfnm{I.}\binits{I.}},
  \bauthor{\bsnm{Aas},~\bfnm{K.}\binits{K.}} \AND
  \bauthor{\bsnm{Frigessi},~\bfnm{A.}\binits{A.}}
(\byear{2010}).
\btitle{On the simplified pair-copula construction -- Simply useful or too
  simplistic?}
\bjournal{Journal of Multivariate Analysis}
\bvolume{101}
\bpages{1296--1310}.
\end{barticle}
\endbibitem

\bibitem[\protect\citeauthoryear{Hoeffding}{1940}]{Hoeffding1940}
\begin{bbook}[author]
\bauthor{\bsnm{Hoeffding},~\bfnm{W.}\binits{W.}}
(\byear{1940}).
\btitle{Ma{\ss}tabinvariante Korrelationstheorie}.
\bseries{Schriften des Mathematischen Instituts und des Instituts f{\"u}r
  Angewandte Mathematik an der Universit{\"a}t Berlin, 5}.
\end{bbook}
\endbibitem

\bibitem[\protect\citeauthoryear{Joe}{1996}]{Joe1996}
\begin{bincollection}[author]
\bauthor{\bsnm{Joe},~\bfnm{H.}\binits{H.}}
(\byear{1996}).
\btitle{Families of $m$-Variate Distributions With Given Margins and $m(m-1)/2$
  Bivariate Dependence Parameters}.
In \bbooktitle{Distributions with Fixed Marginals and Related Topics, IMS
  Lecture Notes -- Monograph Series}
(\beditor{\bfnm{L.}\binits{L.}~\bsnm{R{\"u}schendorf}},
  \beditor{\bfnm{B.}\binits{B.}~\bsnm{Schweizer}} \AND
  \beditor{\bfnm{M.~D.}\binits{M.~D.}~\bsnm{Taylor}}, eds.)
\bpages{120--141}.
\end{bincollection}
\endbibitem

\bibitem[\protect\citeauthoryear{Joe}{1997}]{Joe1997}
\begin{bbook}[author]
\bauthor{\bsnm{Joe},~\bfnm{Harry}\binits{H.}}
(\byear{1997}).
\btitle{{M}ultivariate Models and Dependence Concepts}.
\bpublisher{Chapman {\&} Hall}, \baddress{London}.
\end{bbook}
\endbibitem

\bibitem[\protect\citeauthoryear{Joe, Li and Nikoloulopoulos}{2010}]{Joe2010}
\begin{barticle}[author]
\bauthor{\bsnm{Joe},~\bfnm{Harry}\binits{H.}},
  \bauthor{\bsnm{Li},~\bfnm{Haijun}\binits{H.}} \AND
  \bauthor{\bsnm{Nikoloulopoulos},~\bfnm{Aristidis~K.}\binits{A.~K.}}
(\byear{2010}).
\btitle{{T}ail dependence functions and vine copulas}.
\bjournal{Journal of Multivariate Analysis}
\bvolume{101}
\bpages{252--270}.
\end{barticle}
\endbibitem

\bibitem[\protect\citeauthoryear{Kojadinovic and
  Holmes}{2009}]{Kojadinovic2009}
\begin{barticle}[author]
\bauthor{\bsnm{Kojadinovic},~\bfnm{I.}\binits{I.}} \AND
  \bauthor{\bsnm{Holmes},~\bfnm{M.}\binits{M.}}
(\byear{2009}).
\btitle{Tests of independence among continuous random vectors based on
  Cram{\'e}r-von Mises functionals of the empirical copula process}.
\bjournal{Journal of Multivariate Analysis}
\bvolume{100}
\bpages{1137--1154}.
\end{barticle}
\endbibitem

\bibitem[\protect\citeauthoryear{{Kraus} and {Czado}}{2017}]{Kraus2017}
\begin{bunpublished}[author]
\bauthor{\bsnm{{Kraus}},~\bfnm{D.}\binits{D.}} \AND
  \bauthor{\bsnm{{Czado}},~\bfnm{C.}\binits{C.}}
(\byear{2017}).
\btitle{{Growing simplified vine copula trees: improving Di{\ss}mann's
  algorithm}}.
\bnote{ArXiv e-prints
  \href{http://arxiv.org/abs/1703.05203}{\nolinkurl{arXiv:1703.05203}}}.
\end{bunpublished}
\endbibitem

\bibitem[\protect\citeauthoryear{Kurowicka and Joe}{2011}]{Kurowicka2011}
\begin{bbook}[author]
\beditor{\bsnm{Kurowicka},~\bfnm{Dorota}\binits{D.}} \AND
  \beditor{\bsnm{Joe},~\bfnm{Harry}\binits{H.}}, eds.
(\byear{2011}).
\btitle{{D}ependence Modeling}.
\bpublisher{World Scientific}, \baddress{Singapore}.
\end{bbook}
\endbibitem

\bibitem[\protect\citeauthoryear{Kurz}{2021}]{Kurz2017}
\begin{bmanual}[author]
\bauthor{\bsnm{Kurz},~\bfnm{Malte~S.}\binits{M.~S.}}
(\byear{2021}).
\btitle{pacotest: Testing for Partial Copulas and the Simplifying Assumption in
  Vine Copulas}
\bnote{URL: \url{https://CRAN.R-project.org/package=pacotest},
  \texttt{R}-Package version 0.4.1}.
\end{bmanual}
\endbibitem

\bibitem[\protect\citeauthoryear{Lavergne and Patilea}{2008}]{Lavergne2008}
\begin{barticle}[author]
\bauthor{\bsnm{Lavergne},~\bfnm{Pascal}\binits{P.}} \AND
  \bauthor{\bsnm{Patilea},~\bfnm{Valentin}\binits{V.}}
(\byear{2008}).
\btitle{Breaking the curse of dimensionality in nonparametric testing}.
\bjournal{Journal of Econometrics}
\bvolume{143}
\bpages{103--122}.
\end{barticle}
\endbibitem

\bibitem[\protect\citeauthoryear{Lee et~al.}{2016}]{Lee2016}
\begin{barticle}[author]
\bauthor{\bsnm{Lee},~\bfnm{Jason~D.}\binits{J.~D.}},
  \bauthor{\bsnm{Sun},~\bfnm{Dennis~L.}\binits{D.~L.}},
  \bauthor{\bsnm{Sun},~\bfnm{Yuekai}\binits{Y.}} \AND
  \bauthor{\bsnm{Taylor},~\bfnm{Jonathan~E.}\binits{J.~E.}}
(\byear{2016}).
\btitle{Exact post-selection inference, with application to the lasso}.
\bjournal{The Annals of Statistics}
\bvolume{44}
\bpages{907--927}.
\end{barticle}
\endbibitem

\bibitem[\protect\citeauthoryear{Mroz, Fuchs and Trutschnig}{2021}]{Mroz2020}
\begin{barticle}[author]
\bauthor{\bsnm{Mroz},~\bfnm{Thomas}\binits{T.}},
  \bauthor{\bsnm{Fuchs},~\bfnm{Sebastian}\binits{S.}} \AND
  \bauthor{\bsnm{Trutschnig},~\bfnm{Wolfgang}\binits{W.}}
(\byear{2021}).
\btitle{{How simplifying and flexible is the simplifying assumption in
  pair-copula constructions -- analytic answers in dimension three and a
  glimpse beyond}}.
\bjournal{Electronic Journal of Statistics}
\bvolume{15}
\bpages{1951--1992}.
\end{barticle}
\endbibitem

\bibitem[\protect\citeauthoryear{Nagler}{2018}]{kdecopula}
\begin{barticle}[author]
\bauthor{\bsnm{Nagler},~\bfnm{Thomas}\binits{T.}}
(\byear{2018}).
\btitle{{kdecopula}: An {R} Package for the Kernel Estimation of Bivariate
  Copula Densities}.
\bjournal{Journal of Statistical Software}
\bvolume{84}
\bpages{1--22}.
\end{barticle}
\endbibitem

\bibitem[\protect\citeauthoryear{Nagler and Czado}{2016}]{Nagler2016}
\begin{barticle}[author]
\bauthor{\bsnm{Nagler},~\bfnm{Thomas}\binits{T.}} \AND
  \bauthor{\bsnm{Czado},~\bfnm{Claudia}\binits{C.}}
(\byear{2016}).
\btitle{Evading the curse of dimensionality in nonparametric density estimation
  with simplified vine copulas}.
\bjournal{Journal of Multivariate Analysis}
\bvolume{151}
\bpages{69--89}.
\end{barticle}
\endbibitem

\bibitem[\protect\citeauthoryear{Nagler et~al.}{2021}]{Schepsmeier2017}
\begin{bmanual}[author]
\bauthor{\bsnm{Nagler},~\bfnm{Thomas}\binits{T.}},
  \bauthor{\bsnm{Schepsmeier},~\bfnm{Ulf}\binits{U.}},
  \bauthor{\bsnm{Stoeber},~\bfnm{Jakob}\binits{J.}},
  \bauthor{\bsnm{Brechmann},~\bfnm{Eike~Christian}\binits{E.~C.}},
  \bauthor{\bsnm{Graeler},~\bfnm{Benedikt}\binits{B.}} \AND
  \bauthor{\bsnm{Erhardt},~\bfnm{Tobias}\binits{T.}}
(\byear{2021}).
\btitle{VineCopula: Statistical Inference of Vine Copulas}
\bnote{URL: \url{https://CRAN.R-project.org/package=VineCopula},
  \texttt{R}-Package version 2.4.2}.
\end{bmanual}
\endbibitem

\bibitem[\protect\citeauthoryear{Newey and McFadden}{1994}]{Newey1994}
\begin{barticle}[author]
\bauthor{\bsnm{Newey},~\bfnm{Whitney~K}\binits{W.~K.}} \AND
  \bauthor{\bsnm{McFadden},~\bfnm{Daniel}\binits{D.}}
(\byear{1994}).
\btitle{Large sample estimation and hypothesis testing}.
\bjournal{Handbook of Econometrics}
\bvolume{4}
\bpages{2111--2245}.
\end{barticle}
\endbibitem

\bibitem[\protect\citeauthoryear{Patton}{2006}]{Patton2006}
\begin{barticle}[author]
\bauthor{\bsnm{Patton},~\bfnm{A.~J.}\binits{A.~J.}}
(\byear{2006}).
\btitle{Modelling asymmetric exchange rate dependence}.
\bjournal{International Economic Review}
\bvolume{47}
\bpages{527--556}.
\end{barticle}
\endbibitem

\bibitem[\protect\citeauthoryear{Patton}{2012}]{Patton2012}
\begin{barticle}[author]
\bauthor{\bsnm{Patton},~\bfnm{Andrew~J.}\binits{A.~J.}}
(\byear{2012}).
\btitle{A review of copula models for economic time series}.
\bjournal{Journal of Multivariate Analysis}
\bvolume{110}
\bpages{4--18}.
\end{barticle}
\endbibitem

\bibitem[\protect\citeauthoryear{Quessy}{2010}]{Quessy2010}
\begin{barticle}[author]
\bauthor{\bsnm{Quessy},~\bfnm{J.~F.}\binits{J.~F.}}
(\byear{2010}).
\btitle{Applications and asymptotic power of marginal-free tests of stochastic
  vectorial independence}.
\bjournal{Journal of Statistical Planning and Inference}
\bvolume{140}
\bpages{3058--3075}.
\end{barticle}
\endbibitem

\bibitem[\protect\citeauthoryear{Schellhase and Spanhel}{2018}]{Schellhase2017}
\begin{barticle}[author]
\bauthor{\bsnm{Schellhase},~\bfnm{Christian}\binits{C.}} \AND
  \bauthor{\bsnm{Spanhel},~\bfnm{Fabian}\binits{F.}}
(\byear{2018}).
\btitle{Estimating non-simplified vine copulas using penalized splines}.
\bjournal{Statistics and Computing}
\bvolume{28}
\bpages{387--409}.
\end{barticle}
\endbibitem

\bibitem[\protect\citeauthoryear{Spanhel and Kurz}{2016a}]{Spanhel2016b}
\begin{barticle}[author]
\bauthor{\bsnm{Spanhel},~\bfnm{F.}\binits{F.}} \AND
  \bauthor{\bsnm{Kurz},~\bfnm{M.~S.}\binits{M.~S.}}
(\byear{2016}a).
\btitle{{The partial copula: Properties and associated dependence measures}}.
\bjournal{{S}tatistics \& {P}robability Letters}
\bvolume{119}
\bpages{76--83}.
\end{barticle}
\endbibitem

\bibitem[\protect\citeauthoryear{Spanhel and Kurz}{2016b}]{Spanhel2016c}
\begin{bunpublished}[author]
\bauthor{\bsnm{Spanhel},~\bfnm{F.}\binits{F.}} \AND
  \bauthor{\bsnm{Kurz},~\bfnm{M.~S.}\binits{M.~S.}}
(\byear{2016}b).
\btitle{Estimating parametric simplified vine copulas under general
  conditions}.
\bnote{unpublished working paper}.
\end{bunpublished}
\endbibitem

\bibitem[\protect\citeauthoryear{Spanhel and Kurz}{2019}]{Spanhel2016}
\begin{barticle}[author]
\bauthor{\bsnm{Spanhel},~\bfnm{Fabian}\binits{F.}} \AND
  \bauthor{\bsnm{Kurz},~\bfnm{Malte~S.}\binits{M.~S.}}
(\byear{2019}).
\btitle{Simplified vine copula models: Approximations based on the simplifying
  assumption}.
\bjournal{Electronic Journal of Statistics}
\bvolume{13}
\bpages{1254--1291}.
\end{barticle}
\endbibitem

\bibitem[\protect\citeauthoryear{St\"{o}ber, Joe and Czado}{2013}]{Stoeber2013}
\begin{barticle}[author]
\bauthor{\bsnm{St\"{o}ber},~\bfnm{J.}\binits{J.}},
  \bauthor{\bsnm{Joe},~\bfnm{H.}\binits{H.}} \AND
  \bauthor{\bsnm{Czado},~\bfnm{C.}\binits{C.}}
(\byear{2013}).
\btitle{Simplified pair copula constructions -- Limitations and extensions}.
\bjournal{Journal of Multivariate Analysis}
\bvolume{119}
\bpages{101--118}.
\end{barticle}
\endbibitem

\bibitem[\protect\citeauthoryear{St{\"o}ber and
  Schepsmeier}{2013}]{Stoeber2013c}
\begin{barticle}[author]
\bauthor{\bsnm{St{\"o}ber},~\bfnm{J.}\binits{J.}} \AND
  \bauthor{\bsnm{Schepsmeier},~\bfnm{U.}\binits{U.}}
(\byear{2013}).
\btitle{Estimating standard errors in regular vine copula models}.
\bjournal{Computational Statistics}
\bvolume{28}
\bpages{2679--2707}.
\end{barticle}
\endbibitem

\bibitem[\protect\citeauthoryear{Tsukahara}{2005}]{Tsukahara2005}
\begin{barticle}[author]
\bauthor{\bsnm{Tsukahara},~\bfnm{H.}\binits{H.}}
(\byear{2005}).
\btitle{Semiparametric estimation in copula models}.
\bjournal{Canadian Journal of Statistics}
\bvolume{33}
\bpages{357--375}.
\end{barticle}
\endbibitem

\end{thebibliography}

\end{document}